\documentclass[1p]{elsarticle}
\usepackage[colorlinks]{hyperref}%  hyperlinks [must be loaded after dropping]
\usepackage{graphicx}
\usepackage{float}
\usepackage{subfigure}% subcaptions for subfigures
\usepackage{subfigmat}% matrices of similar subfigures,
\usepackage{amsmath}
\usepackage{amssymb}
\usepackage{bm}

\usepackage{silence}
\usepackage{etoolbox}
\makeatletter
\patchcmd{\ps@pprintTitle}% <cmd>
  {Preprint submitted}% <search>
  {Updated manuscript submitted}% <replace>
  {}{}% <succes><failure>
\makeatother
\journal{arXiv}

\DeclareMathOperator{\sign}{sign}

\begin{document}

\begin{frontmatter}

\title{High-Order Localized Dissipation Weighted Compact Nonlinear Scheme for Shock- and Interface-Capturing in Compressible Flows}

\author[AA_address]{Man Long Wong}

\author[AA_address,ME_address]{Sanjiva K. Lele}

\address[AA_address]{Department of Aeronautics and Astronautics, Stanford University, Stanford, CA 94305, USA}
\address[ME_address]{Department of Mechanical Engineering, Stanford University, Stanford, CA 94305, USA}

\begin{abstract}
Simulations of single- and multi-species compressible flows with shock waves and discontinuities are conducted using a weighted compact nonlinear scheme (WCNS) with a newly developed sixth order localized dissipative interpolation. In smooth regions, the scheme applies the central nonlinear interpolation with minimum dissipation to resolve fluctuating flow features while in regions containing discontinuities and high wavenumber features, the scheme suppresses spurious numerical oscillations by hybridizing the central interpolation with the more dissipative upwind-biased nonlinear interpolation. In capturing material interfaces between species of different densities, a quasi-conservative five equation model that can conserve mass of each species is used to prevent pressure oscillations across the interfaces. Compared to upwind-biased interpolations with classical nonlinear weights~\cite{jiang1995efficient,deng2000developing} and improved weights~\cite{borges2008improved}, and the interpolation with adaptive central-upwind weights for scale-separation~\cite{hu2011scale}, it is shown that WCNS with the proposed localized dissipative interpolation has better performance to simultaneously capture discontinuities and resolve smooth features.
\end{abstract}

\begin{keyword}
weighted compact nonlinear scheme (WCNS), weighted essentially non-oscillatory (WENO) interpolation, high-order method, shock-capturing, interface-capturing, multi-species flows, localized dissipation
\end{keyword}

\end{frontmatter}

\section{Introduction}

In direct numerical simulation (DNS) or large eddy simulation (LES) of high speed turbulent flows, the coexistence of discontinuities and turbulent features poses a challenge to obtain accurate and stable solutions. In simulations of flows involving discontinuities such as shock waves and material interfaces between fluids, Gibbs phenomenon or spurious oscillations appear in the solutions near the discontinuities if the computations are conducted without any regularization. A common way to cure the spurious oscillations is to smear the discontinuities with the addition of certain amount of numerical dissipation. However, the addition of dissipation can damp the small-scale turbulent eddies whose kinetic energy or amplitude is much smaller than the energy-bearing features in the turbulent field. In previous decades, lots of high-order accurate shock-capturing schemes were developed using different methodologies. In regions around discontinuities, these schemes add numerical dissipation locally while in smooth flow regions, these shock-capturing methods adaptively become less dissipative to preserve turbulent features.

Among high-order shock-capturing methods, a popular family of schemes is the weighted essentially non-oscillatory (WENO) schemes first introduced by Jiang and Shu~\cite{jiang1995efficient}. These schemes are famous for their robustness to capture discontinuities without spurious oscillations and their ability to achieve arbitrarily high formal order of accuracy in smooth flows. Nevertheless, in the comparison with other high-order shock capturing schemes~\cite{johnsen2010assessment}, it was shown that these schemes dissipate turbulent fluctuations significantly due to upwind-biased flux reconstructions. Over the years, various versions of the WENO schemes have been proposed to improve the excessively dissipative nature of the schemes. WENO-M~\cite{henrick2005mapped} and WENO-Z~\cite{borges2008improved} schemes minimize the excessive dissipation of traditional WENO scheme at critical points using improved nonlinear weighting functions. The WENO-SYMOO~\cite{martin2006bandwidth} and WENO-CU6~\cite{hu2010adaptive} schemes include a downwind stencil to optimize the resolution and dissipation for turbulence simulations.

Another well-known high-resolution and computationally efficient method to compute turbulence problems is the family of central compact schemes developed by Lele~\cite{lele1992compact}. Through Fourier analysis, it was shown that these compact schemes provide better resolution for small-scale waves than explicit finite difference schemes of the same order of accuracy. Owing to their non-dissipative nature, these schemes have to be used conjointly with filters to suppress high wavenumber spurious oscillations. Despite the application of filters, spurious oscillations may still occur if there are large gradients in the solutions. Two approaches were developed to regularize compact schemes for the simulation of compressible and multi-species flows in the presence of discontinuities.

In the first approach, localized artificial fluid transport properties were used to stabilize solutions near regions containing discontinuities. Cook and Cabot~\cite{cook2004high, cook2005hyperviscosity} were the first to successfully demonstrate the use of artificial shear and bulk viscosities to stabilize solutions near shocks in compact schemes. Since then, the methodology was extended to the simulations of multi-species flows~\cite{cook2007artificial, fiorina2007artificial, shankar2010numerical} and flows in generalized curvilinear coordinates~\cite{kawai2008localized}.

Another approach to capture discontinuities in compact schemes is to incorporate WENO limiting technique into compact schemes. In Pirozzoli's work~\cite{pirozzoli2002conservative}, a conservative formulation was proposed to hybridize a compact upwind numerical flux explicitly with the WENO flux. The hybrid method was later further improved by Ren et al.~\cite{ren2003characteristic}. On the other side, Deng et al.~\cite{deng2000developing} developed the weighted compact nonlinear schemes (WCNS's) by integrating WENO interpolation implicitly into the cell-centered compact schemes. Nonomura et al.~\cite{nonomura2007increasing} and Zhang et al.~\cite{zhang2008development} later increased the order of accuracy of WCNS's. Recently, Liu et al.~\cite{liu2015new} proposed a new family of WCNS's with hybrid linear weights for WENO interpolation. The hybrid weighted interpolation method extends the adaptive central-upwind nonlinear weighting technique designed by Hu et al.~\cite{hu2010adaptive} for WENO-CU6 scheme. It was shown that WCNS's with hybrid weighted interpolation have more localized dissipation than the classical WENO schemes. Although WCNS's are variants of WENO schemes, they have several advantages over the latter schemes: (1) WCNS's generally have higher resolution than the WENO schemes at the same order of accuracy so they can capture high wavenumber waves better; (2) WCNS's are also more flexible in the choice of flux splitting methods since they maintain high order of accuracy even if flux difference splitting methods such as the HLLC or Roe methods are used for computing fluxes at cell midpoints, while the regular finite difference WENO schemes will reduce to second order accuracy under these conditions for multi-dimensional problems; (3) different explicit or implicit compact finite difference methods can be chosen for the computation of flux derivatives in WCNS's.

Unlike simulations for single-species flows, additional effort has to be made to prevent the appearance of numerical instabilities at material interfaces in the simulations of multi-component flows. Instead of conservative variables, Johnsen et al.~\cite{johnsen2006implementation} showed that WENO reconstructions should be carried out on primitive variables to maintain pressure equilibrium at material interfaces. Furthermore, a non-conservative advection equation to describe the material interfaces is essential for stability at material interfaces. In order to maintain high order of accuracy with HLLC Riemann solver during flux reconstruction, they solved the quasi-conservative system with the more expensive finite volume WENO scheme instead of finite difference WENO schemes. Since using flux difference splitting does not degenerate the formal order of accuracy of the WCNS's, Nonomura et al.~\cite{nonomura2012numerical} applied the same interface capturing technique on WCNS's and reduced a significant amount of computational cost compared to finite volume WENO scheme. However, the classical upwind-biased WENO interpolation method used by them in WCNS is known to be very dissipative and not well-suited for turbulent flow simulations. Besides, the four-equation model used by them does not conserve mass of each species.

In this paper, we propose a new form of nonlinear weights for WENO interpolation in WCNS's that can introduce dissipation more locally around shock waves and discontinuities. Instead of following Liu et al.~\cite{liu2015new} to interpolate the fluxes, we follow Johnsen et al.~\cite{johnsen2006implementation}, Coralic et al.~\cite{coralic2014finite}, and Nonomura et al.~\cite{nonomura2012numerical} to interpolate characteristic variables projected from the primitive variables to prevent spurious pressure oscillations at material interfaces. Furthermore, we choose the five-equation model developed by Allaire et al.~\cite{allaire2002five} and follow Coralic et al.~\cite{coralic2014finite} to solve the equations using an improved HLLC type Riemann solver for interface capturing. Through different test problems, it is shown that the proposed interpolation method can well capture both small-scale fluctuating features and sharp discontinuities.

\section{Governing equations}

\subsection{Single-species flows}
The Euler system of equations for simulating single-species, inviscid, non-conducting, and compressible flows is given by:
\begin{equation}
\begin{aligned}
	\frac{\partial{\rho}}{\partial{t}} + \frac{\partial}{\partial{x_j}} \left(\rho u_j \right) &= 0, \\
    \frac{\partial{\rho u_i}}{\partial{t}} + \frac{\partial}{\partial{x_j}} \left( \rho u_i u_j + p\delta_{ij} \right) &= 0, \\
    \frac{\partial{E}}{\partial{t}} + \frac{\partial}{\partial{x_j}} \left( u_j \left( E + p \right) \right) &= 0,
\end{aligned}
\end{equation}
\noindent where $\rho$, $u_i$, $p$, and $E$ are the density, velocity vector, pressure, and total energy per unit volume of the fluid respectively. $E = \rho (e + u_i u_i/2)$, where $e$ is the specific internal energy. The system of equations is closed with the ideal gas equation of state:
\begin{equation} \label{eq:ideal_EOS}
	p = \left( \gamma - 1 \right) \left( E - \frac{\rho u_i u_i}{2} \right),
\end{equation}
\noindent where $\gamma$ is the ratio of specific heats of the fluid.

\subsection{Multi-species flows}
To model two-fluid flows, the seven-equation model proposed by Baer and Nunziato~\cite{baer1986two} is the most complete model. In Baer and Nunziato's model, conservation equations of the mass, momentum and energy are solved for each species and an additional transport equation is solved to describe the topology of the fluid-fluid interface. However, their model is computationally very expensive and arguably retains redundant information. The simplest family of models to describe two-fluid flows is the four-equation model which consists of equations of mass, momentum and energy for the mixture of fluids as a whole and one transport equation. In order to suppress pressure oscillations across material interfaces, different quantities were proposed for the transport equation in non-conservative advection form. Abgrall~\cite{abgrall1996prevent} and Shyue~\cite{shyue1998efficient} respectively suggested $1/(\gamma - 1)$ or $Y$ to be solved in the transport equation for pressure equilibrium across material interfaces, where $\gamma$ is the ratio of specific heats of the mixture and $Y$ is the mass fraction of one of the species. However, Abgrall's model has a technical problem that interfaces cannot be described if both fluids have the same value of $\gamma$ and Shyue's model has a strong assumption that molecular masses of the two fluids are the same. Worse still, neither of the models conserves mass of each species discretely. Another family of reduced models that are able to conserve the mass of each species and maintain pressure equilibrium at interfaces is the five-equation model. The five-equation model proposed by Allaire et al.~\cite{allaire2002five} for two immiscible, inviscid, and non-conducting fluids in the following form is used in present work:
\begin{equation}
\begin{aligned}
	\frac{\partial{Z_1 \rho_1}}{\partial{t}} + \frac{\partial}{\partial{x_j}} \left( Z_1 \rho_1 u_j \right) &= 0, \\
    \frac{\partial{Z_2 \rho_2}}{\partial{t}} + \frac{\partial}{\partial{x_j}} \left( Z_2 \rho_2 u_j \right) &= 0, \\
    \frac{\partial{\rho u_i}}{\partial{t}} + \frac{\partial}{\partial{x_j}} \left( \rho u_i u_j + p\delta_{ij} \right) &= 0, \\
    \frac{\partial{E}}{\partial{t}} + \frac{\partial}{\partial{x_j}} \left( u_j \left( E + p \right) \right) &= 0, \\
        \frac{\partial{Z_1}}{\partial{t}} + u_j \frac{\partial{Z_1}}{\partial{x_j}} &= 0,
\end{aligned}
\end{equation}

\noindent where $\rho_1$ and $\rho_2$ are the densities of fluids 1 and 2 respectively. $\rho$, $u_i$, $p$, and $E$ are the density, velocity vector, pressure, and total energy per unit volume of the mixture respectively. $Z_1$ is the volume fraction of fluid 1. The volume fractions of the two fluids $Z_1$ and $Z_2$ are related by:
\begin{equation}
	Z_2 = 1 - Z_1.
\end{equation}
The ideal equation of state given by equation~\eqref{eq:ideal_EOS} is used to close the system. By using the isobaric assumption, we are able to derive an explicit mixture rule for the ratio of specific heats $\gamma$ of the mixture:
\begin{equation}
	\frac{1}{\gamma - 1} = \frac{Z_1}{\gamma_{1} - 1} + \frac{Z_2}{\gamma_{2} - 1},
\end{equation}

\noindent where $\gamma_1$ and $\gamma_2$ are the ratios of specific heats of fluids 1 and 2 respectively. In the absence of surface tension, the isobaric assumption is consistent with pressure equilibrium across material interfaces.

The transport equation of volume fraction is solved in advection form. Following the approach proposed by Johnsen et al.~\cite{johnsen2006implementation} and extended by Coralic et al.~\cite{coralic2014finite}, the following mathematically equivalent form of the advection equation is used for the adaptation of a HLLC-type Riemann solver to compute fluxes at midpoints between cell nodes:
\begin{equation}
	\frac{\partial{Z_1}}{\partial{t}} + \frac{\partial}{\partial{x_j}} \left( Z_1 u_j \right) = Z_1 \frac{\partial u_j}{\partial{x_j}}.
\end{equation}

\section{Numerical methods}

\subsection{Scheme formulation}
The governing equations of both single-species flows and two-species flows with five-equation models can be written in vector notation. In a three-dimensional (3D) space, we have:
\begin{equation} \label{eq:conservative_form}
	\frac{\partial{\bm{Q}}}{\partial{t}} + \frac{\partial{\bm{F(Q)}}}{\partial{x}} + \frac{\partial{\bm{G(Q)}}}{\partial{y}} +
\frac{\partial{\bm{H(Q)}}}{\partial{z}}= \bm{S(Q)},
\end{equation}
where $\bm{Q}$, $\bm{F}$, $\bm{G}$, $\bm{H}$, and $\bm{S}$ are the vectors of conservative variables, fluxes in the $x$, $y$, and $z$ directions, and sources, respectively.  For single-species flow, $\bm{Q} = (\rho, \rho u, \rho v, \rho w, E)^T$, $\bm{F} = (\rho u, \rho u^2 + p, \rho u v, \rho u w, u(E + p))^T$, $\bm{G} = (\rho v, \rho vu, \rho v^2 + p, \rho vw, v(E + p))^T$, $\bm{H} = (\rho w, \rho wu, \rho wv, \rho w^2 + p, w(E + p))^T$, and $\bm{S} = \bm{0}$. For two-species flow with five-equation model, $\bm{Q} = (Z_1 \rho_1, Z_2 \rho_2, \rho u, \rho v, \rho w, E, Z_1)^T$, $\bm{F} = (Z_1 \rho_1 u, Z_2 \rho_2 u, \rho u^2 + p, \rho u v, \rho u w, u(E + p), Z_1 u)^T$, $\bm{G} = (Z_1 \rho_1 v,  Z_2 \rho_2 v, \rho vu,$ $\rho v^2 + p, \rho vw, v(E + p), Z_1 v)^T$, $\bm{H} = (Z_1 \rho_1 w, Z_2 \rho_2 w, \rho wu, $ $\rho wv, \rho w^2 + p, w(E + p), Z_1 w)^T$, and $\bm{S} = (0, 0, 0, 0, 0, 0, Z_1 \nabla \cdot \bm{u})$. $u$, $v$, and $w$ are the components of velocity $\bm{u}$ in the $x$, $y$, and $z$ directions respectively. The one-dimensional (1D) and two-dimensional (2D) governing equations are only special cases of the 3D equations.

For simplicity, we only consider a scalar conservation law with source term in a 3D domain $[x_a, x_b]\times[y_a, y_b]\times[z_a,z_b]$ in this section:
\begin{equation}
	\frac{\partial Q}{\partial t} + \frac{\partial F(Q)}{\partial x} + \frac{\partial G(Q)}{\partial y} + \frac{\partial H(Q)}{\partial z} = S(Q),
\end{equation}
\noindent where $Q$, $F$, $G$, $H$, and $S$ denote the scalar conservative variable, fluxes in the $x$, $y$, and $z$ directions, and source term respectively. If the domain is discretized uniformly into a Cartesian grid with $N_x \times N_y \times N_z$ points, we have the domain covered by cells $I_{i,j,k}=\left[x_{i-\frac{1}{2}},x_{i+\frac{1}{2}}\right]\times\left[y_{j-\frac{1}{2}},y_{j+\frac{1}{2}}\right]\times\left[z_{k-\frac{1}{2}},z_{k+\frac{1}{2}}\right]$ for $1 \leq i \leq N_x$, $1 \leq j \leq N_y$, $1 \leq k \leq N_z$, where:
\begin{equation}
  x_{i+\frac{1}{2}} = x_a + i \Delta x, \quad
  y_{j+\frac{1}{2}} = y_a + j \Delta y, \quad
  z_{k+\frac{1}{2}} = z_a + k \Delta z,
\end{equation}
\noindent and
\begin{equation}
	\Delta x = \frac{x_b - x_a}{N_x}, \quad \Delta y = \frac{y_b - y_a}{N_y}, \quad \Delta z = \frac{z_b - z_a}{N_z}.
\end{equation}
\noindent The semi-discrete finite difference scheme can be written as:
\begin{equation}
	\left. \frac{\partial Q}{\partial t} \right|_{i,j,k} + \left. \widehat{\frac{\partial F}{\partial x}} \right|_{i,j,k} + \left. \widehat{\frac{\partial G}{\partial y}} \right|_{i,j,k} + \left. \widehat{\frac{\partial H}{\partial z}} \right|_{i,j,k} = \left. \widehat{S} \right|_{i,j,k},
\end{equation}
\noindent where $\partial Q / \partial t \big| _{i,j,k} = \partial Q_{i,j,k} / \partial t$, $\widehat{\partial F / \partial x} \big| _{i,j,k}$, $\widehat{\partial G / \partial y} \big| _{i,j,k}$, $\widehat{\partial H / \partial z} \big|_{i,j,k}$, and $\widehat{S} |_{i,j,k}$ are derivative of solution $Q$, approximations of the spatial flux derivatives in the $x$, $y$, and $z$ directions, and source term at grid point $(x_i, y_j, z_k)$ where:
\begin{equation}
	x_i = \frac{x_{i-\frac{1}{2}} + x_{i+\frac{1}{2}}}{2}, \quad
    y_j = \frac{y_{j-\frac{1}{2}} + y_{j+\frac{1}{2}}}{2}, \quad
    z_k = \frac{z_{k-\frac{1}{2}} + z_{k+\frac{1}{2}}}{2}.
\end{equation}

\subsection{Midpoint-and-node-to-node finite difference scheme}

Traditional WCNS's have the drawback that they are less robust than WENO schemes due to higher propensity to blow up compared to WENO schemes~\cite{nonomura2013robust} as the density or pressure become negative, or when the mass fractions or volume fractions are outside the bound between zero and one during either the nonlinear interpolation or the numerical time stepping processes. To overcome the first difficulty, first order interpolation can be used when the potential for the interpolated density or pressure to become negative, or mass fractions or volume fractions to become outside the bound with a tolerance has been detected. Regarding the second problem due to time stepping, Nonomura et al.~\cite{nonomura2013robust} proposed a family of robust midpoint-and-node-to-node differencing (MND) schemes (up to tenth order accurate) for the  computation of the first order derivative $\widehat{ \partial F / \partial x } \big| _{i,j,k}$. In this paper, the sixth order explicit MND scheme is used. The corresponding approximation of the derivative of the flux in the $x$ direction is given by:

\begin{equation}
\begin{aligned}
	\widehat{\frac{\partial F}{\partial x}} \bigg|_{i,j,k} =& 
    \frac{1}{\Delta x}  \left[
      \frac{3}{2} \left(\tilde{F}_{i+\frac{1}{2},j,k} - \tilde{F}_{i-\frac{1}{2},j,k} \right)
      - \frac{3}{10} \left(F_{i+1,j,k} - F_{i-1,j,k} \right) \right. \\
     & \left. + \frac{1}{30} \left(\tilde{F}_{i+\frac{3}{2},j,k} - \tilde{F}_{i-\frac{3}{2},j,k} \right)
    \right],
\end{aligned}
\end{equation}

\noindent where $\tilde{F}_{i+\frac{1}{2},j,k}$ are fluxes approximated at midpoints between cell nodes by nonlinear WENO interpolations discussed in next few sections and $F_{i,j,k}$ are the fluxes at cell nodes. The approximations of the derivatives in the y and z directions are similar. It was shown by Nonomura et al. that this compact formulation of finite difference scheme with WENO interpolation is numerically more stable than the original implicit or explicit compact finite difference schemes used by Deng et al.~\cite{deng2000developing}

\subsection{Classical upwind-biased (JS) nonlinear interpolation}
\begin{figure}[!ht]
\centering
\includegraphics[width=0.8\textwidth]{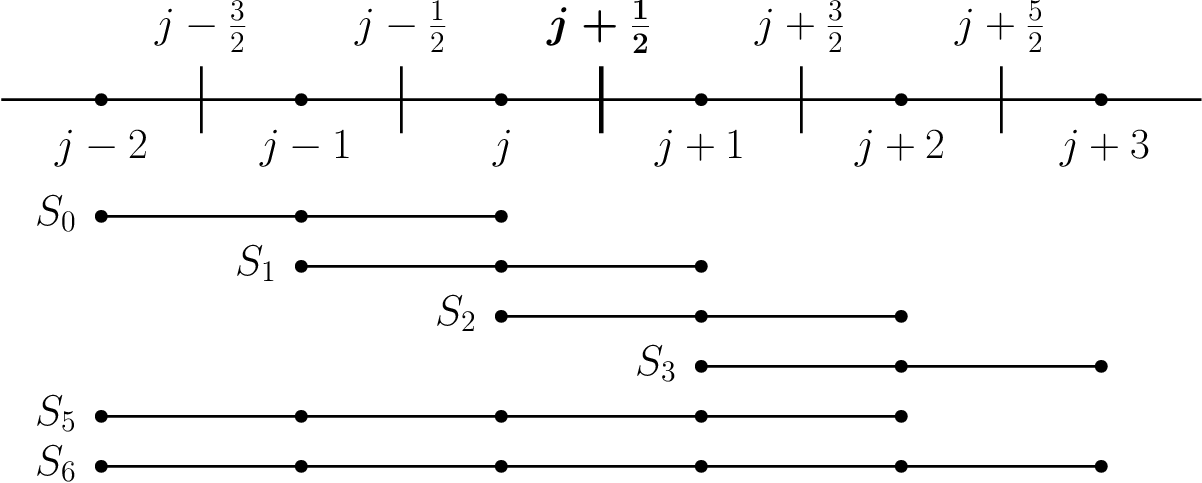}
\caption{Different stencils for approximating variable $u$ at the cell midpoint $j+1/2$.}
\label{fig:stencils}
\end{figure}

Classical WCNS's proposed by Deng et al.~\cite{deng2000developing} approximates fluxes at midpoint between cell nodes with a fifth order upwind-biased WENO interpolation originated from the nonlinear weighting technique in Jiang and Shu's~\cite{jiang1995efficient} fifth order WENO scheme (WENO5-JS). The upwind-biased WENO interpolation, which is called JS interpolation, is a fifth order interpolation nonlinearly weighted from three different third order interpolations on sub-stencils, $S_0$, $S_1$, and $S_2$, which are shown in figure~\ref{fig:stencils}. For simplicity, we only present the interpolation of variables on the left side of the cell midpoint at $x_{j+\frac{1}{2}}$ in a 1D domain in this paper. The three third order interpolations of a variable $u$ are given by:
\begin{equation}
\begin{aligned} \label{eq:upwind_biased_stencils}
	\tilde{u}_{j+\frac{1}{2}}^{(0)} &= \frac{1}{8}\left(3u_{j-2} - 10u_{j-1} + 15u_{j} \right), \\
    \tilde{u}_{j+\frac{1}{2}}^{(1)} &= \frac{1}{8}\left(-u_{j-1} + 6u_{j} + 3u_{j+1} \right), \\
    \tilde{u}_{j+\frac{1}{2}}^{(2)} &= \frac{1}{8}\left(3u_{j} + 6u_{j+1} - u_{j+2} \right),
\end{aligned}
\end{equation}
where $\tilde{u}_{j+\frac{1}{2}}^{(k)}$ are approximated values at cell midpoint from different sub-stencils and $u_j$ are the values at cell nodes. The variable $u$ can either be fluxes, conservative variables, primitive variables or variables that are projected to the characteristic fields. In Johnsen et al.~\cite{johnsen2006implementation} and Nonomura et al.~\cite{nonomura2012numerical}, it is shown that WENO reconstruction and WENO interpolation of primitive variables can suppress pressure oscillations at material interfaces. Moreover, projecting variables to the local characteristic fields before reconstruction and interpolation can improve the numerical stability at discontinuities. As a result, the primitive variables projected to the characteristic fields are employed in the interpolation process in this work. If the complete five-point stencil $S_5$ shown in figure~\ref{fig:stencils} is used, a fifth order upwind-biased linear interpolation can be formed:
\begin{equation} \label{eq:linear_upwind_interpolation}
	\tilde{u}_{j+\frac{1}{2}}^{\mathrm{upwind}} = \frac{1}{128} \left(3u_{j-2} - 20u_{j-1} + 90u_j + 60u_{j+1} - 5u_{j+2} \right).
\end{equation}

\noindent The fifth order upwind-biased linear interpolation can also be constructed linearly from the three third order interpolations:
\begin{equation}
	\tilde{u}_{j+\frac{1}{2}}^{\mathrm{upwind}} = \sum\limits_{k=0}^{2} d_k^{\mathrm{upwind}} \tilde{u}_{j+\frac{1}{2}}^{(k)},
\end{equation}

\noindent where $d_k^{\mathrm{upwind}}$ are the linear weights. $d_k^{\mathrm{upwind}}$ are given by:
\begin{equation} \label{eq:linear_upwind_weights}
	d_0^{\mathrm{upwind}} = \frac{1}{16}, \:
    d_1^{\mathrm{upwind}} = \frac{10}{16}, \:
    d_2^{\mathrm{upwind}} = \frac{5}{16}.
\end{equation}

\noindent A MND scheme with the upwind-biased linear interpolation in equation~\eqref{eq:linear_upwind_interpolation} will generate oscillations around shock waves and discontinuities. Based on the nonlinear weighting idea in WENO schemes~\cite{deng2000developing}, the nonlinear version of upwind-biased interpolation was first used in the classical WCNS to capture shock waves and discontinuities:
\begin{equation}
	\tilde{u}_{j+\frac{1}{2}} = \sum\limits_{k=0}^{2} \omega_k \tilde{u}_{j+\frac{1}{2}}^{(k)},
\end{equation}
where $\omega_k$ are the nonlinear weights. The nonlinear weights use the formulations by Jiang and Shu~\cite{jiang1995efficient}:
\begin{equation} \label{eq:JS5_nonlinear_weights}
	\omega_k = \frac{\alpha_k}{\sum\limits_{k=0}^{2}\alpha_k}, \:
    \alpha_k = \frac{d_k^{\mathrm{upwind}}}{\left(\beta_k + \epsilon \right)^p}, \: k = 0, 1, 2,
\end{equation}
where $p$ and $\beta_k$ are a positive integer and smoothness indicators, respectively. $\epsilon = 1.0\mathrm{e}{-40}$ is a small constant and is also used in other equations in this work to prevent division by zero. $p = 2$ is chosen in this paper for the nonlinear weights. The smoothness indicators are defined by:
\begin{equation}
	\beta_k = \sum^{2}_{l=1} \int^{x_{j+\frac{1}{2}}}_{x_{j-\frac{1}{2}}} \Delta x^{2l-1} \left( \frac{\partial^{l}}{\partial x^l} \tilde{u}^{(k)}(x) \right)^2 dx, \: k = 0, 1, 2,
\end{equation}
\noindent where $\tilde{u}^{(k)}(x)$ are the Lagrange interpolating polynomials from stencils $S_k$. After integration, the smoothness indicators for the sub-stencils are given by~\cite{zhang2008development}:
\begin{equation}
\begin{aligned}
	\beta_0 =& \frac{1}{3} \left[u_{j-2}\left(4u_{j-2} - 19u_{j-1} + 11u_j\right) + u_{j-1}\left(25u_{j-1} - 31u_j\right) + 10u_j^2\right], \\
    \beta_1 =& \frac{1}{3} \left[u_{j-1}\left(4u_{j-1} - 13u_j + 5u_{j+1}\right) + 13u_j\left(u_j - u_{j+1}\right) + 4u_{j+1}^2 \right], \\
    \beta_2 =& \frac{1}{3} \left[u_j\left(10u_j - 31u_{j+1} + 11u_{j+2}\right) + u_{j+1}\left(25u_{j+1}  - 19u_{j+2}\right) + 4u_{j+2}^2 \right].
\end{aligned}
\end{equation}

\subsection{Improved upwind-biased (Z) nonlinear interpolation}
The JS nonlinear weights are known to be excessively dissipative in both smooth regions and regions near discontinuities~\cite{henrick2005mapped, borges2008improved}. An improved version of upwind-biased nonlinear weights was proposed by Borges et al.~\cite{borges2008improved} for reconstruction in WENO scheme:
\begin{equation} \label{eq:Z5_nonlinear_weights}
	\omega_k = \frac{\alpha_k}{\sum\limits_{k=0}^{2}\alpha_k}, \:
    \alpha_k = d_k^{\mathrm{upwind}} \left(1 + \left( \frac{\tau_5}{\beta_k + \epsilon} \right)^p \right), \: k = 0, 1, 2,
\end{equation}

\begin{equation}
	\tau_5 = \left| \beta_2 - \beta_0 \right|.
\end{equation}

\noindent $\tau_5$ is a fifth order reference smoothness indicator. The nonlinear weights can also be used in WENO interpolation besides WENO reconstruction. The WENO interpolation with the improved upwind-biased nonlinear weights is fifth order accurate and is referred to as Z interpolation in this work. It is shown~\cite{borges2008improved} that WENO scheme with the improved weights, WENO5-Z, is less dissipative than classical WENO5-JS scheme to capture shock waves. Besides, the nonlinear weights lead to smaller loss in accuracy at critical points compared to classical weights. It was shown that at critical points where only the first order derivative vanishes, WENO5-JS becomes third order accurate but WENO5-Z with $p=1$ and $p=2$ has fourth order and the optimal fifth order accuracy respectively. In this paper, $p=2$ is employed in all test problems.

\subsection{Central nonlinear interpolations}

Although the fifth order upwind-biased WENO interpolations are robust in capturing shock waves and discontinuities, they have the drawback of being dissipative in smooth regions because of the upwind-bias. Hu et al.~\cite{hu2010adaptive,hu2011scale} included a downwind sub-stencil, $S_3$ shown in figure~\ref{fig:stencils}, in the WENO reconstruction to form sixth order adaptive central-upwind WENO schemes (WENO6-CU, WENO6-CU-M1, WENO6-CU-M2). In WENO interpolation, the downwind third order interpolation is given by:
\begin{equation} \label{eq:downwind_stencil}
	\tilde{u}_{j+\frac{1}{2}}^{(3)} = \frac{1}{8}\left(15u_{j+1} - 10u_{j+2} + 3u_{j+3} \right).
\end{equation}

\noindent The sixth order central linear interpolation on the complete six-point stencil $S_6$ as shown in figure~\ref{fig:stencils} is given by:
\begin{equation}
    \tilde{u}_{j+\frac{1}{2}}^{\mathrm{central}} = \frac{1}{256} \left(3u_{j-2} - 25u_{j-1} + 150u_j + 150u_{j+1} - 25u_{j+2} + 3u_{j+3} \right).
\end{equation}

\noindent The sixth order central linear interpolation can be formulated as a linear combination of the four third order interpolations in equations~\eqref{eq:upwind_biased_stencils} and \eqref{eq:downwind_stencil}:
\begin{equation}
	\tilde{u}_{j+\frac{1}{2}}^{\mathrm{central}} = \sum\limits_{k=0}^{3} d_k^{\mathrm{central}} \tilde{u}_{j+\frac{1}{2}}^{(k)},
\end{equation}

\noindent where the linear weights $d_k^{\mathrm{central}}$ are given by:
\begin{equation} \label{eq:linear_central_weights}
	d_0^{\mathrm{central}} = \frac{1}{32}, \:
    d_1^{\mathrm{central}} = \frac{15}{32}, \:
    d_2^{\mathrm{central}} = \frac{15}{32}, \:
    d_3^{\mathrm{central}} = \frac{1}{32}.
\end{equation}

\noindent Direct application of the sixth order central linear interpolation in WCNS will generate spurious oscillations around shock waves and discontinuities. Suggested by Hu et al.~\cite{hu2010adaptive}, sixth order nonlinear WENO interpolation can be constructed by introducing the following nonlinear weights:
\begin{equation}
	\tilde{u}_{j+\frac{1}{2}} = \sum\limits_{k=0}^{3} \omega_k \tilde{u}_{j+\frac{1}{2}}^{(k)}.
\end{equation}

\noindent The nonlinear weights $\omega_k$ in WENO6-CU~\cite{hu2010adaptive}, WENO6-CU-M1, and WENO-CU-M2~\cite{hu2011scale} are all derivations of the following form:
\begin{equation} \label{eq:CU6_nonlinear_weights_1}
	\omega_k = \frac{\alpha_k}{\sum\limits_{k=0}^{3}\alpha_k}, \:
    \alpha_k = d_k^{\mathrm{central}} \left( C + \frac{\tau_6}{\beta_k + \epsilon} \right)^q, \: k = 0, 1, 2, 3,
\end{equation}
where $\beta_k$ and $\tau_6$ are the smoothness indicators and reference smoothness indicator respectively. The smoothness indicator for the downwind stencil in WENO interpolation utilizes the six-point stencil for the sixth order central linear interpolation~\cite{liu2015new}:
\begin{equation}
\begin{aligned}
	\beta_3 = \beta_6 &= \sum^{5}_{l=1} \int^{x_{j+\frac{1}{2}}}_{x_{j-\frac{1}{2}}} \Delta x^{2l-1} \left( \frac{\partial^{l}}{\partial x^l} \tilde{u}^{(6)}(x) \right)^2 dx\\
	&= \frac{1}{232243200} \left[u_{j-2}\left(525910327u_{j-2} - 4562164630u_{j-1} + 7799501420u_j \right.\right. \\
    & \quad \left.\left. - 6610694540u_{j+1} + 2794296070u_{j+2} - 472758974u_{j+3}\right) \right. \\ 
    & \quad \left. + 5u_{j-1}\left(2146987907u_{j-1} - 7722406988u_j + 6763559276u_{j+1} \right. \right. \\ 
    & \quad \quad \left. \left. - 2926461814u_{j+2} + 503766638u_{j+3} \right) \right. \\
    & \quad \left. + 20u_j\left(1833221603u_j-3358664662u_{j+1}+1495974539u_{j+2} \right. \right. \\
    & \quad \quad \left. \left. -263126407u_{j+3}\right) \right. \\
    & \quad \left. + 20u_{j+1} \left( 1607794163u_{j+1} - 1486026707u_{j+2} + 268747951u_{j+3} \right) \right. \\
    & \quad \left. + 5u_{j+2} \left(1432381427u_{j+2}-536951582u_{j+3}\right) + 263126407u_{j+3}^2\right].
\end{aligned}
\end{equation}

\noindent $\tau_6$ is a sixth order reference smoothness indicator:
\begin{equation}
	\tau_6 = \left| \beta_3 - \beta_{avg} \right|,
\end{equation}
where
\begin{equation}
	\beta_{avg} = \frac{1}{8} \left( \beta_0 + 6\beta_1 + \beta_2 \right).
\end{equation}

In this work, we suggest another form of nonlinear weights for central interpolation, which is the central element of the proposed localized dissipative interpolation:
\begin{equation} \label{eq:CU6_nonlinear_weights_2}
	\omega_k = \frac{\alpha_k}{\sum\limits_{k=0}^{3}\alpha_k}, \:
    \alpha_k = d_k^{\mathrm{central}} \left( C + \left( \frac{\tau_6}{\beta_k + \epsilon} \right)^{q} \right), \: k = 0, 1, 2, 3.
\end{equation}

Nonlinear weights in the form proposed by Hu et al.~\cite{hu2010adaptive}  (equation~\eqref{eq:CU6_nonlinear_weights_1}) is labeled as version A and the new modification (equation~\eqref{eq:CU6_nonlinear_weights_2}) is labeled as version B in this work. Both versions of nonlinear weights have the same set of user-defined parameters $C$ and $q$. In either version, nonlinear weight for the pure downwind stencil $S_3$ is much smaller than others in regions near discontinuities as $\beta_3$ becomes much larger than $\beta_0$, $\beta_1$, and $\beta_2$. Hu and Adams~\cite{hu2011scale} showed that both $C$ and $q$ have significant effects on the scale-separation capabilities of the WENO6 schemes. In their study, they adopted the approach of Taylor et al.~\cite{taylor2007optimization} to study  the numerical dissipation incurred in the interpolation of a sinusoidal function $u(x) = \sin(kx + \psi)$ for $0 \leq x \leq 2\pi$, where $\psi$ is a constant phase shift. If the function is discretized uniformly with $N$ points such that $x_j = j \Delta x$, $\forall j \in \{ 0, \: 1, \: \dots, \: N-1 \}$, we get:
\begin{equation}
	u_j = \sin(k j \Delta x + \psi) = \sin(j \phi + \psi),
\end{equation}
where $\phi = k\Delta x$ is the reduced wavenumber. As suggested by Hu and Adams~\cite{hu2011scale}, the numerical dissipation introduced in fifth order or sixth order WENO reconstructions and interpolations can be estimated through the non-dimensional dissipation, $\epsilon_d$:
\begin{equation}
	\epsilon_d = \frac{1}{N} \sum_{j=0}^{N-1}\left[ \sum_{r=0}^{3} (d_r^{\mathrm{central}} - \omega_r(x=x_j) )^2 \right] .
\end{equation}

\noindent $\epsilon_d$ represents the discrepancy between the nonlinear weights and linear weights. The larger the discrepancy, the more the numerical dissipation is introduced in the interpolation process. In this paper, $\epsilon_d$ is ensemble averaged over 20 test functions with different phase shifts $\psi$ equally spaced in $[0, 2\pi)$ to obtain statistically converged results.

\begin{figure}[!ht]
\centering
\subfigure[Version A]{%
\includegraphics[height=0.36\textwidth]{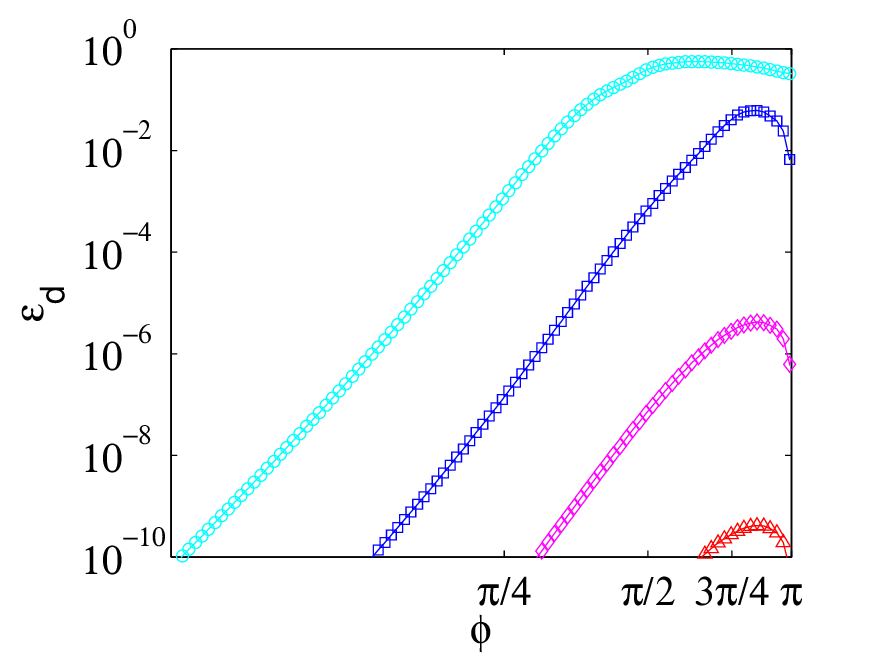}
\label{fig:C_effect}}
\subfigure[Version B]{%
\includegraphics[height=0.36\textwidth]{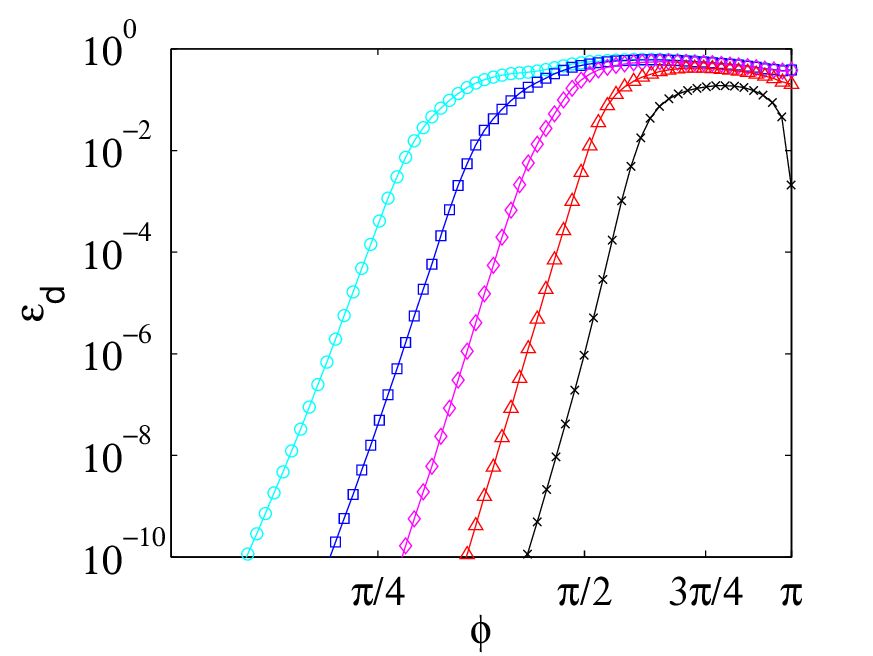}
\label{fig:C_effect_improved}}
\caption{Comparison of effects of $C$ on scale-separation between two different versions of nonlinear central interpolations. Cyan circles: $C = 1.0\mathrm{e}{1}$, $q = 4$; blue squares: $C = 1.0\mathrm{e}{3}$, $q = 4$; magenta diamonds: $C = 1.0\mathrm{e}{5}$, $q = 4$; red triangles: $C = 1.0\mathrm{e}{7}$, $q = 4$; black crosses: $C = 1.0\mathrm{e}{9}$, $q = 4$ ($\epsilon_d$ is too small to be shown in version A).}
\label{fig:compare_central_interpolations_C_effect}
\end{figure}

\begin{figure}[!ht]
\centering
\subfigure[Version A]{%
\includegraphics[height=0.36\textwidth]{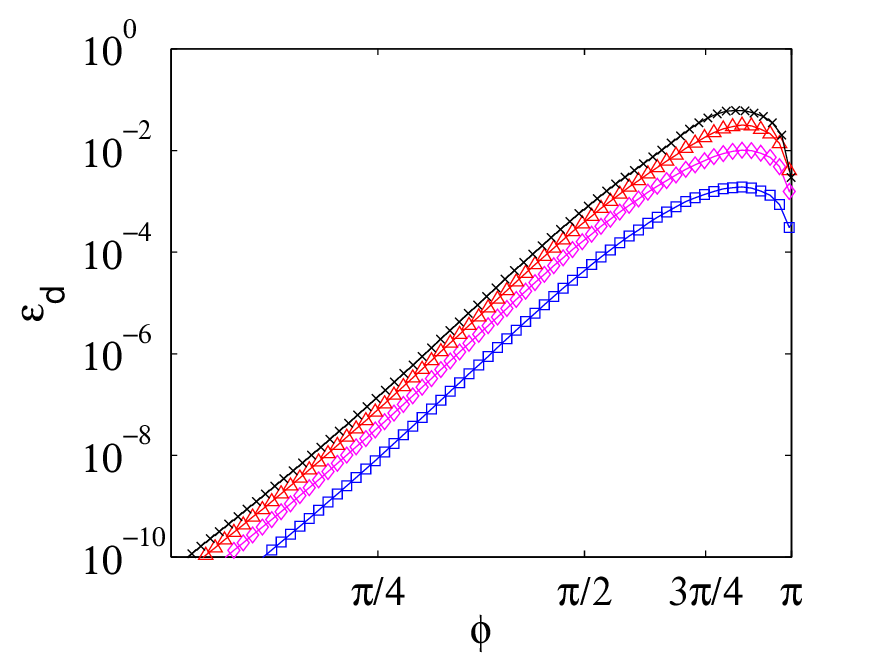}
\label{fig:q_effect}}
\subfigure[Version B]{%
\includegraphics[height=0.36\textwidth]{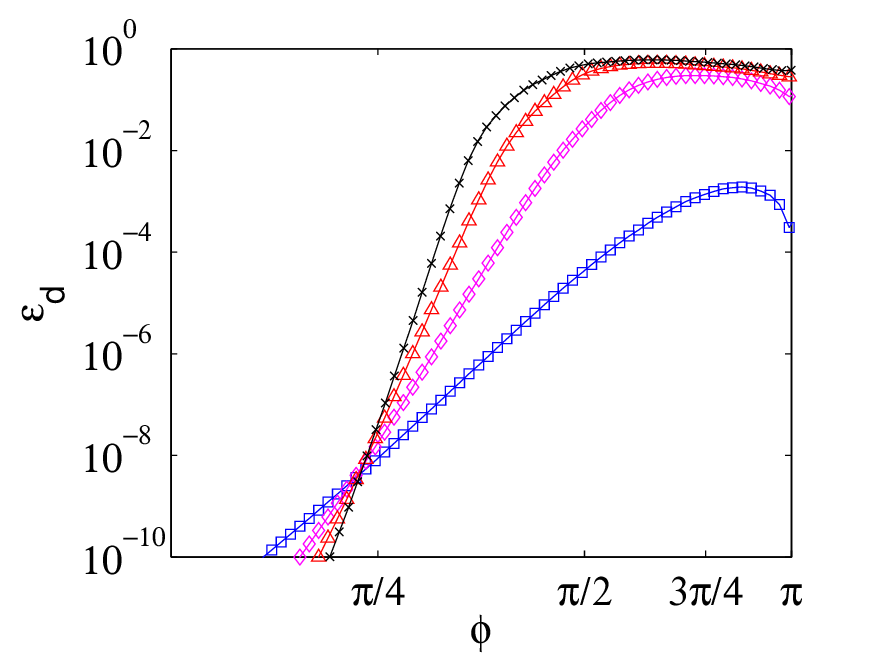}
\label{fig:q_effect_improved}}
\caption{Comparison of effects of $q$ on scale-separation between two different versions of nonlinear central interpolations. Blue squares: $C = 1.0\mathrm{e}{3}$, $q = 1$; magenta diamonds: $C = 1.0\mathrm{e}{3}$, $q = 2$; red triangles: $C = 1.0\mathrm{e}{3}$, $q = 3$; black crosses: $C = 1.0\mathrm{e}{3}$, $q = 4$.}
\label{fig:compare_central_interpolations_q_effect}
\end{figure}

Figures~\ref{fig:compare_central_interpolations_C_effect} and \ref{fig:compare_central_interpolations_q_effect} show the relations between numerical dissipation $\epsilon_d$ and reduced wavenumber $\phi$ for the two versions of nonlinear central interpolations at different values of $C$ and $q$. In figures~\ref{fig:C_effect} and ~\ref{fig:C_effect_improved}, the effects of $C$ between versions A and B of nonlinear weights are compared at $q=4$. It can be seen that as the values of $C$ are increased, the dissipation curves of both versions of nonlinear weights shift to the right. This means dissipation is added more locally at high wavenumber features. Another observation is that the peak of $\epsilon_d$ in version A decreases more significantly than that in version B when the dissipation curve shifts. This suggests that version B of nonlinear weights has more robust scale-separation behavior for fine-tuning. Figure~\ref{fig:q_effect} and figure~\ref{fig:q_effect_improved} compare the effects of $q$ on numerical dissipation between both versions of nonlinear weights respectively at constant value of $C=1.0\mathrm{e}{3}$. It can be seen that changing the value of $q$ has the same effect as $C$ to shift the dissipation curve in version A of nonlinear weights. On the other hand, increasing the value of $q$ in version B increases the slope of the curve dramatically which is equivalent to having sharper cutoff for scale-separation. In short, version B of nonlinear weights allows sharper scale-separation between dissipated and non-dissipated features than version A.

\subsection{Modified adaptive central-upwind (CU-M2) nonlinear interpolation}
Both versions of nonlinear weights given by equation~\eqref{eq:CU6_nonlinear_weights_1} and equation~\eqref{eq:CU6_nonlinear_weights_2} respectively are not numerically dissipative enough in large gradient regions to regularize the solutions. Hu and Adams~\cite{hu2011scale} proposed the following modified version A of nonlinear weights to capture strong shocks when large value of $C$ is used for scale-separation:
\begin{equation} \label{eq:CU6_M2_nonlinear_weights}
	\omega_k = \frac{\alpha_k}{\sum\limits_{k=0}^{3}\alpha_k}, \:
    \alpha_k = d_k^{\mathrm{central}} \left( C + \frac{\tau_6}{\beta_k + \chi^{-1} \Delta x^2}
    \frac{\beta_{avg} + \chi \Delta x^2}{\beta_k + \chi \Delta x^2} \right)^{q}, \: k = 0, 1, 2, 3,
\end{equation}

\noindent where $q$ is a positive integer and $\chi =  1.0\mathrm{e}{8}$ is a large positive dimensional constant. Interpolation with the nonlinear weights in equation~\eqref{eq:CU6_M2_nonlinear_weights} is called CU-M2 interpolation in this paper. Although it was shown that WENO scheme with the nonlinear weights in equation~\eqref{eq:CU6_M2_nonlinear_weights} can capture a Mach 10 shock~\cite{hu2011scale}, this form of nonlinear weights has the dismerit that $\chi$ is a dimensional parameter that makes the overall method scale-variant.

\subsection{Localized dissipative (LD) nonlinear interpolation} \label{section:LD_interpolation}

In this section, we propose a nonlinear interpolation that hybridizes the upwind-biased Z interpolation with version B of central interpolation. The hybrid interpolation has both the advantage of upwind-biased interpolation to provide good numerical stability in regions containing shocks and high wavenumber features, and the advantage of central interpolation to add minimal numerical dissipation in smooth regions to preserve small features.

A switch is required to turn on the hybridization of the central interpolation with the Z interpolation when a region containing non-smooth features is detected. Taylor et al.~\cite{taylor2007optimization} proposed two relative sensors to distinguish smooth and non-smooth regions. They are respectively relative total variation indicator and relative smoothness indicator. The relative total variation indicator $R_{\mathrm{TV}}$ is given as:
\begin{equation}
	R_{\mathrm{TV}} = \frac{\max_{0 \leq k \leq 3} \mathrm{TV}_k}{\min_{0 \leq k \leq 3} \mathrm{TV}_k + \epsilon},
\end{equation}
where $\mathrm{TV}_k$'s are defined as the total variation of the variable $u$ in different sub-stencils:
\begin{equation}
	\mathrm{TV}_{k} = \sum^{2}_{l=1} \left| u_{j+k+l-2} - u_{j+k+l-3} \right|, \: k = 0, 1, 2, 3.
\end{equation}

\noindent The relative smoothness indicator $R_\beta$ is defined similarly based on smoothness indicators:
\begin{equation}
	R_\beta = \frac{\max_{0 \leq k \leq 3} \beta_k}{\min_{0 \leq k \leq 3} \beta_k + \epsilon}.
\end{equation}

\begin{figure}[!ht]
\centering
\subfigure[Relative total variation indicator]{%
\includegraphics[width=0.48\textwidth]{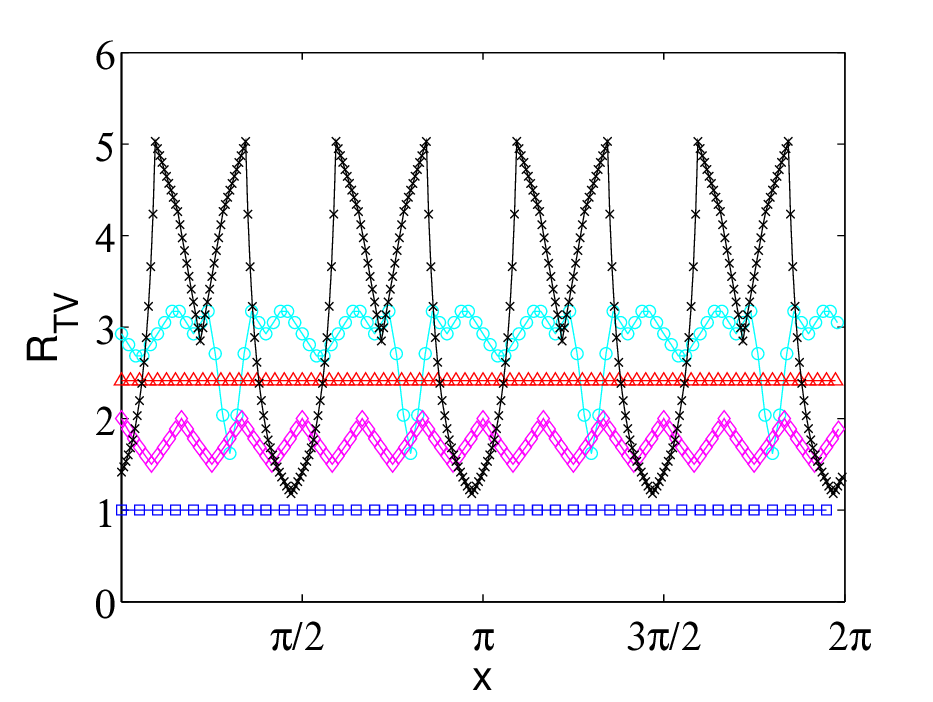}
\label{fig:R_TV}}
\subfigure[Relative smoothness indicator]{%
\includegraphics[width=0.48\textwidth]{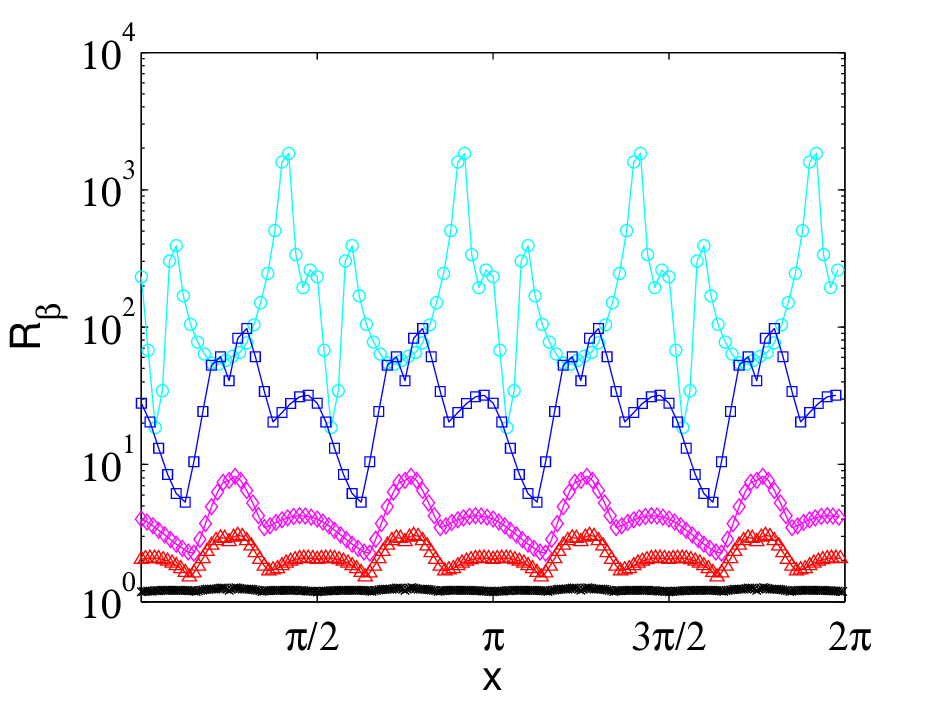}
\label{fig:R_beta}}
\subfigure[Relative reference smoothness indicator]{%
\includegraphics[width=0.48\textwidth]{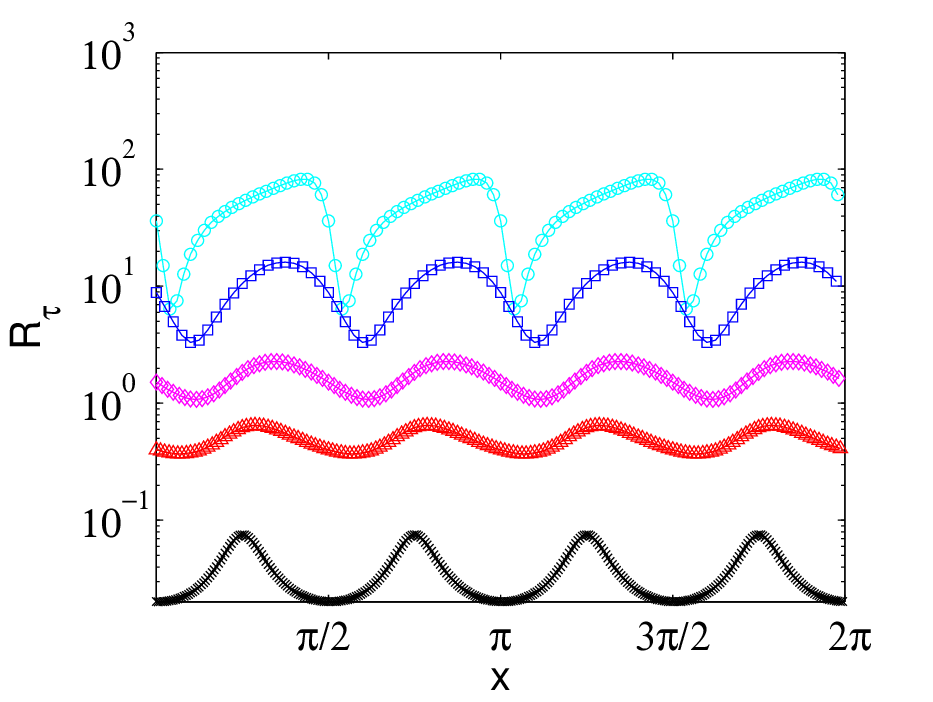}
\label{fig:R_tau}}
\caption{Relative total variation indicator, relative smoothness indicator, and relative reference smoothness indicator for a smooth sinusoidal function sampled with different points per wavelength (PPW's). Cyan circles: 2.5 PPW ($\phi = 0.8 \pi$); blue squares: 4 PPW ($\phi = 0.5\pi$); magenta diamonds: 6 PPW ($\phi = 0.33\pi$); red triangles: 8 PPW ($\phi = 0.25\pi$); black crosses: 16 PPW ($\phi = 0.125\pi$).}
\label{fig:compare_TV_beta_tau}
\end{figure}

\begin{figure}[!ht]
 \centering
	\includegraphics[width=0.6\textwidth]{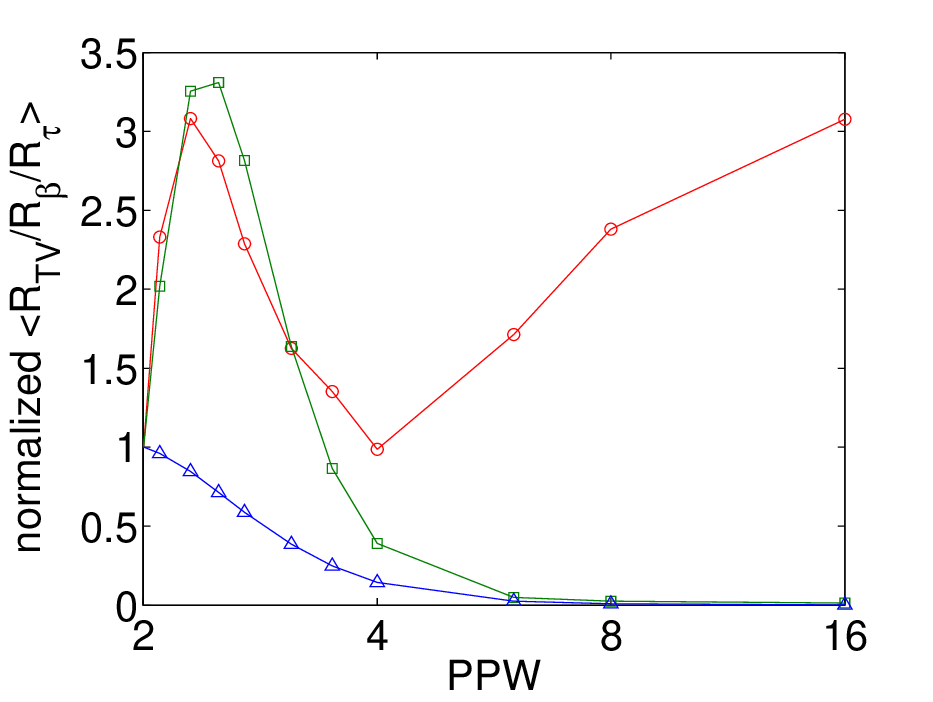}
	\caption{Normalized averaged values of different relative sensors against points per wavelength (PPW). The averaged values are normalized by averaged values at PPW=2. Red circles: relative total variation indicator; green squares: relative smoothness indicator; blue triangles: relative reference smoothness indicator.}
    \label{fig:compare_relative_sensors}
\end{figure}

By design, both relative indicators are scale-invariant. In figure~\ref{fig:compare_TV_beta_tau}, the values of $R_{\mathrm{TV}}$ and $R_\beta$ against $x$ for a sinusoidal wave $u(x) = \sin(2x)$ for $0 \leq x \leq 2\pi$ sampled with different points per wavelength (PPW) or equivalently inverse of reduced wavenumber $\phi$ are plotted. Figure~\ref{fig:compare_relative_sensors} shows the average of the values over $x$ from the relative sensors at different PPW's. The averaged sensor values are normalized by the averaged value from the corresponding sensor at $\textnormal{PPW}=2$. From the plot, it is observed that the normalized averaged value of $R_{\mathrm{TV}}$ is not monotonic. We cannot use a threshold to distinguish low and high wavenumber features. Hence, relative TV indicator is not suitable for distinguishing smooth and non-smooth regions. On the other hand, relative smoothness indicator has better performance as averaged $R_\beta$ decreases monotonically with increasing PPW after $\textnormal{PPW}=2.5$. However, the averaged value increases from $\textnormal{PPW}=2$ (Nyquist limit) to $\textnormal{PPW}=2.5$.

We also propose a relative indicator called relative reference smoothness indicator $R_\tau$, which is defined as:
\begin{equation}
	R_\tau = \frac{\tau_6}{\beta_{avg} + \epsilon}.
\end{equation}

\noindent From figure~\ref{fig:compare_relative_sensors}, it can be seen that as we increases PPW from 2, averaged value of $R_\tau$ deceases monotonically. Besides, we can see from figure~\ref{fig:compare_TV_beta_tau} that $R_\tau$ has better scale-separation capability compared to $R_\beta$ as there is less overlapping between neighboring curves for waves of different PPW's in $R_\tau$ compared to that in $R_\beta$. Therefore, we decide to use the relative reference smoothness indicator to identify non-smooth regions for the hybridization of nonlinear upwind-biased and central interpolations.

The proposed hybrid nonlinear WENO interpolation has form given by:

\begin{equation} \label{eq:LD_nonlinear_weights}
 \omega_k = \begin{cases} 
  \sigma \omega^{\mathrm{upwind}}_k + (1 - \sigma) \omega^{\mathrm{central}}_k,
   &\mbox{if } R_\tau > \alpha^{\tau}_{RL}, \\
  \omega^{\mathrm{central}}_k,
   & \mbox{otherwise },
 \end{cases}
 , \: k = 0, 1, 2, 3,
\end{equation}

\noindent where $\omega^{\mathrm{upwind}}_k$ and $\omega^{\mathrm{central}}_k$ are nonlinear weights in equation~\eqref{eq:Z5_nonlinear_weights} and equation~\eqref{eq:CU6_nonlinear_weights_2} respectively. $\omega^{\mathrm{upwind}}_3$ is always set to be zero. $\alpha^{\tau}_{RL}$ is a user-defined constant. $0 \leq \sigma \leq 1$ is a value given by a sensor to control the contributions of upwind-biased and central interpolations. $\sigma$ should be close to one in regions near discontinuities and high wavenumber features. In this paper, the following formulation is used to compute $\sigma$:
\begin{equation}
	\sigma_{j+\frac{1}{2}} = \max \left( \sigma_{j}, \sigma_{j+1} \right),
\end{equation}
where $\sigma_j$ is defined as:
\begin{eqnarray}
    \sigma_j &=& \frac{\left| \Delta u _{j+\frac{1}{2}} - \Delta u _{j-\frac{1}{2}} \right|}{\left|\Delta u _{j+\frac{1}{2}} \right| + \left| \Delta u _{j-\frac{1}{2}} \right| + \epsilon}, \\
    \Delta u _{j+\frac{1}{2}} &=& u_{j+1} - u_{j}.
\end{eqnarray}
The hybrid interpolation is called LD interpolation in this paper because of the numerically localized dissipative (LD) nature of the interpolation discussed in next few sections. Note that LD interpolation in this paper has a form different from the localized dissipation interpolation with hybrid linear weights proposed by Wong and Lele~\cite{wong2016improved} (denoted as HW interpolation in this paper) in following aspects:
\begin{itemize}
\item The hybridization in LD interpolation is performed on nonlinear weights from nonlinear upwind-biased and central interpolations, while that in HW interpolation is operated on linear weights.
\item The relative reference smoothness indicator $R_\tau$ is introduced in this paper for LD interpolation to identify non-smooth regions, while a combination of relative indicators $R_{\mathrm{TV}}$ and $R_\beta$ is used in HW interpolation. The improved performance of $R_\tau$ over $R_{\mathrm{TV}}$ and $R_\beta$ in identifying non-smooth regions is discussed above.
\item A new form of $\sigma$ is introduced for hybridization in LD interpolation. This new sensor can switch the nonlinear weights from hybrid central-upwind to purely upwind-biased ones in regions containing discontinuities or odd-even oscillations. The switching function in HW interpolation is found to be incapable in identifying odd-even oscillations.
\end{itemize}
In this paper we mainly focus on the comparison of LD interpolation with JS, Z and CU-M2 interpolations but not the comparison between two versions of localized interpolations (LD vs. HW). The main purpose of this paper is to examine the effect of localized numerical dissipation in high wavenumber range and discontinuities on compressible flow problems. By comparing the results of our new scheme with those of other state-of-the art schemes from recent literature, we highlight the superior performance of the proposed scheme.

\subsection{Comparison of scale-separation capabilities}

\begin{figure}[!ht]
\centering
\includegraphics[width=0.6\textwidth]{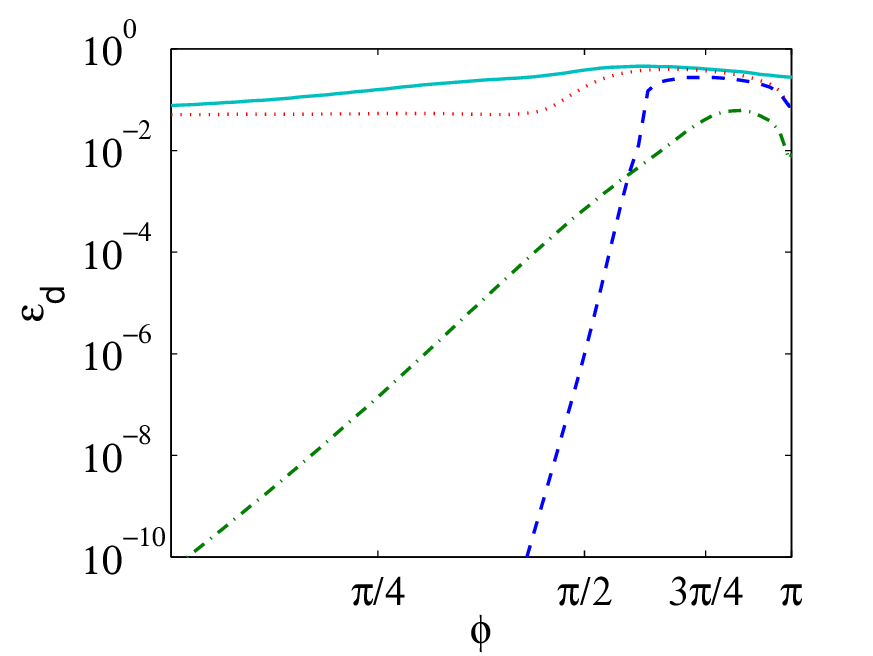}
\caption{Scale-separation capabilities of different WENO interpolation methods. Cyan solid line: JS interpolation; red dotted line: Z interpolation; green dash-dotted line: CU-M2 interpolation; blue dashed line: LD interpolation.}
\label{fig:compare_scale_separation}
\end{figure}

Table~\ref{table:parameters} shows the parameter values chosen for the various interpolation methods in this work. In figure~\ref{fig:compare_scale_separation}, relations between numerical dissipation and reduced wavenumber of different interpolation methods are compared. The parameters in CU-M2 nonlinear weights follow the suggested values by Hu and Adams~\cite{hu2011scale}. From the figure, we can see that both classical JS and improved Z interpolation methods add dissipation of the same order of magnitude to sinusoidal waves over a large range of wavenumber due to their upwind-biased nature ($\omega_3$ is always zero). This implies that in either of the methods the same extent of numerical dissipation is used to damp features over a wide range of scales. Unlike JS and Z interpolations, both CU-M2 and LD interpolations exhibit scale-separation as the amount of dissipation added decreases dramatically with decreasing wavenumber. This suggests that turbulent features in the small wavenumber range are better captured by these two methods compared to the upwind-biased interpolations. In high wavenumber range where dispersion errors of numerical schemes are large, both methods are designed on purpose to introduce a significant amount of dissipation to regularize the solutions.

\begin{table}[!ht]
  \begin{center}
  \begin{tabular}{ | c | c | c | c | c |}
    \hline
    \textbf{Interpolation} & \multicolumn{4}{c|}{\textbf{Parameter values}} \\
    \cline{2-5}
    \textbf{methods} & $p$ & $q$ & $C$ & $\alpha^\tau_{RL}$\\[0.1pc] \hline \hline
    \textbf{JS}    & $2$ & $-$ & $-$ & $-$ \\[0.1pc] \hline
    \textbf{Z}     & $2$ & $-$ & $-$ & $-$ \\[0.1pc] \hline
    \textbf{CU-M2} & $2$ & $4$ & $1000.0$ & $-$ \\[0.1pc] \hline
    \textbf{LD}    & $2$ & $4$ & $1.0\mathrm{e}{9}$ & $35.0$ \\[0.1pc] \hline
  \end{tabular}
  \caption{Parameters for different interpolation methods.}
  \label{table:parameters}
  \end{center}
\end{table}

Comparing the two sixth order interpolation methods CU-M2 and LD, the latter method clearly shows better scale-separation capability because of the sharper cutoff between low and high wavenumber range that can be seen at $\phi \approx  \pi/2$. This localized dissipative behavior follows the decoupled scale-separation effects of $C$ and $q$ discussed in previous section. Moreover, LD interpolation introduces much more numerical dissipation in the high wavenumber range from the hybridization of the central interpolation with the upwind-biased interpolation in non-smooth regions. This suggests that the LD interpolation should have better preservation of turbulent features from low wavenumber range up to medium wavenumber range and also numerical stability in regions containing high wavenumber features compared to CU-M2 interpolation.

\subsection{HLLC-HLL Riemann solver}

In a 1D domain, the WENO interpolations approximate $\bm{Q}^{L}_{j+\frac{1}{2}}$ and $\bm{Q}^{R}_{j+\frac{1}{2}}$, which are values of the conservative variables $\bm{Q}$ on the left and right sides of the cell midpoints respectively. The fluxes at cell midpoints can then be obtained by a HLLC Riemann solver~\cite{toro1994restoration} using methods by Johnsen et al.~\cite{johnsen2006implementation} and Coralic et al~\cite{coralic2014finite}. However, it is well-known that HLLC Riemann solver can give rise to numerical instability near shocks in some multi-dimensional problems because of insufficient numerical dissipation for stabilization when the shock normal direction does not align well with the grid normal surface direction. To improve numerical stability, the HLLC Riemann solver is hybridized with the more dissipative HLL Riemann solver~\cite{harten1983upstream} in the way proposed by Huang et al.~\cite{huang2011cures} in regions where shock normals do not align well with the grid normals. We improve their method so that the hybridization is only carried out around shock waves. To detect shock waves, the Ducros-like sensor designed by Larsson et al.~\cite{larsson2007effect}:
\begin{equation} \label{eq:Larsson_switch}
	s = \frac{-\theta}{|\theta| + \sqrt{\omega_i \omega_i} + \epsilon}
\end{equation}
is computed at every grid point and timestep, where $\theta = \partial u_j / \partial x_j$ is the rate of dilatation and $\omega_i = \epsilon_{ijk} \partial u_k / \partial x_j$ is the vorticity, where $\epsilon_{ijk}$ is the Levi-Civita tensor. The hybrid HLLC-HLL Riemann solver is used when $s > 0.65$. When the shock sensor detects the presence of shock waves, the hybridization of the fluxes is implemented in the equations of densities, momentum that is tangential to the cell interface and volume fractions. For example, the HLLC-HLL flux for the 3D two-species Euler system with five-equation model in the $x$ direction is given by:
\begin{equation}
	\begin{aligned}
    	\mathbf{F}_{\textnormal{HLLC-HLL}}(1) &= \tilde{\alpha}_1 \mathbf{F}_{\textnormal{HLLC}}(1) + \tilde{\alpha}_2 \mathbf{F}_{\textnormal{HLL}}(1), \\
        \mathbf{F}_{\textnormal{HLLC-HLL}}(2) &= \tilde{\alpha}_1 \mathbf{F}_{\textnormal{HLLC}}(2) + \tilde{\alpha}_2 \mathbf{F}_{\textnormal{HLL}}(2), \\
        \mathbf{F}_{\textnormal{HLLC-HLL}}(3) &= \mathbf{F}_{\textnormal{HLLC}}(3), \\
        \mathbf{F}_{\textnormal{HLLC-HLL}}(4) &= \tilde{\alpha}_1 \mathbf{F}_{\textnormal{HLLC}}(4) + \tilde{\alpha}_2 \mathbf{F}_{\textnormal{HLL}}(4), \\
        \mathbf{F}_{\textnormal{HLLC-HLL}}(5) &= \tilde{\alpha}_1 \mathbf{F}_{\textnormal{HLLC}}(5) + \tilde{\alpha}_2 \mathbf{F}_{\textnormal{HLL}}(5), \\
        \mathbf{F}_{\textnormal{HLLC-HLL}}(6) &= \mathbf{F}_{\textnormal{HLLC}}(6), \\
        \mathbf{F}_{\textnormal{HLLC-HLL}}(7) &= \tilde{\alpha}_1 \mathbf{F}_{\textnormal{HLLC}}(7) + \tilde{\alpha}_2 \mathbf{F}_{\textnormal{HLL}}(7),
    \end{aligned}
\end{equation}

\noindent where $\mathbf{F}_{\textnormal{HLLC-HLL}}$, $\mathbf{F}_{\textnormal{HLL}}$, and $\mathbf{F}_{\textnormal{HLLC}}$ are fluxes computed from HLLC-HLL, HLL, and HLLC Riemann solvers respectively. The weights $\tilde{\alpha}_1$ and $\tilde{\alpha}_2$ similar to those suggested by Huang et al.~\cite{huang2011cures} are used:
\begin{equation} \label{eq:HLLC_HLL_betas}
\begin{aligned}
\alpha_1 &=
\begin{cases}
    1, &\mbox{if } \sqrt{\left( u_R - u_L \right)^2 + \left( v_R - v_L \right)^2 + \left( w_R - w_L \right)^2} < \epsilon, \\
    \frac{\left| u_R - u_L \right|}{\sqrt{\left( u_R - u_L \right)^2 + \left( v_R - v_L \right)^2 + \left( w_R - w_L \right)^2}}, &\mbox{otherwise},
\end{cases} \\
\alpha_2 &= \sqrt{1 - \alpha_1^{2}}, \\
\tilde{\alpha}_1 &= \frac{1}{2} + \frac{1}{2} \frac{\alpha_1}{\alpha_1 + \alpha_2}, \\
\tilde{\alpha}_2 &= 1 - \tilde{\alpha}_1.
\end{aligned}
\end{equation}

The weights $\tilde{\alpha}_1$ and $\tilde{\alpha}_2$ are designed in the way such that when the shock normal direction is aligned with the surface normal direction, the hybrid flux is purely the HLLC flux. When the shock normal direction is perpendicular to the surface normal direction, HLL flux adds dissipation by sharing the same weight as the HLLC flux. In 1D problems, the HLLC-HLL Riemann solver is reduced to the regular HLLC Riemann solver since the shock normal direction is always perpendicular to the grid surface normal.

The MND WCNS's with the JS, Z, CU-M2, and LD interpolations are called WCNS5-JS, WCNS6-CU-M2, and WCNS6-LD respectively in this paper. The numbers in the names represent the formal orders of accuracy of the corresponding schemes in smooth regions that are verified in section \ref{convergence_study}.

\subsection{Approximation of source term}
The source term $Z_1 \nabla \cdot \bm{u}$ in the volume fraction equation can be approximated by the sixth order finite difference scheme like the convective fluxes. The approximation to the source term in a 3D space has the following form:
\begin{equation}
\begin{aligned}
	& (Z_1 \nabla \cdot \bm{u})|_{i,j,k} = \\
    & Z_1 |_{i,j,k} \times
    \\
    \bigg\lbrace & \frac{1}{\Delta x}  \left[
      \frac{3}{2} \left(\tilde{u}_{i+\frac{1}{2},j,k} - \tilde{u}_{i-\frac{1}{2},j,k} \right)
      - \frac{3}{10} \left(u_{i+1,j,k} - u_{i-1,j,k} \right)
     + \frac{1}{30} \left(\tilde{u}_{i+\frac{3}{2},j,k} - \tilde{u}_{i-\frac{3}{2},j,k} \right)
    \right] + 
    \\
    & \frac{1}{\Delta y}  \left[
      \frac{3}{2} \left(\tilde{v}_{i,j+\frac{1}{2},k} - \tilde{v}_{i,j-\frac{1}{2},k} \right)
      - \frac{3}{10} \left(v_{i,j+1,k} - v_{i,j-1,k} \right)
     + \frac{1}{30} \left(\tilde{v}_{i,j+\frac{3}{2},k} - \tilde{v}_{i,j-\frac{3}{2},k} \right) \right] +
    \\
    & \frac{1}{\Delta z}  \left[
      \frac{3}{2} \left(\tilde{w}_{i,j,k+\frac{1}{2}} - \tilde{w}_{i,j,k-\frac{1}{2}} \right)
      - \frac{3}{10} \left(w_{i,j,k+1} - w_{i,j,k-1} \right)
     + \frac{1}{30} \left(\tilde{w}_{i,j,k+\frac{3}{2}} - \tilde{w}_{i,j,k-\frac{3}{2}} \right) \right] \bigg\rbrace,
\end{aligned}
\end{equation}

\noindent where $\tilde{u}_{i+\frac{1}{2},j,k}$, $\tilde{v}_{i+\frac{1}{2},j,k}$, and $\tilde{w}_{i+\frac{1}{2},j,k}$ are components of velocity numerically approximated at midpoints between cell nodes and $u_{i,j,k}$, $v_{i,j,k}$, and $w_{i,j,k}$ are velocity components at cell nodes. The approximated velocity components at midpoints between cell nodes are consistent with the HLLC fluxes and the computation is discussed in appendix \ref{velocity_appendix}.

\section{Approximate dispersion relation (ADR)}

This section discusses the dispersion and dissipation characteristics of WCNS with different interpolation methods. In the study of dispersion and dissipation characteristics of WENO schemes and WCNS's, Fourier analysis is usually carried out on the linear counterparts~\cite{deng2000developing, martin2006bandwidth, nonomura2007increasing, zhang2008development, nonomura2013robust} by replacing the nonlinear weights with linear weights. However, the nonlinearity in the shock-capturing schemes has a very significant effect on their characteristics. So far there are still no analytical methods developed to study the spectral behavior of nonlinear schemes. However, the modified wavenumber of the nonlinear schemes can be obtained numerically using the approximate dispersion relation (ADR) technique of Pirozzoli~\cite{pirozzoli2006spectral}.

\begin{figure}[!ht]
\centering
\includegraphics[width=0.6\textwidth]{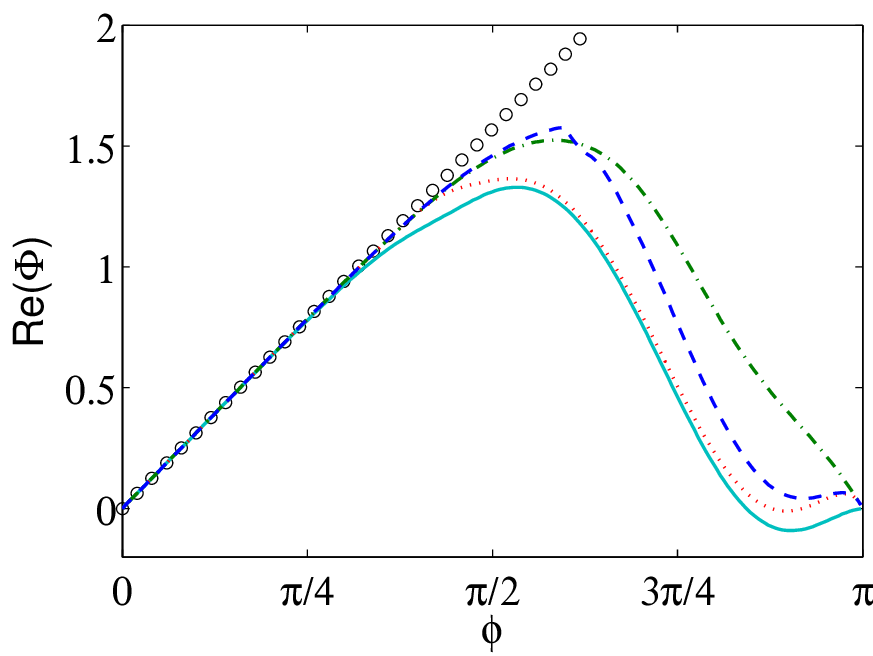}
\caption{Dispersion characteristics of different WCNS's. Cyan solid line: WCNS5-JS; red dotted line: WCNS5-Z; green dash-dotted line: WCNS6-CU-M2; blue dashed line: WCNS6-LD; black circles: spectral.}
\label{fig:compare_dispersion_errors}
\end{figure}

\begin{figure}[!ht]
\centering
\includegraphics[width=0.6\textwidth]{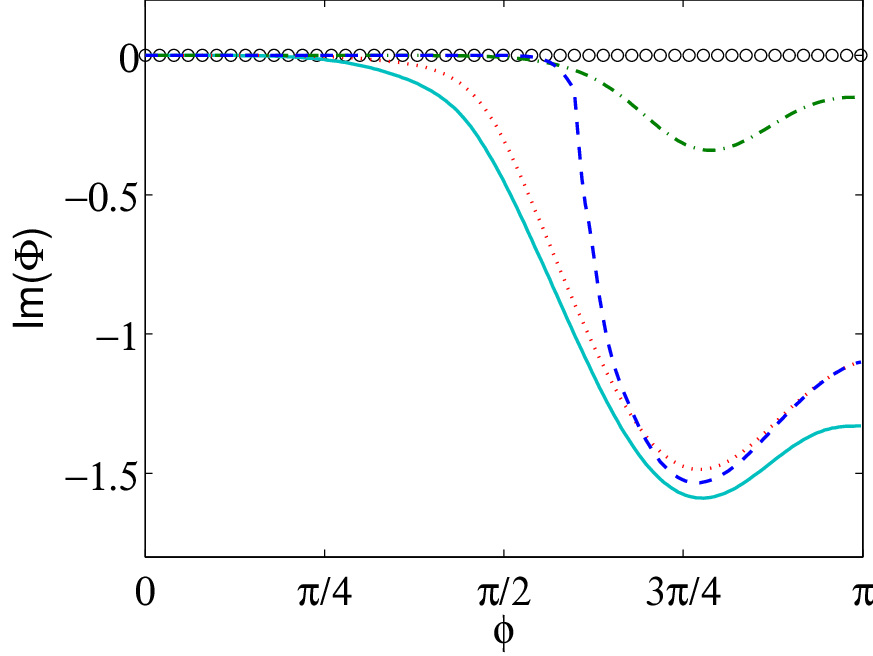}
\caption{Dissipation characteristics of different WCNS's. Cyan solid line: WCNS5-JS; red dotted line: WCNS5-Z; green dash-dotted line: WCNS6-CU-M2; blue dashed line: WCNS6-LD; black circles: spectral.}
\label{fig:compare_dissipation_errors}
\end{figure}

The real (Re($\Phi$)) and imaginary (Im($\Phi$)) parts of the modified wavenumber $\Phi$ at different wavenumber $\phi$ computed using the ADR technique~\cite{pirozzoli2006spectral} are shown in figures~\ref{fig:compare_dispersion_errors} and \ref{fig:compare_dissipation_errors} respectively. From figure~\ref{fig:compare_dissipation_errors}, we can compare the resolutions of various methods with the exact solution. The closer the curve to that of spectral method, the higher is the resolution of the scheme. It can be noticed that WCNS5-JS has the worst resolution. WCNS5-Z has improved performance over the WCNS5-JS in resolution but clearly that both WCNS6-CU-M2 and WCNS6-LD schemes have much better resolution than the upwind-biased schemes because of the inclusion of downwind stencil. In figure~\ref{fig:compare_dissipation_errors}, the dissipation of different schemes is compared through the imaginary part of modified wavenumber. Larger amount of numerical dissipation leads to more negative imaginary part of modified wavenumber. It can be seen that both WCNS6-CU-M2 and WCNS6-LD have more localized dissipation at high wavenumber region compared to WCNS5-JS and WCNS5-Z. WCNS6-LD has larger dissipation for stabilization at high wavenumber region where the dispersion error is also high. The dissipation of WCNS6-LD approaches that of WCNS5-Z at the Nyquist limit and this is expected due to the hybridization between interpolations.

\begin{figure}[!ht]
\centering
\includegraphics[width=0.6\textwidth]{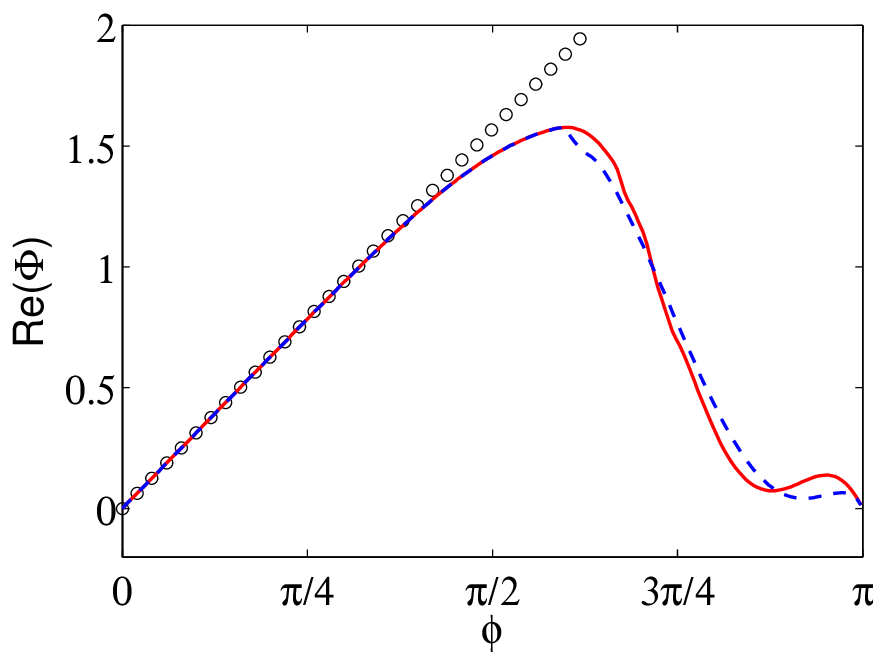}
\caption{Dispersion characteristics of WCNS6-HW and WCNS6-LD. Red solid line: WCNS6-HW; blue dashed line: WCNS6-LD; black circles: spectral.}
\label{fig:compare_dispersion_errors_LD}
\end{figure}

\begin{figure}[!ht]
\centering
\includegraphics[width=0.6\textwidth]{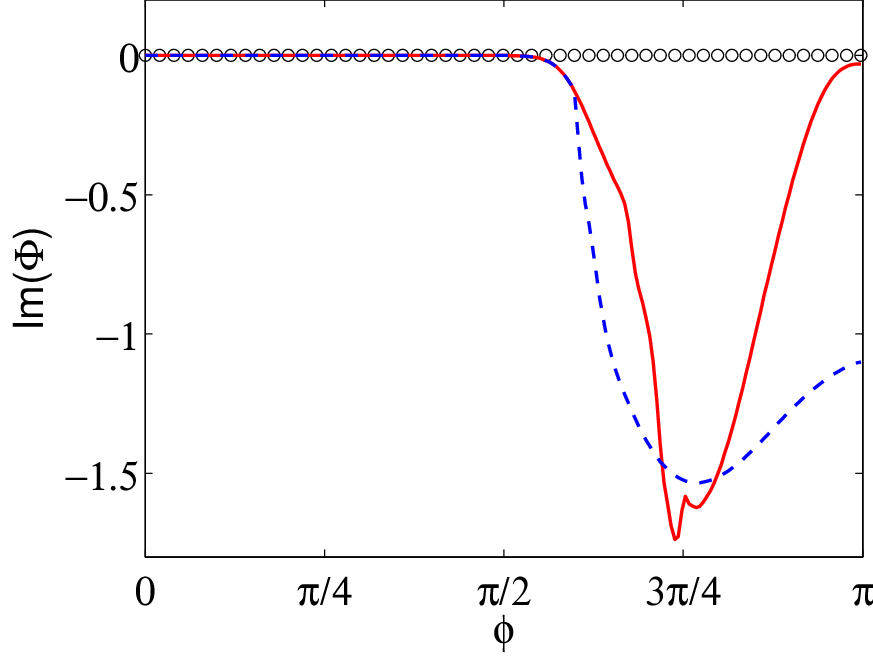}
\caption{Dissipation characteristics of WCNS6-HW and WCNS6-LD. Red solid line: WCNS6-HW; blue dashed line: WCNS6-LD; black circles: spectral.}
\label{fig:compare_dissipation_errors_LD}
\end{figure}

The dispersion and dissipation characteristics of WCNS6-LD are also compared with the WCNS with HW interpolation (WCNS6-HW) in figures~\ref{fig:compare_dispersion_errors_LD} and ~\ref{fig:compare_dissipation_errors_LD}. The parameter settings in WCNS6-HW follow those in Wong and Lele~\cite{wong2016improved} except $C=1.0e9$ to match that of WCNS6-LD. The major difference between the two interpolation methods is that WCNS6-HW has zero dissipation at Nyquist limit while WCNS6-LD has the same amount of dissipation as WCNS6-Z at the limit because of the changes discussed in section~\ref{section:LD_interpolation}. This difference can improve the numerical stability of WCNS6-LD around extremely high wavenumber features.

\section{Numerical tests}
In this section, WCNS's using different interpolation methods with settings in table~\ref{table:parameters} are compared for various test problems with different number of dimensions. The third order total variation diminishing Runge--Kutta scheme~\cite{shu1989efficient} (RK-TVD) is used for time integration.

\subsection{Convergence study} \label{convergence_study}
To verify the formal order of accuracy of both conservation and advection equations in each scheme, advections of density and volume fraction disturbances are used as test problems for single-species and multi-species flows respectively.

For the single-species advection, the initial conditions in a 1D periodic domain $\left[-1, 1 \right)$ and a 2D periodic domain $[-1, 1)\times[-1, 1)$ are respectively given by:
\begin{eqnarray*}
	\left( \rho, u, p \right)
    &=&
    \left(1 + 0.5 \sin \left( \pi x \right) , 1, 1 \right), \\
	\left( \rho, u, v, p \right)
    &=&
    \left(1 + 0.5 \sin \left[ \pi \left( x + y \right) \right], 1, 1, 1 \right).
\end{eqnarray*}

\noindent Since the velocity and pressure are constant, the problems are basically linear advections of density disturbances. Therefore, the exact solutions are given respective by:
\begin{eqnarray*}
	\left( \rho, u, p \right)
    &=&
    \left(1 + 0.5 \sin \left[ \pi \left( x - t \right) \right], 1, 1 \right), \\
	\left( \rho, u, v, p \right)
    &=&
    \left(1 + 0.5 \sin \left[ \pi \left( x + y - 2t \right) \right] , 1, 1, 1 \right).
\end{eqnarray*}

\noindent Simulations using different schemes are conducted up to $t = 2$ with mesh refinements from $N = 4$ to $N = 128$ in each direction. All simulations are run with very small constant time steps in order to observe the spatial order of accuracy of different numerical schemes. $\Delta t / \Delta x = 0.01$ and $\Delta t / \Delta x = 0.005$ are chosen for 1D and 2D flows.

Tables~\ref{table:1D_L2_error_and_rate_of_convergence_single_species} and \ref{table:2D_L2_error_and_rate_of_convergence_single_species} show the $L_2$ errors and the computed rates of convergence of density for the 1D and 2D single-species advection problems from different schemes at $t=2$. From the tables, we can see that all schemes achieve the expected rates of convergence in the conservation equations. WCNS5-JS and WCNS5-Z are essentially fifth order accurate while both WCNS6-CU-M2 and WCNS6-LD also achieve the desired sixth order accuracy. Comparing the $L_2$ errors among the schemes, it is clear that WCNS's with CU-M2 and LD interpolations lead to smaller numerical errors over the upwind-biased JS and Z interpolations. The difference between WCNS6-CU-M2 and WCNS6-LD interpolations is almost unnoticeable in these two smooth flow problems.

\begin{table}[!ht]
\centering
\begin{tabular}{| c | c c | c c | c c | c c|}
	\hline
    Number & \multicolumn{2}{c|}{WCNS5-JS} & \multicolumn{2}{c|}{WCNS5-Z} & \multicolumn{2}{c|}{WCNS6-CU-M2} & \multicolumn{2}{c|}{WCNS6-LD} \\
    \cline{2-9}
    of points& error & order & error & order & error & order & error & order \\
    \hline
    4 & 4.695e-01 & & 4.194e-01 & & 2.379e-01 & & 2.188e-01 &  \\
    8 & 6.338e-02 & 2.89 & 1.795e-02 & 4.55 & 5.167e-03 & 5.53 & 5.115e-03 & 5.42 \\
    16 & 4.025e-03 & 3.98 & 5.484e-04 & 5.03 & 8.838e-05 & 5.87 & 8.830e-05 & 5.86 \\
    32 & 1.390e-04 & 4.86 & 1.710e-05 & 5.00 & 1.415e-06 & 5.97 & 1.415e-06 & 5.96 \\
    64 & 4.263e-06 & 5.03 & 5.364e-07 & 4.99 & 2.224e-08 & 5.99 & 2.224e-08 & 5.99 \\
    128 & 1.310e-07 & 5.02 & 1.680e-08 & 5.00 & 3.484e-10 & 6.00 & 3.484e-10 & 6.00 \\
    \hline
\end{tabular}
\caption{$L_2$ errors and orders of convergence of density for the 1D single-species advection problem from different schemes at $t = 2$.}
\label{table:1D_L2_error_and_rate_of_convergence_single_species}
\end{table}

\begin{table}[!ht]
\centering
\begin{tabular}{| c | c c | c c | c c | c c|}
	\hline
    Number & \multicolumn{2}{c|}{WCNS5-JS} & \multicolumn{2}{c|}{WCNS5-Z} & \multicolumn{2}{c|}{WCNS6-CU-M2} & \multicolumn{2}{c|}{WCNS6-LD} \\
    \cline{2-9}
    of points& error & order & error & order & error & order & error & order \\
    \hline
    $4^2$ & 7.286e-01 & & 7.292e-01 & & 6.467e-01 & & 6.036e-01 &  \\
    $8^2$ & 1.687e-01 & 2.11 & 5.050e-02 & 3.85 & 1.461e-02 & 5.47 & 1.447e-02 & 5.38 \\
    $16^2$ & 1.029e-02 & 4.03 & 1.555e-03 & 5.02 & 2.500e-04 & 5.87 & 2.498e-04 & 5.86 \\
    $32^2$ & 3.856e-04 & 4.74 & 4.869e-05 & 5.00 & 4.002e-06 & 5.97 & 4.001e-06 & 5.96 \\
    $64^2$ & 1.204e-05 & 5.00 & 1.530e-06 & 4.99 & 6.291e-08 & 5.99 & 6.291e-08 & 5.99 \\
    $128^2$ & 3.726e-07 & 5.01 & 4.795e-08 & 5.00 & 9.855e-10 & 6.00 & 9.855e-10 & 6.00 \\
    \hline
\end{tabular}
\caption{$L_2$ errors and orders of convergence of density for the 2D single-species advection problem from different schemes at $t = 2$.}
\label{table:2D_L2_error_and_rate_of_convergence_single_species}
\end{table}

For multi-species advection, the initial conditions in a 1D periodic domain $\left[-1, 1 \right)$ and a 2D periodic domain $[-1, 1)\times[-1, 1)$ are respectively given by:
\begin{eqnarray*}
	\left( \rho_1, \rho_2, u, p, Z_1 \right)
    &=&
    \left(2, 1, 1, 1, 0.5 + 0.25 \sin \left( \pi x \right) \right), \\
	\left( \rho_1, \rho_2, u, v, p, Z_1 \right)
    &=&
    \left(2, 1, 1, 1, 1, 0.5 + 0.25 \sin \left[ \pi (x + y) \right] \right).
\end{eqnarray*}

\noindent The exact solutions are given respectively by:
\begin{eqnarray*}
	\left( \rho_1, \rho_2, u, p, Z_1 \right)
    &=&
    \left(2, 1, 1, 1, 0.5 + 0.25 \sin \left[ \pi (x - t) \right] \right), \\
	\left( \rho_1, \rho_2, u, v, p, Z_1 \right)
    &=&
    \left(2, 1, 1, 1, 1, 0.5 + 0.25 \sin \left[ \pi (x + y - 2t) \right] \right).
\end{eqnarray*}

\noindent Similar to the single-species advection problems, simulations by different schemes are conducted up to $t = 2$ with mesh refinements from $N = 4$ to $N = 128$ in each direction. The ratios of specific heats are 1.6 and 1.4 respectively for the two gases. $\Delta t / \Delta x = 0.01$ and $\Delta t / \Delta x = 0.005$ are used for 1D and 2D flows respectively.

Tables~\ref{table:1D_L2_error_and_rate_of_convergence_multispecies} and \ref{table:2D_L2_error_and_rate_of_convergence_multispecies} show the $L_2$ errors and the computed rates of convergence of volume fraction $Z_1$ respectively of the 1D and 2D multi-species advection problems from various schemes at $t=2$. It can be seen that all schemes achieve the expected rates of convergence for the non-conservative advection equation in smooth multi-species flows.

\begin{table}[!ht]
\centering
\begin{tabular}{| c | c c | c c | c c | c c|}
	\hline
    Number & \multicolumn{2}{c|}{WCNS5-JS} & \multicolumn{2}{c|}{WCNS5-Z} & \multicolumn{2}{c|}{WCNS6-CU-M2} & \multicolumn{2}{c|}{WCNS6-LD} \\
    \cline{2-9}
    of points& error & order & error & order & error & order & error & order \\
    \hline
    4 & 2.347e-01 & & 2.097e-01 & & 1.190e-01 & & 1.094e-01 &  \\
    8 & 3.169e-02 & 2.89 & 8.975e-03 & 4.55 & 2.583e-03 & 5.53 & 2.558e-03 & 5.42 \\
    16 & 2.013e-03 & 3.98 & 2.742e-04 & 5.03 & 4.419e-05 & 5.87 & 4.415e-05 & 5.86 \\
    32 & 6.951e-05 & 4.86 & 8.548e-06 & 5.00 & 7.074e-07 & 5.97 & 7.073e-07 & 5.96 \\
    64 & 2.132e-06 & 5.03 & 2.682e-07 & 4.99 & 1.112e-08 & 5.99 & 1.112e-08 & 5.99 \\
    128 & 6.550e-07 & 5.02 & 8.398e-09 & 5.00 & 1.743e-10 & 6.00 & 1.743e-10 & 6.00 \\
    \hline
\end{tabular}
\caption{$L_2$ errors and orders of convergence of volume fraction for the 1D multi-species advection problem from different schemes at $t = 2$.}
\label{table:1D_L2_error_and_rate_of_convergence_multispecies}
\end{table}

\begin{table}[!ht]
\centering
\begin{tabular}{| c | c c | c c | c c | c c|}
	\hline
    Number & \multicolumn{2}{c|}{WCNS5-JS} & \multicolumn{2}{c|}{WCNS5-Z} & \multicolumn{2}{c|}{WCNS6-CU-M2} & \multicolumn{2}{c|}{WCNS6-LD} \\
    \cline{2-9}
    of points& error & order & error & order & error & order & error & order \\
    \hline
    $4^2$ & 3.650e-01 & & 3.646e-01 & & 3.234e-01 & & 3.018e-01 &  \\
    $8^2$ & 8.426e-02 & 2.12 & 2.498e-02 & 3.87 & 7.307e-03 & 5.47 & 7.234e-03 & 5.38 \\
    $16^2$ & 5.126e-03 & 4.04 & 7.752e-04 & 5.01 & 1.250e-04 & 5.87 & 1.249e-04 & 5.86 \\
    $32^2$ & 1.918e-04 & 4.74 & 2.419e-05 & 5.00 & 2.001e-06 & 5.97 & 2.001e-06 & 5.96 \\
    $64^2$ & 5.973e-06 & 5.01 & 7.592e-07 & 4.99 & 3.146e-08 & 5.99 & 3.146e-08 & 5.99 \\
    $128^2$ & 1.846e-07 & 5.02 & 2.377e-08 & 5.00 & 4.932e-10 & 6.00 & 4.932e-10 & 6.00 \\
    \hline
\end{tabular}
\caption{$L_2$ errors and orders of convergence of volume fraction for the 2D multi-species advection problem from different schemes at $t = 2$.}
\label{table:2D_L2_error_and_rate_of_convergence_multispecies}
\end{table}

\subsection{Single-species test problems}

\subsubsection{Sod shock tube}
This is a 1D shock tube problem introduced by Sod~\cite{sod1978survey}. The problem consists of the propagation of a shock wave, a contact discontinuity and an expansion fan. The initial conditions are given by:
\begin{equation*}
\begin{aligned}
	\left( \rho, u, p \right)
    =
    \begin{cases}
    	\left(1, 0, 1 \right), &\mbox{$x < 0$}, \\
    	\left(0.125, 0, 0.1 \right), &\mbox{$x \geq 0$}. \\
    \end{cases}
\end{aligned}
\end{equation*}

\noindent The ratio of specific heats $\gamma$ is 1.4. The computational domain has size $x \in \left[-0.5, 0.5 \right]$. Simulations are performed with constant time steps $\Delta t = 0.002$ on a uniform grid composed of 100 grid points where $\Delta x = 0.01$.

Comparison of different fields from the numerical solutions to the exact solution at $t = 0.2$ is shown in figure~\ref{fig:compare_Sod_rho}. All schemes can capture the contact discontinuity without spurious oscillations. However, only WCNS5-JS, WCNS5-Z and WCNS6-LD can capture the shock wave well, while WCNS6-CU-M2 produces a very large overshoot near the shock in solutions of different fields. This may be due to inadequate introduction of numerical dissipation at the shock for stabilization. The more localized dissipative nature of WCNS6-LD allows it to capture the shock wave more sharply compared to WCNS5-JS and WCNS5-Z.

\begin{figure}[!ht]
\centering
\subfigure[Global density profile]{%
\includegraphics[width=0.48\textwidth]{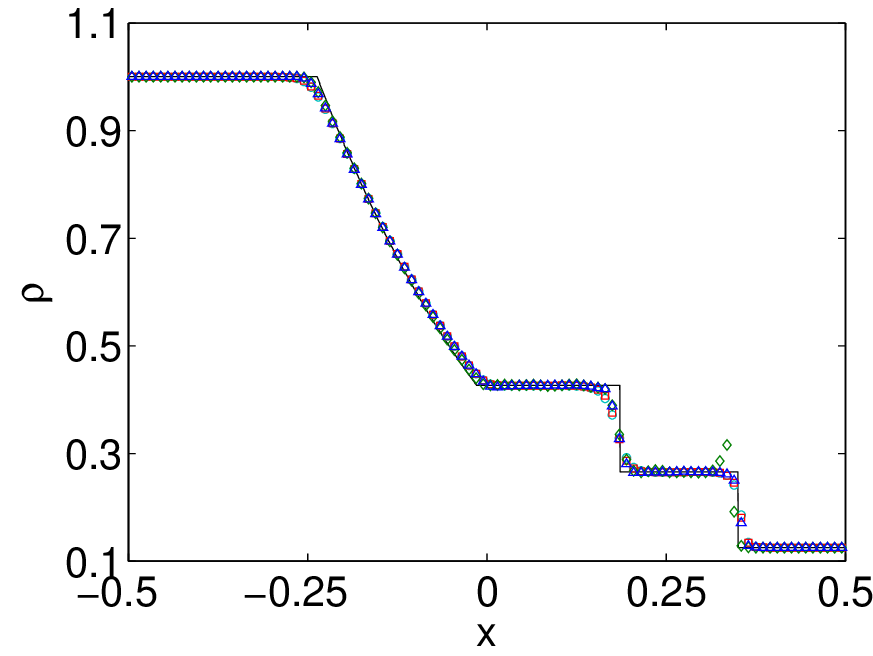}
\label{fig:compare_Sod_rho_global}}
\subfigure[Local density profile]{%
\includegraphics[width=0.48\textwidth]{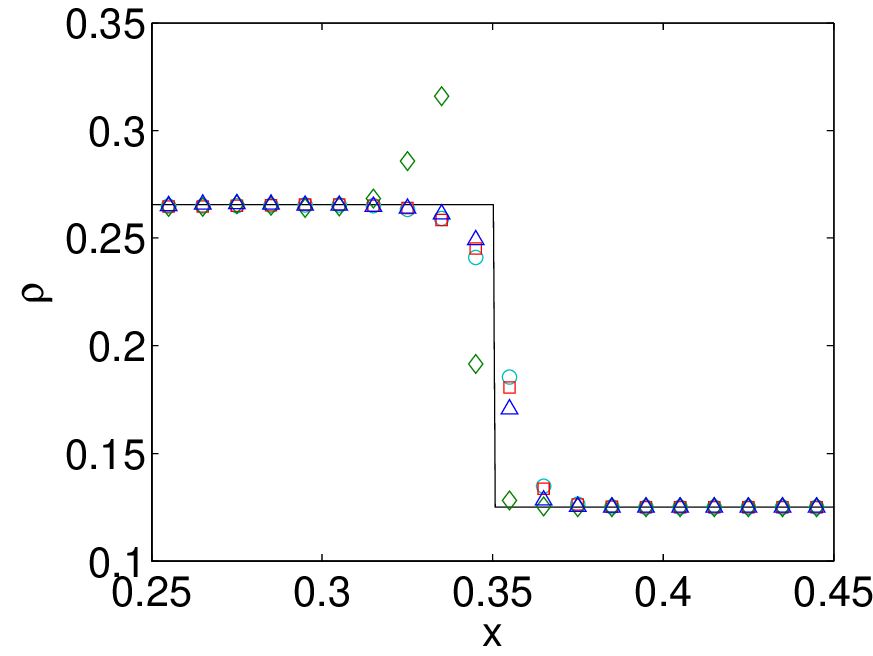}
\label{fig:compare_Sod_rho_local}}
\subfigure[Velocity profile]{%
\includegraphics[width=0.48\textwidth]{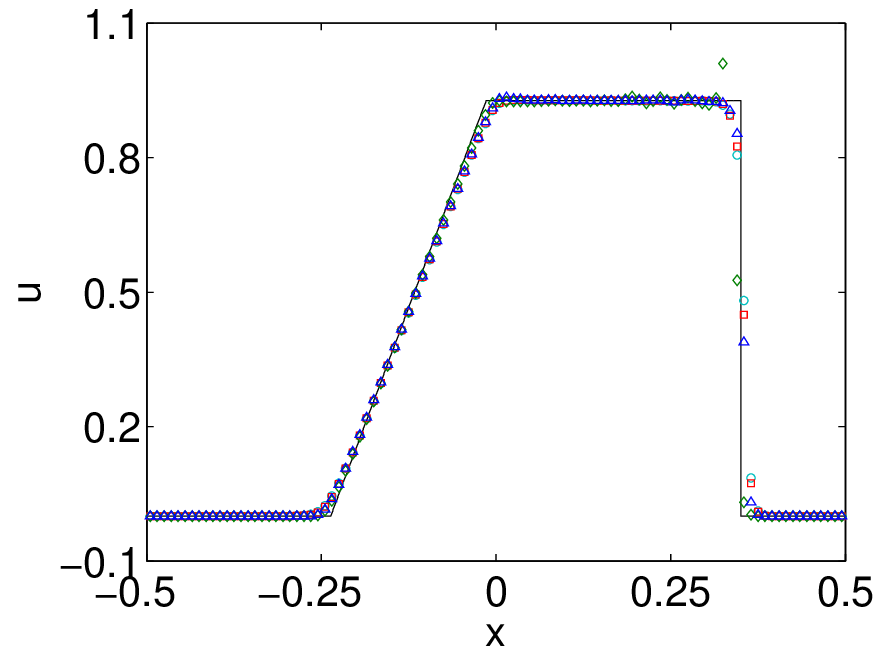}
\label{fig:compare_Sod_u_global}}
\subfigure[Pressure profile]{%
\includegraphics[width=0.48\textwidth]{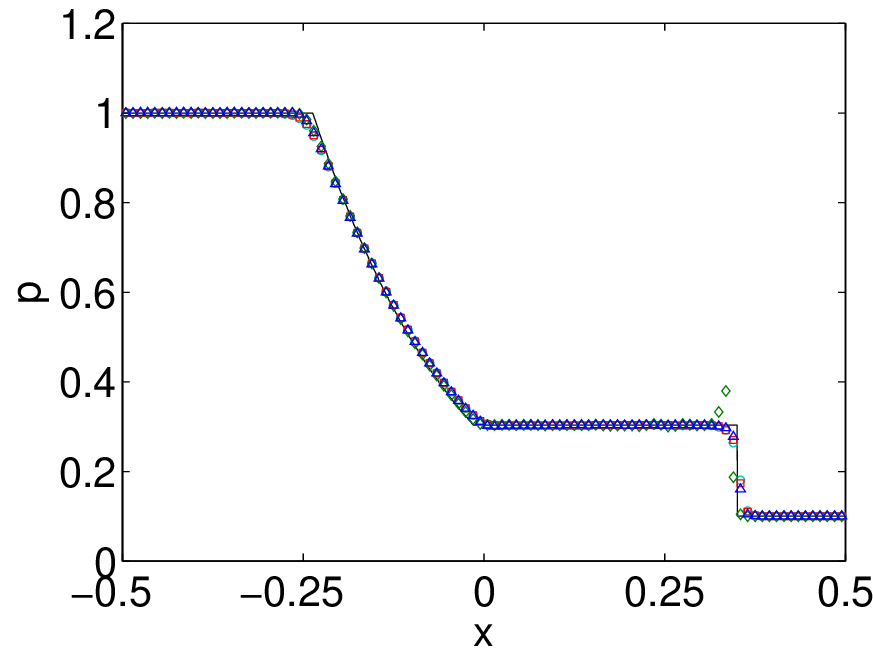}
\label{fig:compare_Sod_p_global}}
\caption{Sod shock tube problem at $t = 0.2$ using different schemes. Black solid line: exact; cyan circles: WCNS5-JS; red squares: WCNS5-Z; green diamonds: WCNS6-CU-M2; blue triangles: WCNS6-LD.}
\label{fig:compare_Sod_rho}
\end{figure}

\iffalse
\begin{figure}[t!]
\centering
\subfigure[Velocity profile]{%
\includegraphics[width=0.48\textwidth]{Sod_WCNS_u_global.eps}
\label{fig:compare_Sod_u_global}}
\subfigure[Pressure profile]{%
\includegraphics[width=0.48\textwidth]{Sod_WCNS_p_global.eps}
\label{fig:compare_Sod_p_local}}
%
\caption{Profiles of Sod shock tube problem at $t = 0.2$ using different schemes. Black solid line: exact; cyan circles: WCNS5-JS; red squares: WCNS5-Z; green diamonds: WCNS6-CU-M2; blue triangles: WCNS6-LD.}
\label{fig:compare_Sod_rho}
\end{figure}
\fi

\subsubsection{Shu--Osher problem}
This 1D problem was first proposed by Shu and Osher~\cite{shu1988efficient}. It involves the interaction of a Mach 3 shock wave with a fluctuation density field. This problem can test the scale-separation capabilities of different schemes in capturing discontinuities and resolving smooth fluctuating waves. The initial conditions are given by:
\begin{equation*}
\begin{aligned}
	\left( \rho, u, p \right)
    =
    \begin{cases}
    	\left(27/7, 4 \sqrt{35}/9, 31/3 \right), &\mbox{$x < -4$}, \\
    	\left(1 + 0.2 \sin{(5x)}, 0, 1 \right), &\mbox{$x \geq -4$}. \\
    \end{cases}
\end{aligned}
\end{equation*}

\noindent The ratio of specific heats $\gamma$ is 1.4. The spatial domain of the problem is $x \in \left[-5, 5 \right]$. Simulations are conducted with constant time steps $\Delta t = 0.004$ on a uniform grid with 200 grid points where $\Delta x = 0.05$.

\begin{figure}[!ht]
\centering
\subfigure[Global density profile]{%
\includegraphics[width=0.48\textwidth]{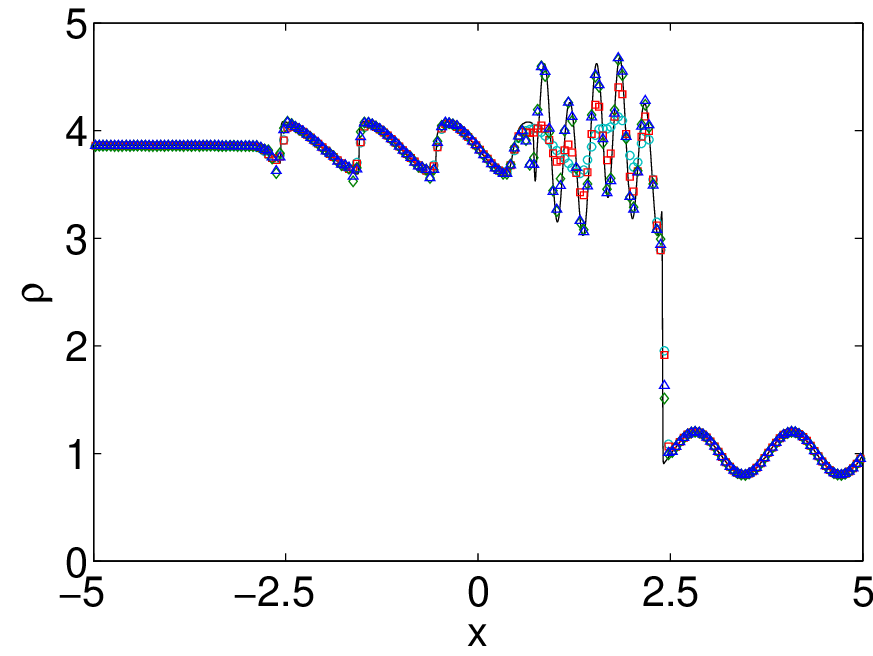}
\label{fig:compare_Shu_Osher_rho_global}}
\subfigure[Local density profile]{%
\includegraphics[width=0.48\textwidth]{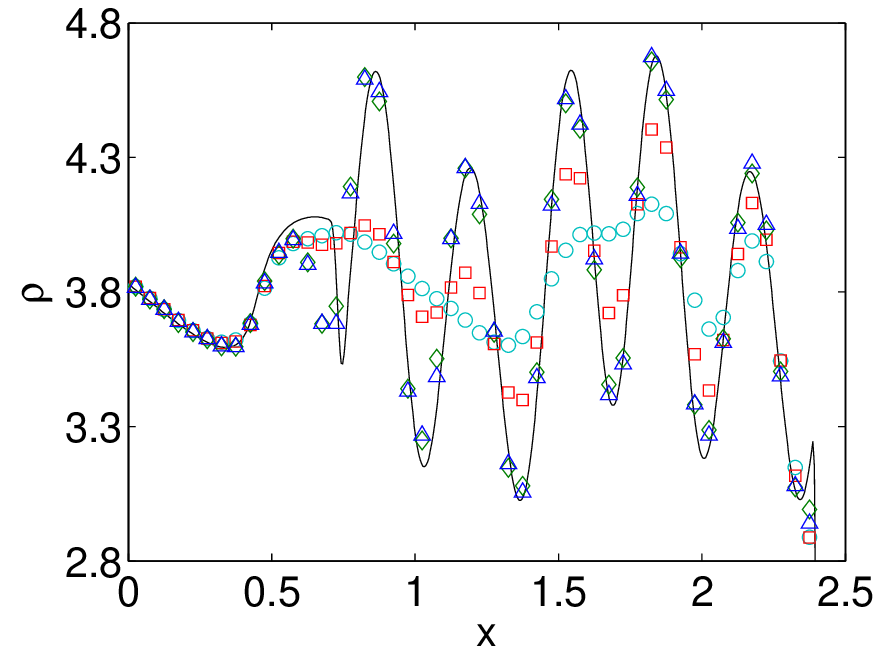}
\label{fig:compare_Shu_Osher_rho_local}}
\caption{Shu--Osher problem at $t = 1.8$ using different schemes. Black solid line: reference; cyan circles: WCNS5-JS; red squares: WCNS5-Z; green diamonds: WCNS6-CU-M2; blue triangles: WCNS6-LD.}
\label{fig:compare_Shu_Osher_rho}
\end{figure}

Figure~\ref{fig:compare_Shu_Osher_rho} shows the comparison between the reference solution and the numerical simulations for density from different schemes at $t = 1.8$. The reference solution is obtained on a 2000-point grid with a seventh order WENO7-JS scheme with global Lax--Friedrichs flux splitting. All of the schemes can capture the shock wave without spurious oscillations. The improved WCNS5-Z resolves the fluctuating waves better compared to WCNS5-JS. Due to smaller numerical diffusivity in smooth regions, both WCNS6-CU-M2 and WCNS6-LD give almost equivalently better results with regards to the resolution of the density waves over the upwind-biased WCNS5-JS and WCNS5-Z.

\subsubsection{Titarev--Toro problem}
Titarev--Toro problem is a more severe extension of the Shu--Osher problem. The initial conditions for this problem are given by:

\begin{equation*}
\begin{aligned}
	\left( \rho, u, p \right)
    =
    \begin{cases}
    	\left(1.515695, 0.523326, 1.805 \right), &\mbox{$x < -4.5$}, \\
    	\left(1 + 0.1 \sin{(20 \pi x)}, 0, 1 \right), &\mbox{$x \geq -4.5$}. \\
    \end{cases}
\end{aligned}
\end{equation*}

\noindent The ratio of specific heats $\gamma$ is 1.4. The spatial domain of the problem is $x \in \left[-5, 5 \right]$. Simulations are conducted with constant time steps $\Delta t = 0.002$ on a uniform grid with 1000 grid points where $\Delta x = 0.01$.

\begin{figure}[!ht]
\centering
\subfigure[Global density profile]{%
\includegraphics[width=0.48\textwidth]{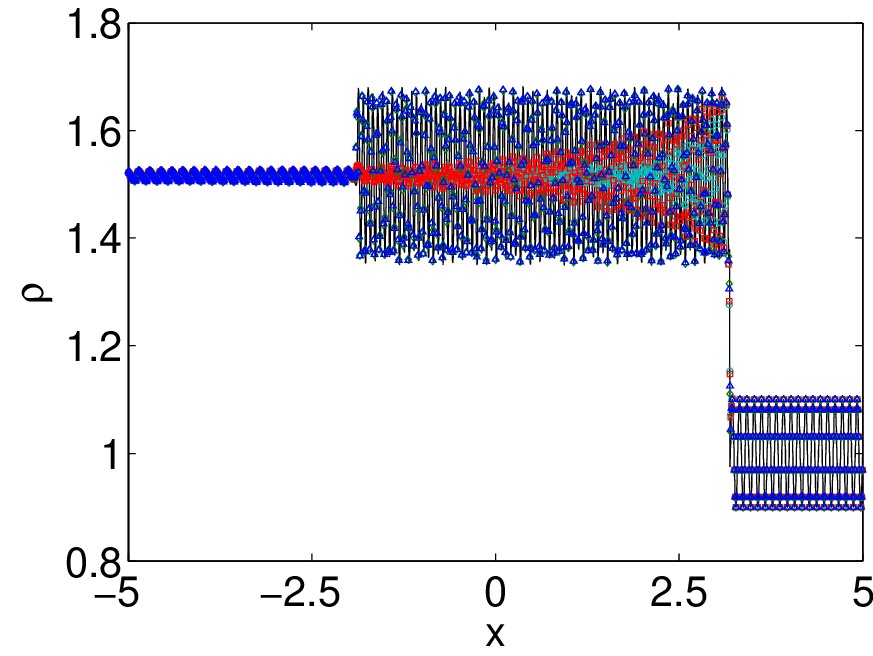}
\label{fig:compare_Titarev_Toro_rho_global}}
\subfigure[Local density profile]{%
\includegraphics[width=0.48\textwidth]{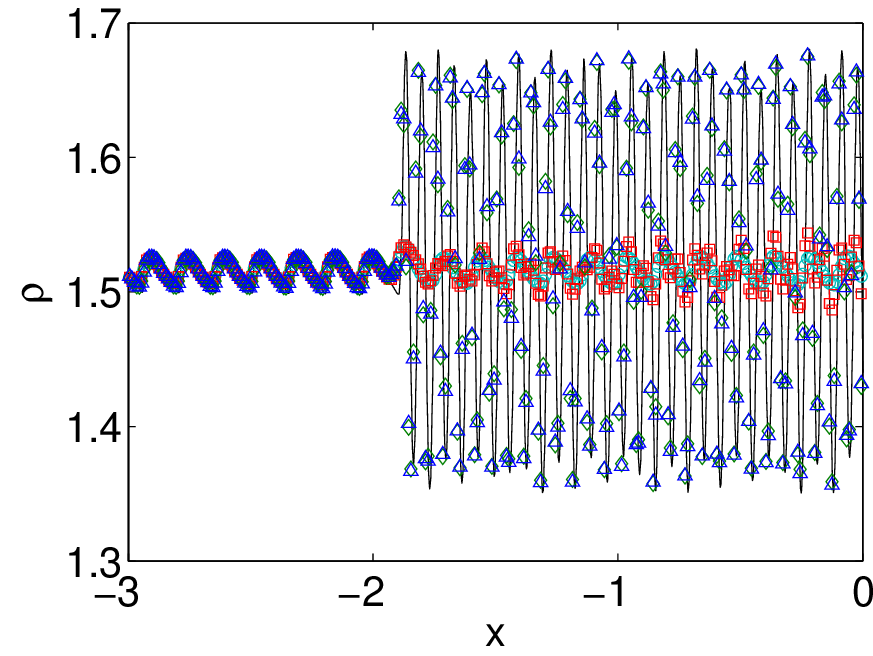}
\label{fig:compare_Titarev_Toro_rho_local}}
\caption{Titarev--Toro problem at $t = 5$ using different schemes. Black solid line: reference; cyan circles: WCNS5-JS; red squares: WCNS5-Z; green diamonds: WCNS6-CU-M2; blue triangles: WCNS6-LD.}
\label{fig:compare_Titarev_Toro_rho}
\end{figure}

Figure~\ref{fig:compare_Titarev_Toro_rho} shows the comparison between the reference solution and the numerical simulations for density from different schemes at $t = 5$. The reference solution is obtained on a 5000-point grid with a seventh order WENO7-JS scheme with global Lax--Friedrichs flux splitting. Similar to the Shu--Osher test case, both WCNS6-CU-M2 and WCNS6-LD resolve the local large amplitude density fluctuations much better than that from both upwind-biased WCNS5-JS and WCNS5-Z.

\subsubsection{Double Mach reflection}
This is a 2D problem with the domain size of $\left[ 0, 4 \right] \times \left[ 0, 1 \right]$ by Woodward and Colella~\cite{woodward1984numerical}. The initial conditions are given by:
\begin{equation*}
\begin{aligned}
	\left( \rho, u, v, p \right)
    =
    \begin{cases}
    	\left(8, 8.25\cos{\left( \frac{\pi}{6} \right)}, -8.25\sin{\left( \frac{\pi}{6} \right)}, 116.5 \right), &\mbox{$x < \frac{1}{6} + \frac{y}{\sqrt{3}}$}, \\
    	\left(1.4, 0, 0, 1 \right), &\mbox{$x \geq \frac{1}{6} + \frac{y}{\sqrt{3}}$}. \\
    \end{cases}
\end{aligned}
\end{equation*}

\noindent A Mach 10 shock initially makes a $60 ^{\circ}$ angle  with the horizontal wall at location $x = 1/6$ of the bottom boundary. As the shock moves and reflects on the wall, a complex shock structure with two triple points and a slip line evolves. The same initial and boundary conditions as Woodward and Colella~\cite{woodward1984numerical} are used. Ahead of the shock is undisturbed stationary air of $\gamma = 1.4$ with density of 1.4 and pressure of 1.0. At the bottom boundary, the conditions in the region $x \in \left[ 0, 1/6 \right]$ are fixed at the post-shock flow conditions and reflecting boundary conditions are used for $x \geq 1/6$. The left boundary is set at the post-shock flow conditions and zero-gradient conditions are applied at the right boundary. Time-dependent conditions are applied on the top boundary to match the movement of the shock wave. The simulations are conducted with constant Courant--Friedrichs--Lewy number, $\textnormal{CFL}=0.5$ until $t = 0.2$.

\begin{figure}[!ht]
\centering
\subfigure[WCNS5-JS]{%
\includegraphics[height=0.29\textwidth]{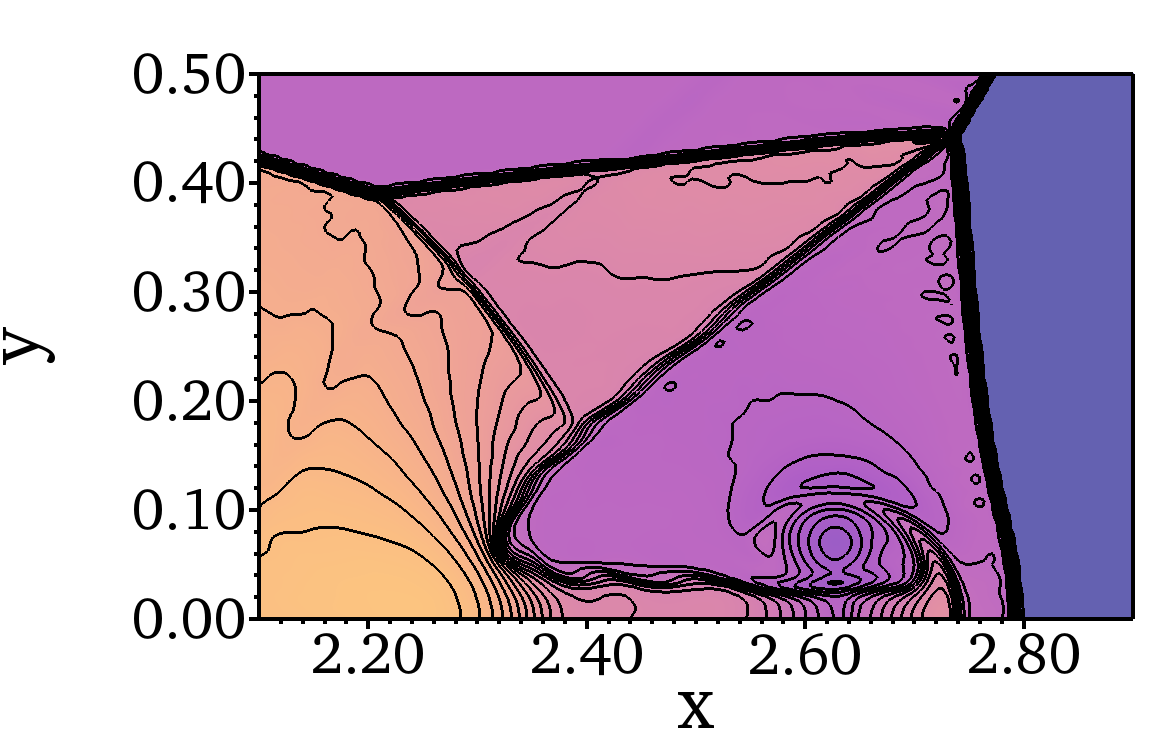}
\label{fig:double_Mach_reflection_JS}}\hspace{0em}%
\subfigure[WCNS5-Z]{%
\includegraphics[height=0.29\textwidth]{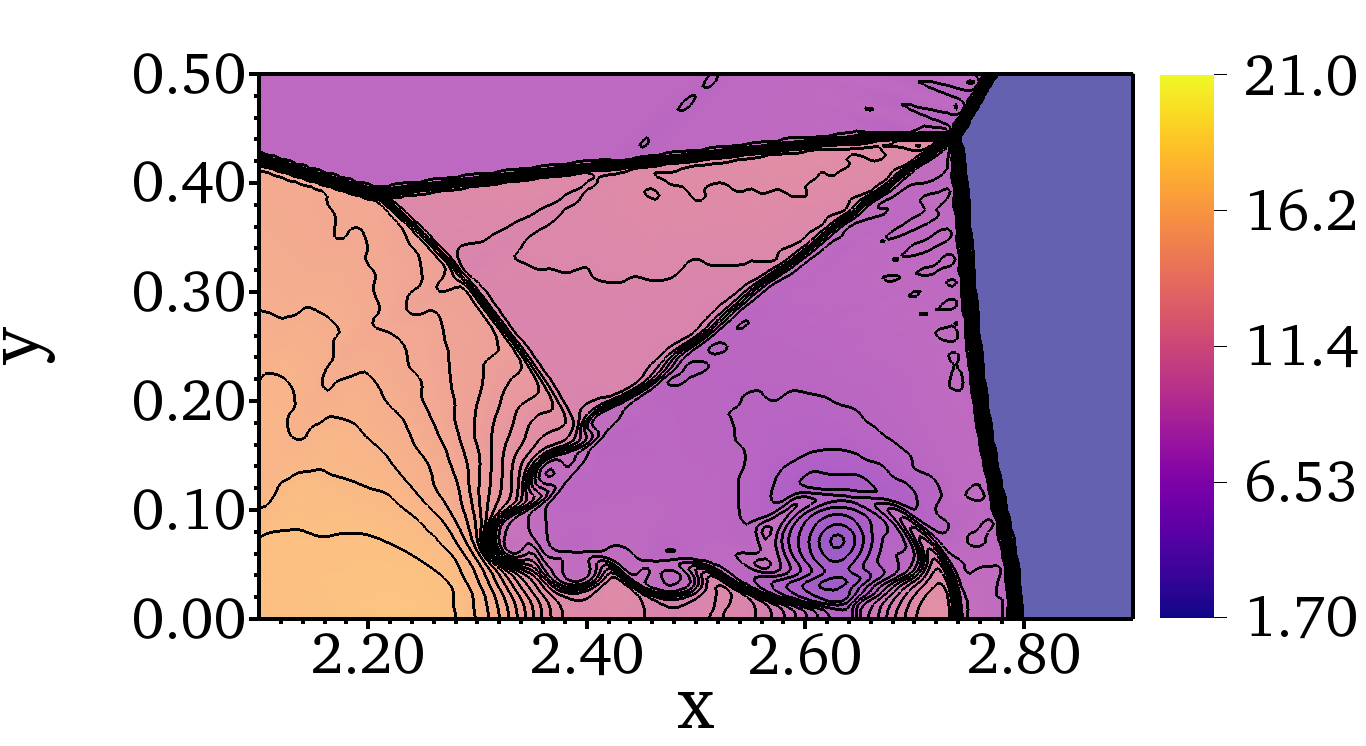}
\label{fig:double_Mach_reflection_Z}}
\subfigure[WCNS6-CU-M2]{%
\includegraphics[height=0.29\textwidth]{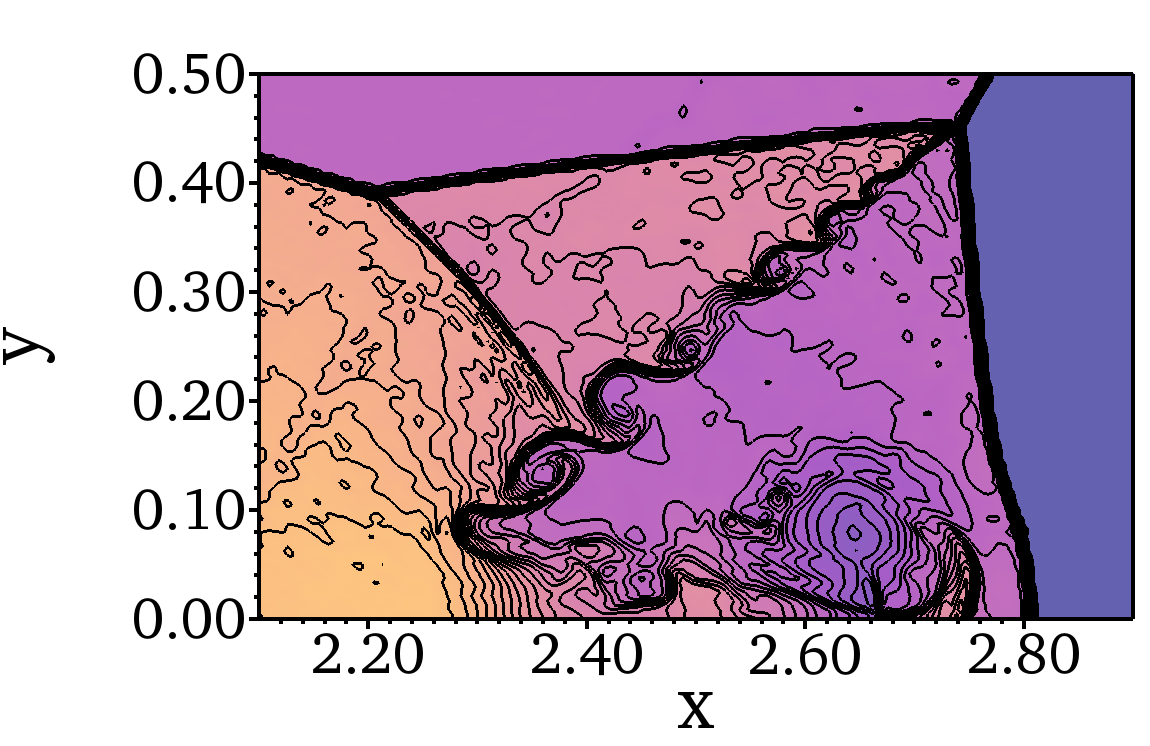}
\label{fig:double_Mach_reflection_CU_M2}}\hspace{0em}%
\subfigure[WCNS6-LD]{%
\includegraphics[height=0.29\textwidth]{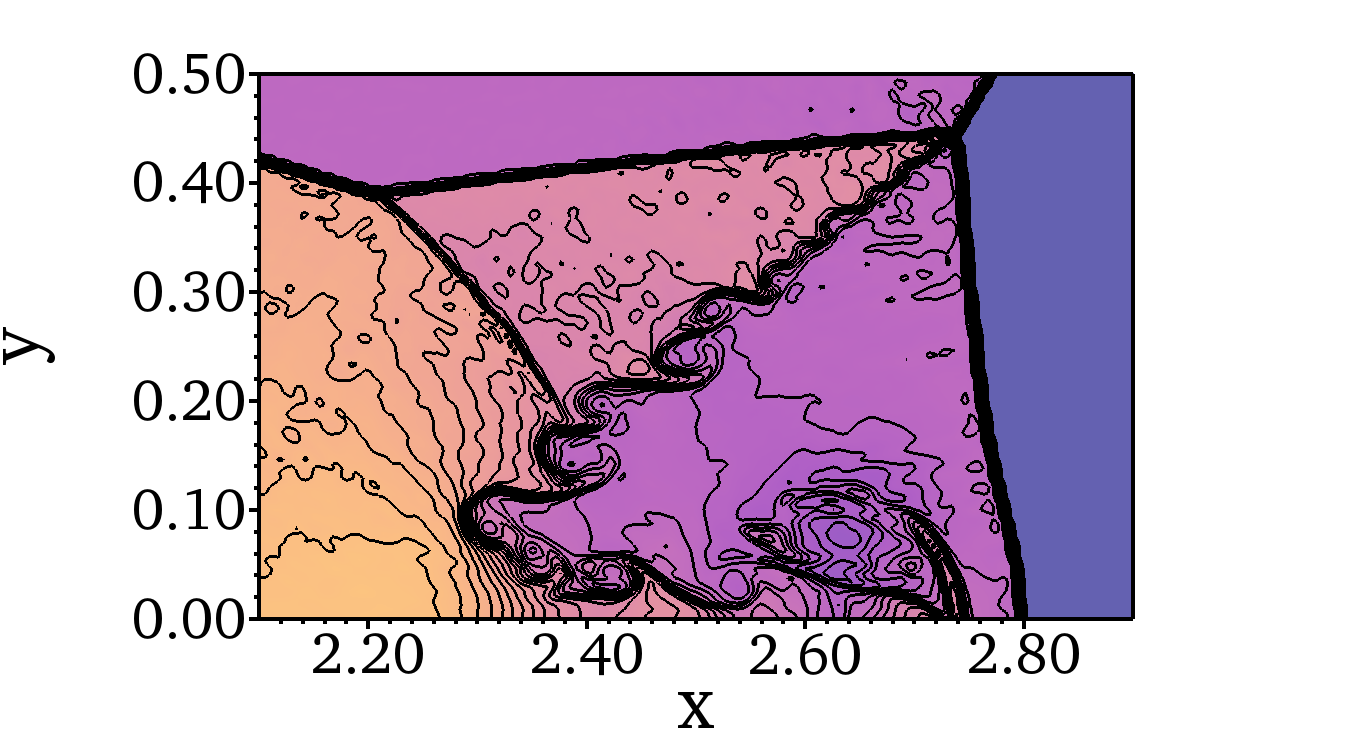}
\label{fig:double_Mach_reflection_LD}}
\caption{30 equally spaced contours of density from 1.7 to 21 at $t = 0.2$ using different schemes in the blown-up region around the Mach stem for the double Mach reflection problem with $\Delta x = \Delta y = 1/240$.}
\label{fig:compare_double_Mach_reflection}
\end{figure}

Since inviscid Euler equations are solved, there is no physical dissipation in this test problem. The vortices along the discontinuous slip line due to Kelvin--Helmholtz instability are only damped by numerical dissipation. More localized numerical dissipation at the slip line can help capture rolled-up vortices with more details along it. From figure~\ref{fig:compare_double_Mach_reflection}, we can notice that with the same mesh resolution of $960 \times 240$, WCNS5-JS is too dissipative to produce any rolled-up vortices along the slip line and using WCNS5-Z only slightly improves the resolution of vortices there. On the other hand, both WCNS6-CU-M2 and WCNS6-LD can capture much more small-scale vortical structures along the slip line as smaller and more localized dissipation is applied at the discontinuity. It is also seen that the Mach stem of WCNS6-CU-M2 is more kinked compare to other WCNS's and this may be due to insufficient addition of numerical dissipation for stability at shock.

\subsubsection{Taylor--Green vortex}
We also consider the 3D Taylor--Green vortex problem to compare the scale-separation capabilities of different methods. The initial conditions of the problem are given by:
\begin{equation*}
	\begin{pmatrix}
		\rho \\
        u \\
        v \\
        w \\
        p \\
	\end{pmatrix}
    =
    \begin{pmatrix}
		1 \\
        \sin{x} \cos{y} \cos{z} \\
        -\cos{x} \sin{y} \cos{z} \\
        0 \\
        100 + \frac{\left( \cos{(2z)} + 2 \right) \left( \cos{(2x)} + \cos{(2y)} \right) - 2}{16}
	\end{pmatrix}.
\end{equation*}

\noindent The ratio of specific heats of the gas $\gamma$ is $5/3$. The domain is periodic with size $\left[0, 2\pi \right)^3$. Two levels of mesh resolutions $32^3$ and $64^3$ are employed. Simulations are conducted until $t = 10$ with a constant $\textnormal{CFL}=0.6$. 

This flow problem is essentially incompressible as the mean pressure is chosen to be very large. This implies that the kinetic energy of the flow is conserved over time. As time evolves, the inviscid Taylor--Green vortex in the initial flow stretches and produces smaller and smaller scale features. This problem can be used as a test to examine the scale-separation ability of different schemes to under-resolved flow. We can also compare the ability of different schemes to preserve kinetic energy and predict the growth of enstrophy.

\begin{figure}[!ht]
\centering
\subfigure[WCNS5-JS]{%
\includegraphics[height=0.46\textwidth]{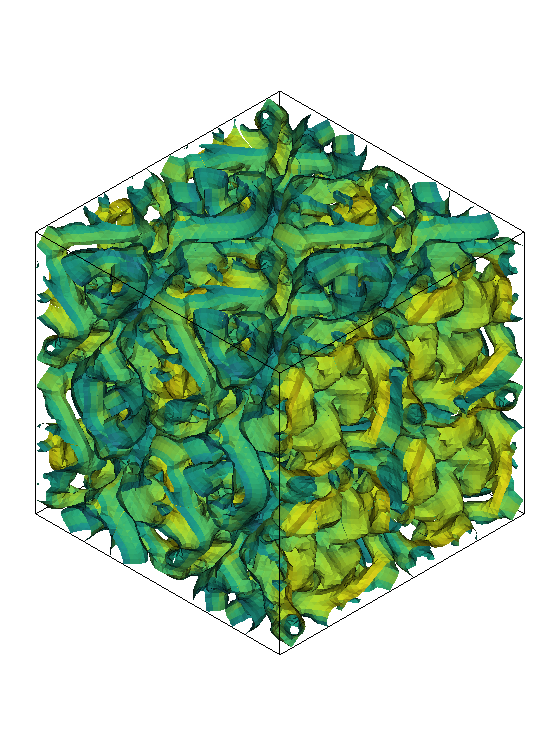}
\label{fig:Tayor_Green_iso_surface_JS}}
\subfigure[WCNS5-Z]{%
\includegraphics[height=0.46\textwidth]{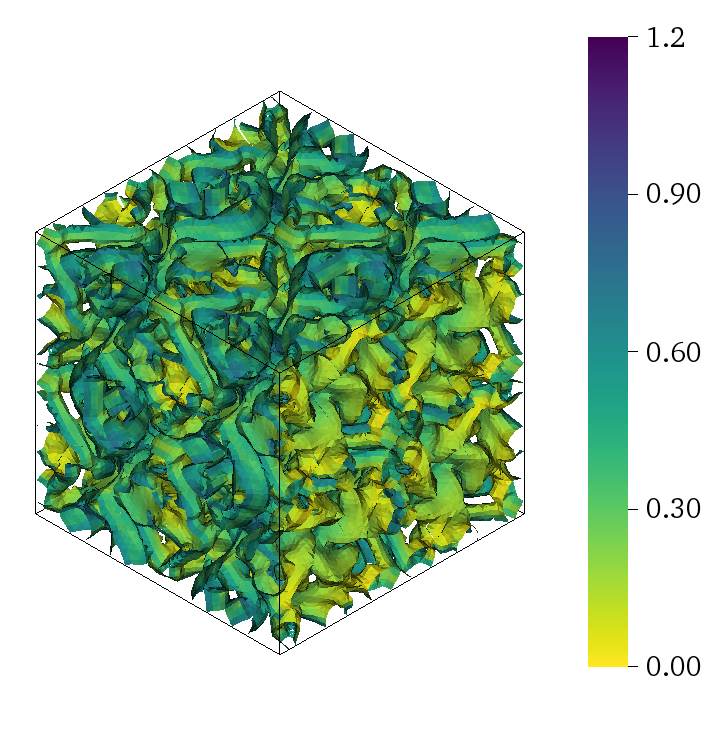}
\label{fig:Tayor_Green_iso_surface_Z}}
\subfigure[WCNS6-CU-M2]{%
\includegraphics[height=0.46\textwidth]{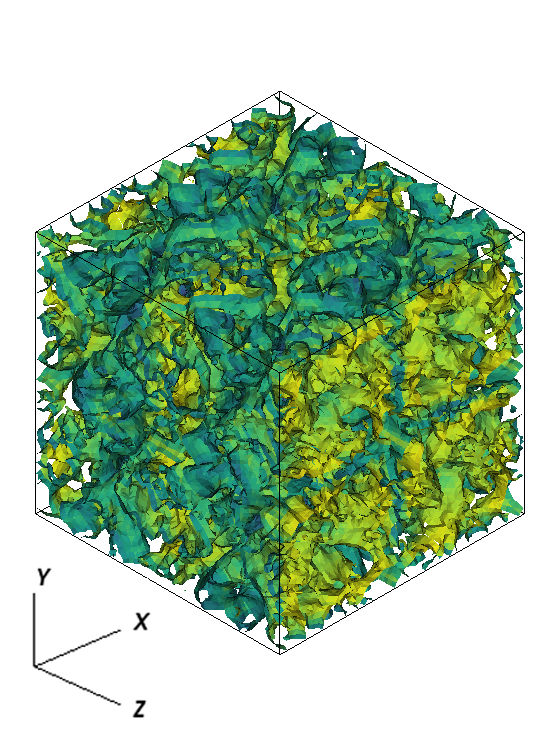}
\label{fig:Tayor_Green_iso_surface_CU_M2}}
\subfigure[WCNS6-LD]{%
\includegraphics[height=0.46\textwidth]{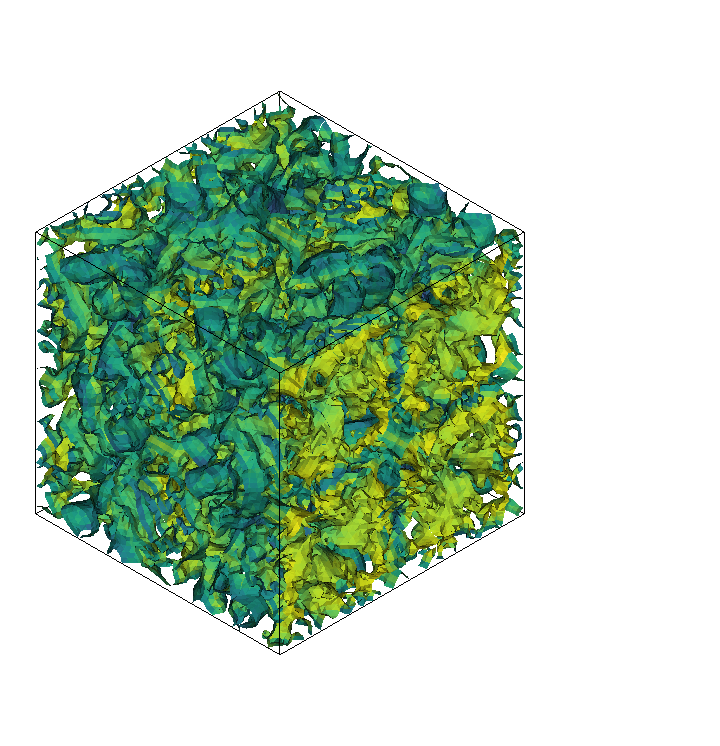}
\label{fig:Tayor_Green_iso_surface_LD}}
\caption{Iso-surfaces of zero Q-criterion, colored by velocity magnitude, for the Taylor--Green vortex problem on the $64^3$ grid at $t = 8$ using different schemes.}
\label{fig:compare_Taylor_Green_iso_surface}
\end{figure}

\begin{figure}[!ht]
\centering
\subfigure[Kinetic energy]{%
\includegraphics[width=0.48\textwidth]{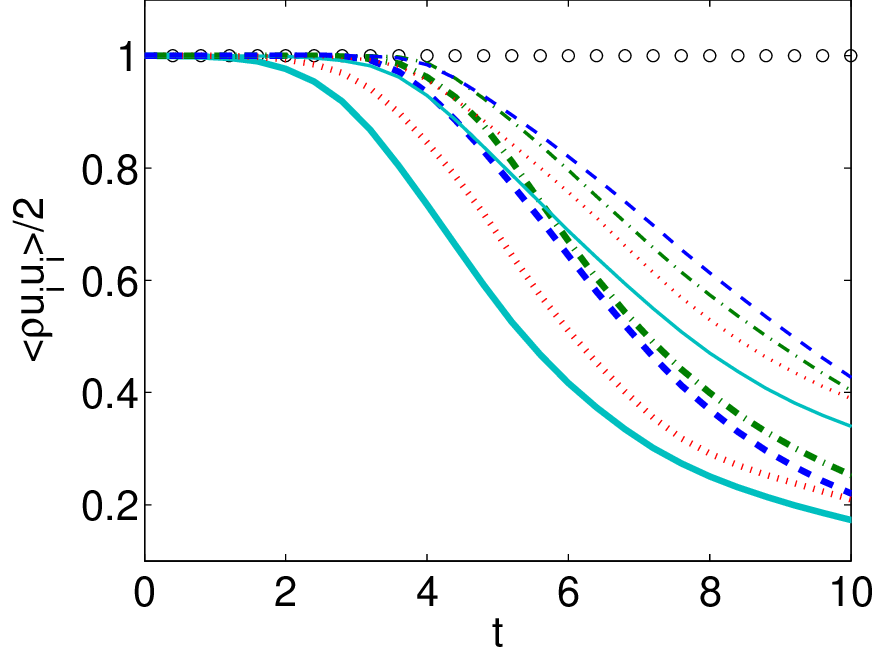}
\label{fig:compare_Taylor_Green_KE}}
\subfigure[Enstrophy]{%
\includegraphics[width=0.48\textwidth]{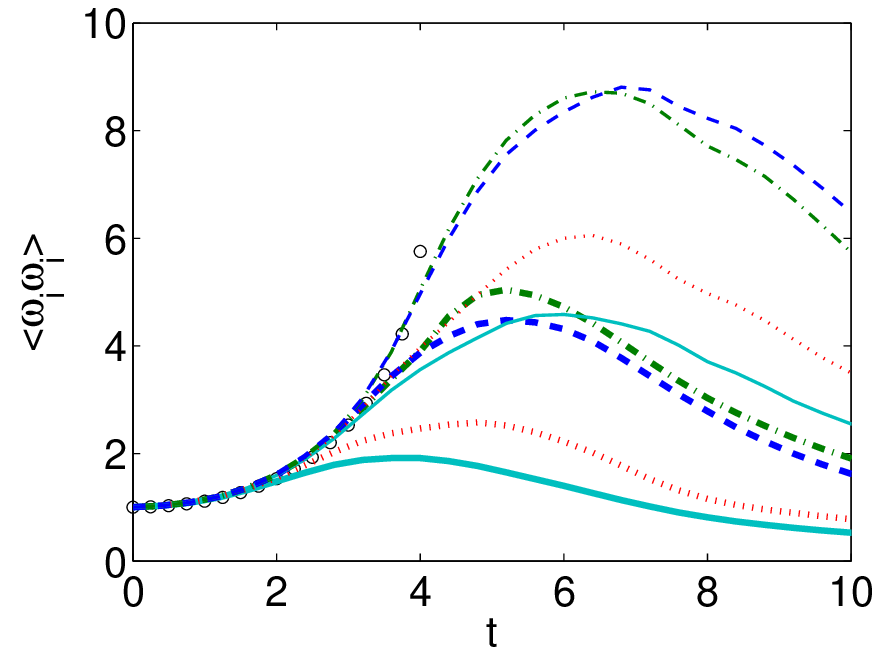}
\label{fig:compare_Taylor_Green_Omega}}
\caption{Time evolution of kinetic energy and enstrophy for the Taylor--Green vortex problem on the $32^3$ and $64^3$ grids, quantities normalized by their initial values. Thick cyan solid line: WCNS5-JS ($32^3$ grid); thick red dotted line: WCNS5-Z ($32^3$ grid); thick green dash-dotted line: WCNS6-CU-M2 ($32^3$ grid); thick blue dashed line: WCNS6-LD ($32^3$ grid); thin cyan solid line: WCNS5-JS ($64^3$ grid); thin red dotted line: WCNS5-Z ($64^3$ grid); thin green dash-dotted line: WCNS6-CU-M2 ($64^3$ grid); thin blue dashed line: WCNS6-LD ($64^3$ grid); black circles: semi-analytical results of Brachet et al.~\cite{brachet1983small}.}
\label{fig:compare_Taylor_Green_vortex}
\end{figure}

Figure~\ref{fig:compare_Taylor_Green_iso_surface} shows the iso-surfaces of zero Q-criterion at $t = 8$ from different schemes. From the iso-surfaces, we can see that both WCNS6-CU-M2 and WCNS6-LD can capture more small-scale features compared to the upwind-biased schemes WCNS5-JS and WCNS5-Z qualitatively. Figure~\ref{fig:compare_Taylor_Green_vortex} shows the temporal evolution of the mean kinetic energy, $\left\langle \rho u_i u_i \right\rangle /2$ and enstrophy, $\left\langle \omega_i \omega_i \right\rangle$, normalized by their initial values, from different schemes. At both mesh resolutions ($32^3$ and $64^3$), we can observe that WCNS5-JS is the worst among all schemes to preserve the kinetic energy over time. The improved WCNS5-Z preserves more kinetic energy but is still worse than both WCNS6-CU-M2 and WCNS6-LD. The kinetic energy of WCNS6-LD starts to decay slightly earlier than WCNS6-CU-M2 at both mesh resolutions but in the case with mesh resolution of $64^3$, the overall dissipation from WCNS6-LD is smaller than that of WCNS6-CU-M2 so its kinetic energy is higher at late times. At both resolutions, WCNS5-Z predicts the growth of enstrophy more accurately than WCNS5-JS but is worse than both sixth order WCNS's. The enstrophy computed by WCNS6-LD starts to deviate from the reference solution at earlier time compared to WCNS6-CU-M2 at both mesh resolutions but in the case with mesh resolution of $64^3$, WCNS6-LD has higher enstrophy value at late times. The difference in the time evolution of quantities between the sixth order schemes is due to the difference in their scale-separation capabilities. WCNS6-LD is more dissipative at very high wavenumber features than WCNS6-CU-M2 so its kinetic energy and enstrophy start to be damped as soon as high wavenumber features appear in the solutions. However, as the numerical dissipation added by WCNS6-LD is more local at high wavenumber features compared to WCNS6-CU-M2, the decay rates of both kinetic energy and enstrophy from WCNS6-LD are smaller as observed in the case with mesh resolution of $64^3$.

\begin{figure}[!ht]
 \centering
	\includegraphics[width=0.6\textwidth]{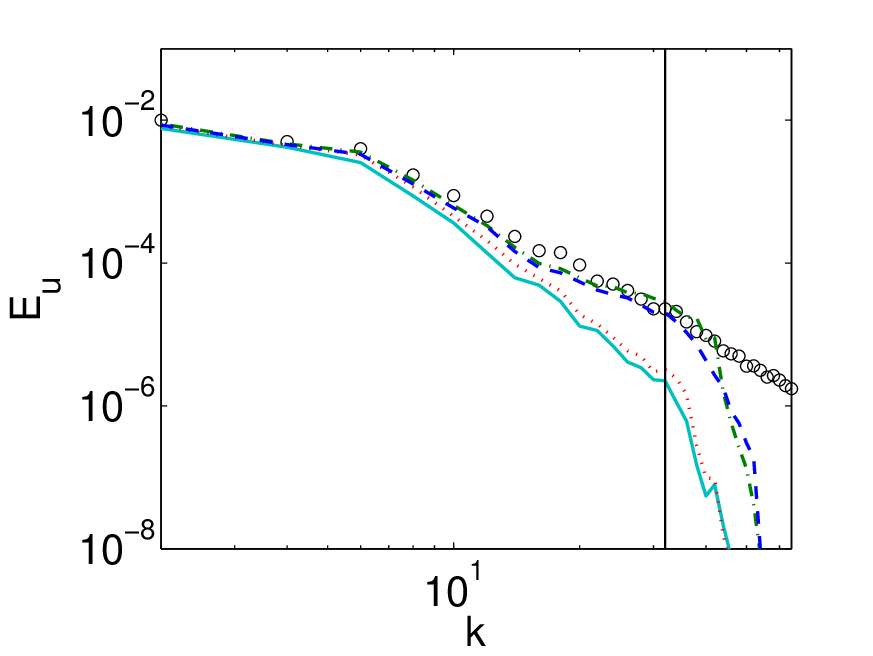}
	\caption{Spectra of $x$ velocity component for the Taylor--Green vortex problem on the $64^3$ grid at $t=5$. Cyan solid line: WCNS5-JS; red dotted line: WCNS5-Z; green dash-dotted: WCNS6-CU-M2; blue dashed line: WCNS6-LD; black circles: Converged spectrum on a $256^3$ grid of WCNS6-LD. The vertical black solid line shows the Nyquist limit of $64^3$ grid.}
    \label{fig:compare_Taylor_Green_vortex_spectra}
\end{figure}

Figure~\ref{fig:compare_Taylor_Green_vortex_spectra} compares the spectra of velocity component in $x$ direction, $E_u$, against angular wavenumber, $k$, at $t = 5$ for cases with mesh resolution $64^3$. Since no analytical spectrum is known at that time, a converged solution computed on a $256^3$ grid from WCNS6-LD is used as the reference solution. It can be seen that the traditional WCNS5-JS performs the poorest because of the lack of scale-separation capability. WCNS5-Z has improvement in velocity spectra over WCNS5-JS but the improvement is very small. Their spectra only agree well with the reference spectrum up to $k \approx 3$ and $k \approx 5$ respectively. As for the sixth order schemes, spectra of WCNS6-CU-M2 and WCNS6-LD compare well with the reference spectrum up to $k \approx 40$ and $k \approx 30$ respectively. In the case of WCNS6-LD, energy transferred from large features to small features is dissipated numerically at the Nyquist limit ($k = 32$) to prevent aliasing. The fact that the spectrum of WCNS6-CU-M2 does not deviate from the reference spectrum after Nyquist limit suggests that it may not introduce enough dissipation to very high wavenumber features to prevent aliasing.

\subsection{Multi-species test problems}

\subsubsection{Isolated material interface advection}
The first multi-species problem is a 1D problem with the advection of an isolated material interface. This problem was introduced by Johnsen et al.~\cite{johnsen2006implementation} and is slightly modified here. The initial conditions are given by:
\begin{equation*}
	\left( \rho, u, p, \gamma \right)
    =
    \begin{cases}
    	\left(10, 0.5, 1/1.4, 1.6 \right), &\mbox{$0.25 \leq x < 0.75$}, \\
        \left(1, 0.5, 1/1.4, 1.4 \right), &\mbox{$x < 0.25$ or $x \geq 0.75$}. \\
    \end{cases}
\end{equation*}

\noindent Periodic conditions are applied at both boundaries. The spatial domain is $x \in \left[0, 1 \right)$ and the final time is at $t = 2$. Simulations are evolved with constant time steps $\Delta t = 0.005$ on a uniform grid with 50 grid points where $\Delta x = 0.02$. The material interface has exactly advected one period at the end of the simulation.

\begin{figure}[!ht]
\centering
\subfigure[Global density profile]{%
\includegraphics[width=0.45\textwidth]{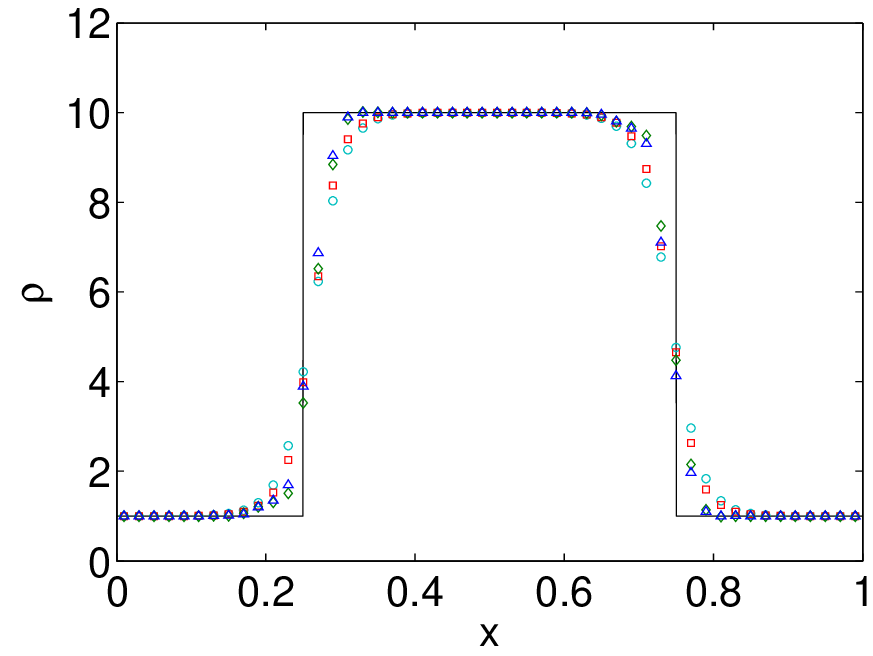}
\label{fig:compare_material_interface_advection_rho_global}}
\subfigure[Local density profile]{%
\includegraphics[width=0.45\textwidth]{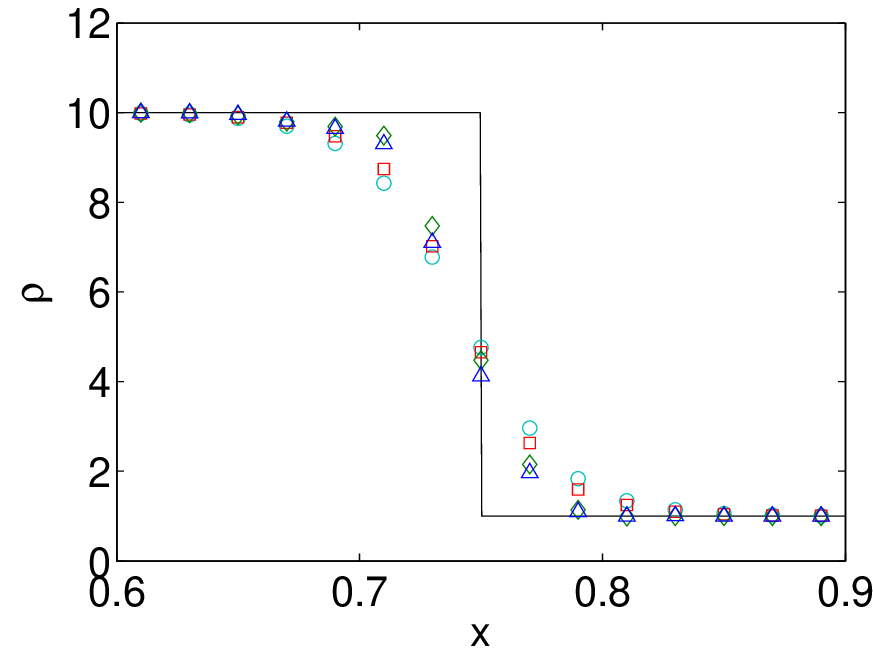}
\label{fig:compare_material_interface_advection_rho_local}}
\caption{Isolated material interface problem at $t = 2$ using different schemes. Black solid line: exact; cyan circles: WCNS5-JS; red squares: WCNS5-Z; green diamonds: WCNS6-CU-M2; blue triangles: WCNS6-LD.}
\label{fig:compare_material_interface_advection_rho}
\end{figure}

\begin{figure}[!ht]
\centering
\subfigure[Velocity]{%
\includegraphics[width=0.45\textwidth]{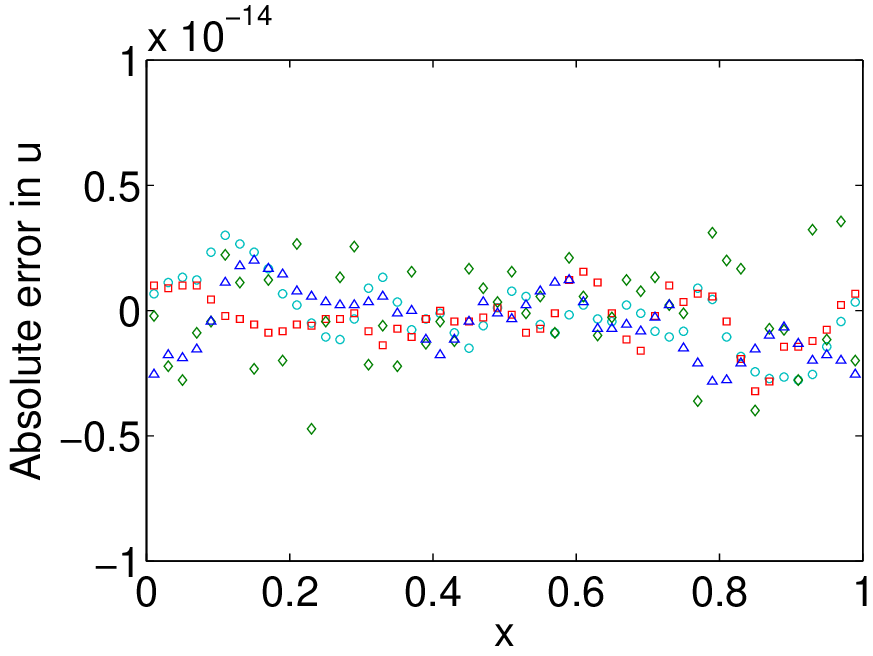}
\label{fig:compare_material_interface_advection_u_global}}
\subfigure[Pressure]{%
\includegraphics[width=0.45\textwidth]{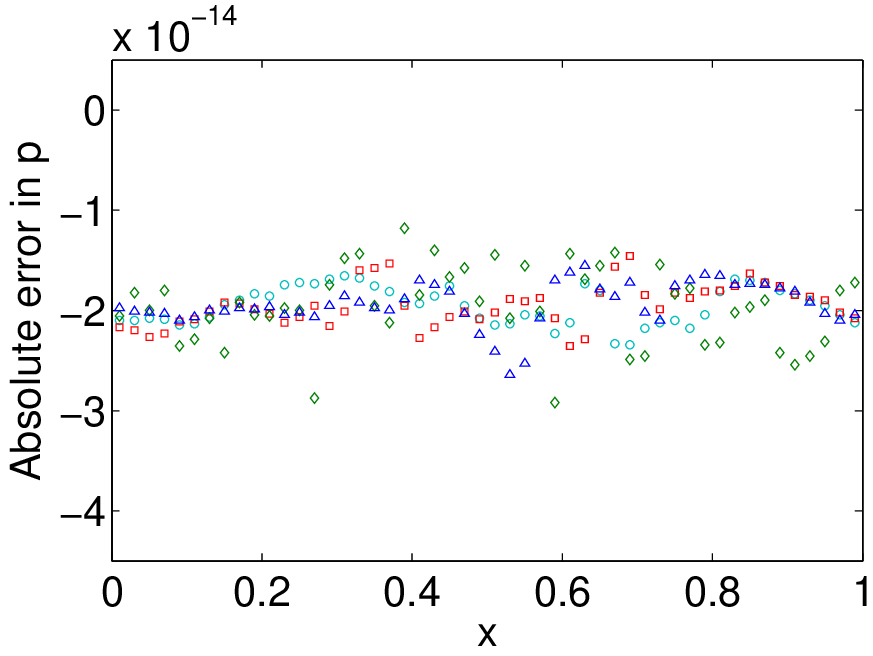}
\label{fig:compare_material_interface_advection_p_global}}
\caption{Errors for the isolated material interface problem at $t = 2$ using different schemes. Cyan circles: WCNS5-JS; red squares: WCNS5-Z; green diamonds: WCNS6-CU-M2; blue triangles: WCNS6-LD.}
\label{fig:compare_material_interface_advection_error}
\end{figure}

The comparison between exact solution and numerical solutions from different schemes for density is shown in figure~\ref{fig:compare_material_interface_advection_rho}. As shown in the density profiles, all schemes can capture the material interfaces at the correct locations without any numerical spurious oscillations. Among the upwind-biased WCNS's, WCNS5-Z gives a thinner numerical interface compared to WCNS5-JS. Both WCNS6-CU-M2 and WCNS6-LD have almost equivalent improvements with regards to the thicknesses of the discontinuities over the upwind-biased WCNS's. From figure~\ref{fig:compare_material_interface_advection_error}, it can be seen that both the velocity and pressure errors are close to machine precision for all of the schemes.

\subsubsection{Multi-species shock tube}
This is a 1D two-species modified Sod shock tube problem introduced by Abgrall and Karni~\cite{abgrall2001computations}. The initial conditions are given by:
\begin{equation*}
	\left( \rho, u, p, \gamma \right)
    =
    \begin{cases}
    	\left(1, 0, 1, 1.4 \right), &\mbox{$x < 0$}, \\
        \left(0.125, 0, 0.1, 1.6 \right), &\mbox{$x \geq 0$}. \\
    \end{cases}
\end{equation*}
\noindent The spatial domain is $x \in \left[-0.5, 0.5 \right]$ and the final time is at $t = 0.2$. Simulations are performed with constant time steps $\Delta t = 0.001$ on a uniform grid composed of 100 grid points where $\Delta x = 0.01$.

\begin{figure}[!ht]
\centering
\subfigure[Global density profile]{%
\includegraphics[width=0.45\textwidth]{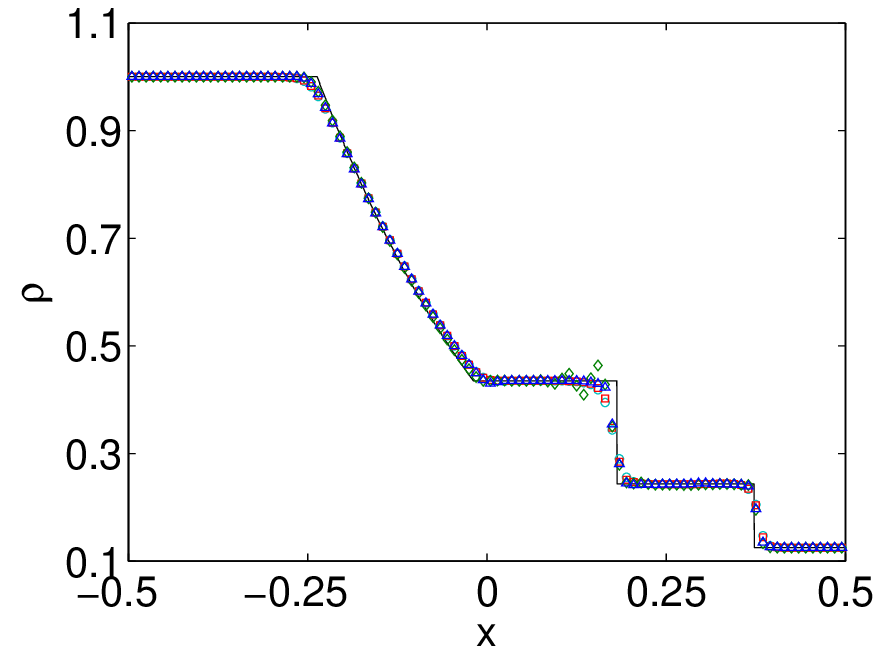}
\label{fig:compare_multicomponent_Sod_rho_global}}
\subfigure[Local density profile]{%
\includegraphics[width=0.45\textwidth]{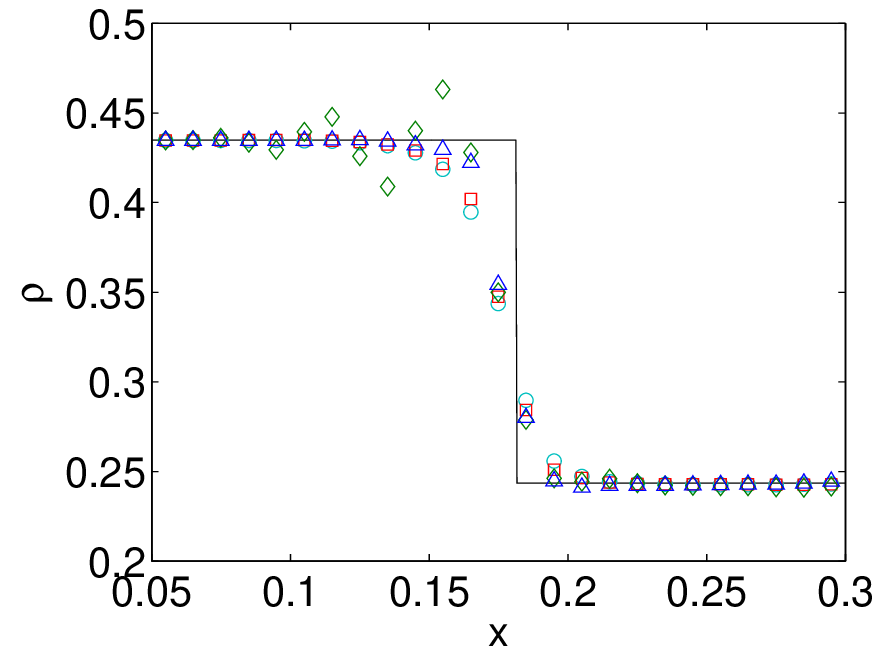}
\label{fig:compare_multicomponent_Sod_rho_local}}
\caption{Multi-species Sod shock tube problem at $t = 0.2$ using different schemes. Black solid line: exact; cyan circles: WCNS5-JS; red squares: WCNS5-Z; green diamonds: WCNS6-CU-M2; blue triangles: WCNS6-LD.}
\label{fig:compare_multicomponent_Sod_rho}
\end{figure}

Figure~\ref{fig:compare_multicomponent_Sod_rho} compares the exact solution with the numerical solutions from different schemes for the density. WCNS5-Z is slightly better than WCNS5-JS in capturing both shock wave and material interface in terms of numerical thicknesses. Both WCNS6-CU-M2 and WCNS6-LD can capture the shock wave with a smaller thickness compared to the upwind-biased WCNS's. However, there are spurious oscillations around the material interface in the solution of WCNS6-CU-M2 but the oscillations are not found in other schemes. The solution of WCNS6-LD contains a sharper material interface over those of the upwind-biased WCNS's.

\subsubsection{One-dimensional shock-curtain interaction}
This is a 1D shock-curtain interaction problem introduced by Abgrall~\cite{abgrall1996prevent}. It consists of a shock wave that is initially at $x  = 0.25$. The shock wave travels in air and moves to the right to interact with a helium curtain in region $0.4 < x < 0.6$. The initial conditions are given by:
\begin{equation*}
	\left( \rho, u, p, \gamma \right)
    =
    \begin{cases}
    	\left(1.3765, 0.3948, 1.57, 1.4 \right), &\mbox{$0 \leq x < 0.25$}, \\
        \left(1, 0, 1, 1.4 \right), &\mbox{$0.25 \leq x < 0.4$ or $0.6 \leq x < 1$}, \\
        \left(0.138, 0, 1, 1.67 \right), &\mbox{$0.4 \leq x < 0.6$}.  \\
    \end{cases}
\end{equation*}

\noindent The spatial domain is $x \in \left[0, 1 \right]$. Simulations are performed with constant time steps $\Delta t = 0.0015$ on a uniform grid with 200 grid points where $\Delta x = 0.005$.

\begin{figure}[!ht]
\centering
\subfigure[Global density profile]{%
\includegraphics[width=0.45\textwidth]{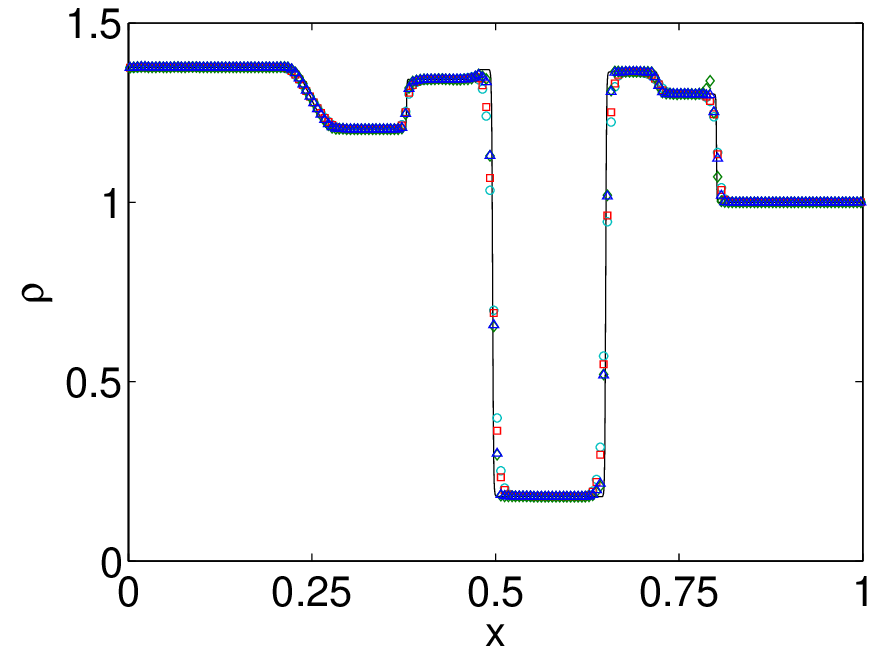}
\label{fig:compare_1D_shock_curtain_interaction_rho_global}}
\subfigure[Local density profile]{%
\includegraphics[width=0.45\textwidth]{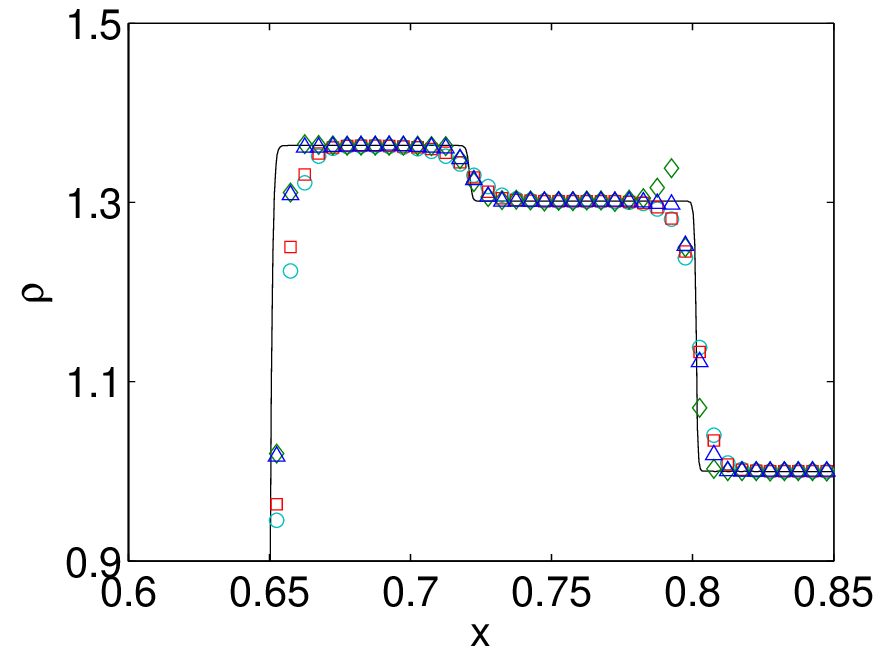}
\label{fig:compare_1D_shock_curtain_interaction_rho_local}}
\caption{Shock-curtain interaction problem at $t = 0.3$ using different schemes. Black solid line: reference; cyan circles: WCNS5-JS; red squares: WCNS5-Z; green diamonds: WCNS6-CU-M2; blue triangles: WCNS6-LD.}
\label{fig:compare_1D_shock_curtain_interaction_rho}
\end{figure}

Figure~\ref{fig:compare_1D_shock_curtain_interaction_rho} shows the comparison between the reference solution and the numerical solutions for density from different schemes at $t = 0.3$. The reference solution is computed with WCNS5-JS on a mesh composed of 2000 grid cells. WCNS5-Z improves WCNS5-JS slightly in capturing the discontinuities. Generally, both WCNS6-CU-M2 and WCNS6-LD have the same level of improvements in capturing discontinuities over the upwind-biased WCNS's. Nevertheless, there is an overshoot around the shock wave at around $x = 0.8$ in the solution of WCNS6-CU-M2 while no observable numerical instability is found in the solution of WCNS6-LD.

\subsubsection{Two-dimensional Richtmyer--Meshkov instability}

\begin{figure}[!ht]
 \centering
	\includegraphics[width=0.9\textwidth]{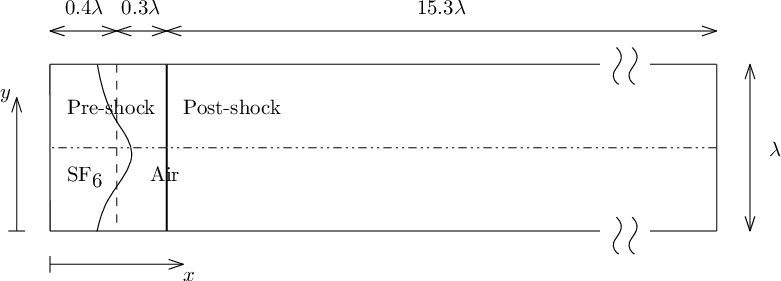}
	\caption{Schematic diagram of initial flow field and computational domain of the Richtmyer--Meshkov instability problem.}
    \label{fig:RMI_settings}
\end{figure}

This is a 2D single-mode Richtmyer--Meshkov instability problem modified by Nonomura et al.~\cite{nonomura2012numerical} from the problem performed experimentally by Brouillette and Sturtevant~\cite{brouillette1994experiments}, and Collins and Jacobs~\cite{collins2002plif}. Figure~\ref{fig:RMI_settings} shows the schematic of the initial flow field and domain. The domain has size $\left[0, 16 \lambda \right] \times \left[ 0, \lambda \right]$ and the initial perturbed interface is located at:
\begin{equation}
	\frac{x}{\lambda} = 0.4 - 0.1 \sin{\left( 2\pi \left( \frac{y}{\lambda} + 0.25 \right) \right)}.
\end{equation}
\noindent The following initial conditions are used:
\begin{equation*}
	\left( \rho, u, v, p, \gamma \right)
    =
    \begin{cases}
    	\left(1, 1.24, 0, 1/1.4, 1.4 \right), &\mbox{for pre-shock air}, \\
        \left(1.4112, 0.8787, 0, 1.6272/1.4, 1.4 \right), &\mbox{for post-shock air}, \\
        \left(5.04, 1.24, 0, 1/1.4, 1.093 \right), &\mbox{for $\mathrm{SF_6}$}.  \\
    \end{cases}
\end{equation*}

\noindent Because of the Cartesian gridding, a sharp interface will numerically trigger secondary instabilities along the interface. The artificial seeding of disturbances can be removed by smoothing the initial material interface with an artificial diffusion layer between the fluids. The diffusion layer is given by:
\begin{equation} \label{eq:smoothing_interface}
\begin{aligned}
 f_{sm} &= \frac{1}{2}(1 + erf(\frac{\Delta D}{C_{i} \sqrt{\Delta x \Delta y}})), \\
 v &= v_{L}(1 - f_{sm}) + v_{R} f_{sm}, \\
\end{aligned}
\end{equation}
\noindent where $v$ are any primitive variables near the initial interface. Subscripts $L$ and $R$ denote the left and right interface conditions. $erf()$ is the error function. $\Delta D$ is the distance from the initial perturbed material interface. $C_{i}$ is a parameter to control the number of grid points across the material interface. The greater the value of $C_{i}$, the thicker is the initial material interface. $C_{i} = 6$ is chosen for the Richtmyer--Meshkov instability problem.

 A grid with $2048 \times 128$ points is employed with $\lambda = 1$, where the grid spacing is $\Delta x = \Delta y = 1/128$. All simulations are conducted with a constant $\textnormal{CFL}=0.5$.

\begin{figure}[!ht]
\centering
\subfigure[$t=5.50$]{%
\includegraphics[width=0.25\textwidth]{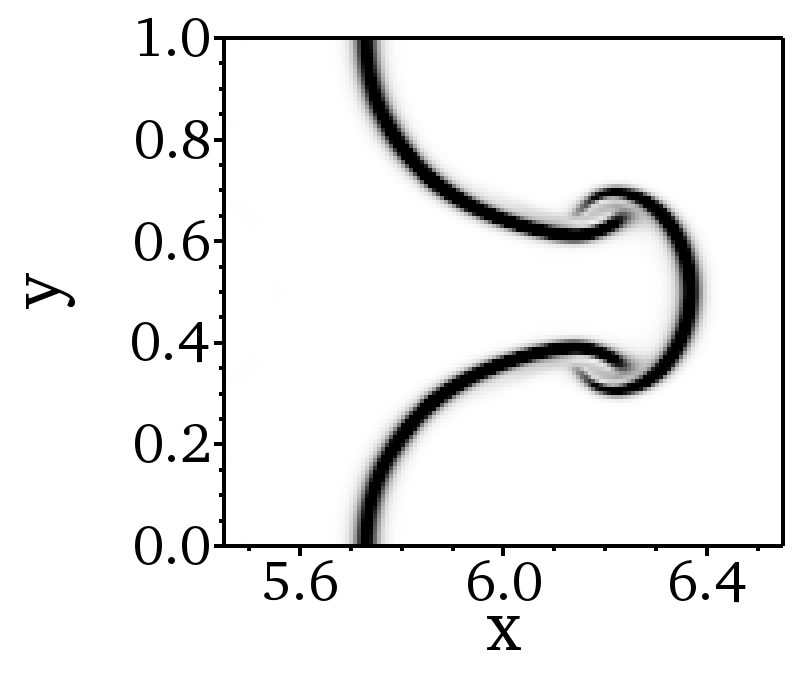}
\label{fig:2D_Richtmyer_Meshkov_5_50_WCNS5_JS}}
\subfigure[$t=8.25$]{%
\includegraphics[width=0.25\textwidth]{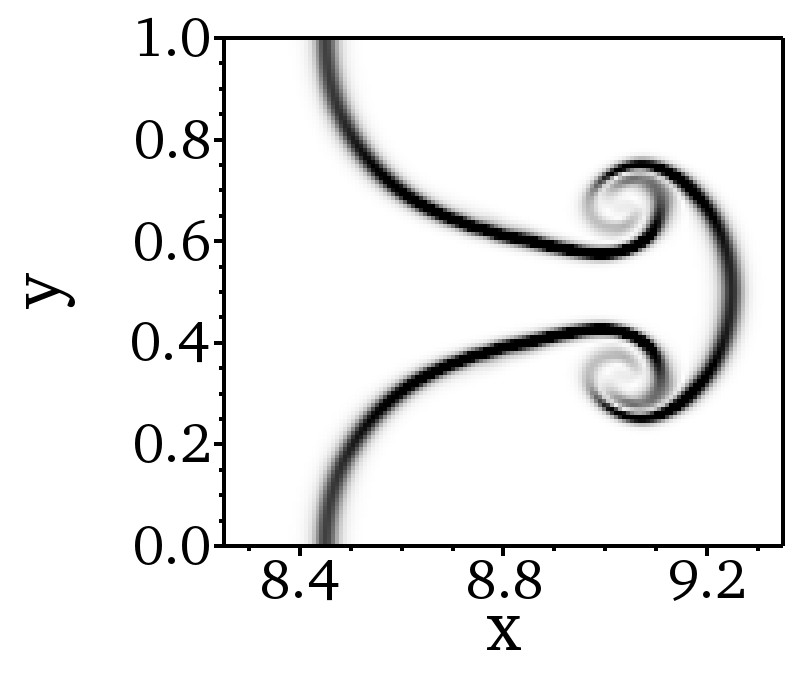}
\label{fig:2D_Richtmyer_Meshkov_8_25_WCNS5_JS}}
\subfigure[$t=11.00$]{%
\includegraphics[width=0.25\textwidth]{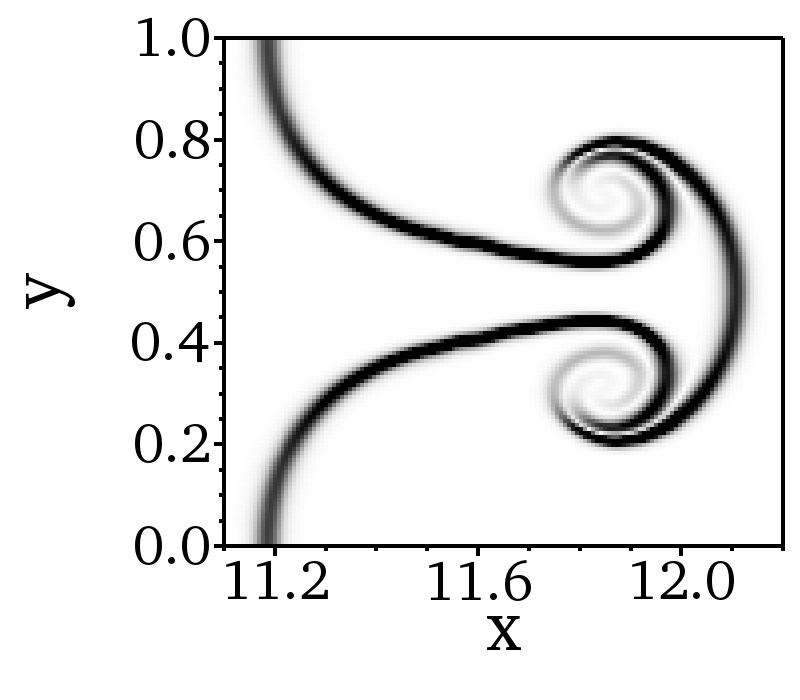}
\label{fig:2D_Richtmyer_Meshkov_11_10_WCNS5_JS}}

\subfigure[$t=5.50$]{%
\includegraphics[width=0.25\textwidth]{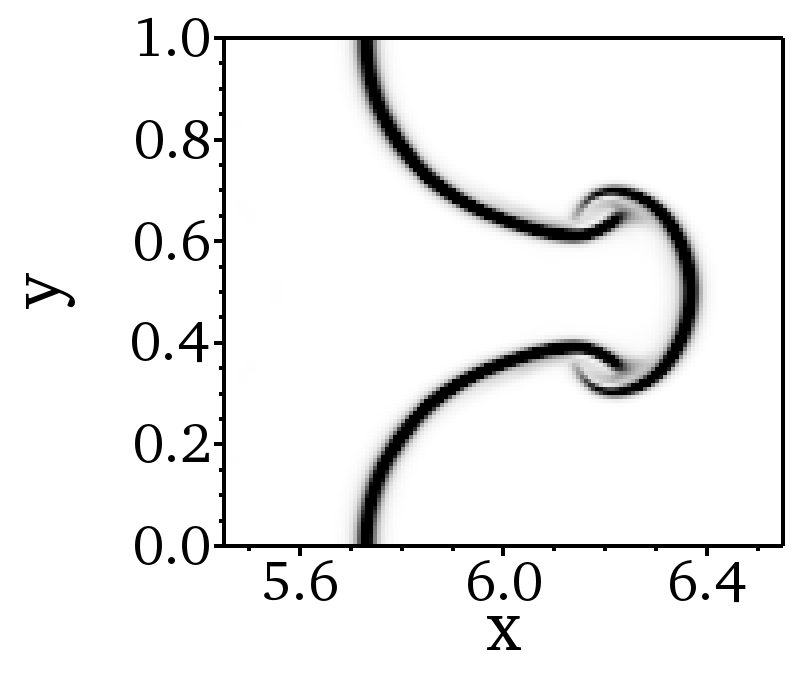}
\label{fig:2D_Richtmyer_Meshkov_5_50_WCNS5_Z}}
\subfigure[$t=8.25$]{%
\includegraphics[width=0.25\textwidth]{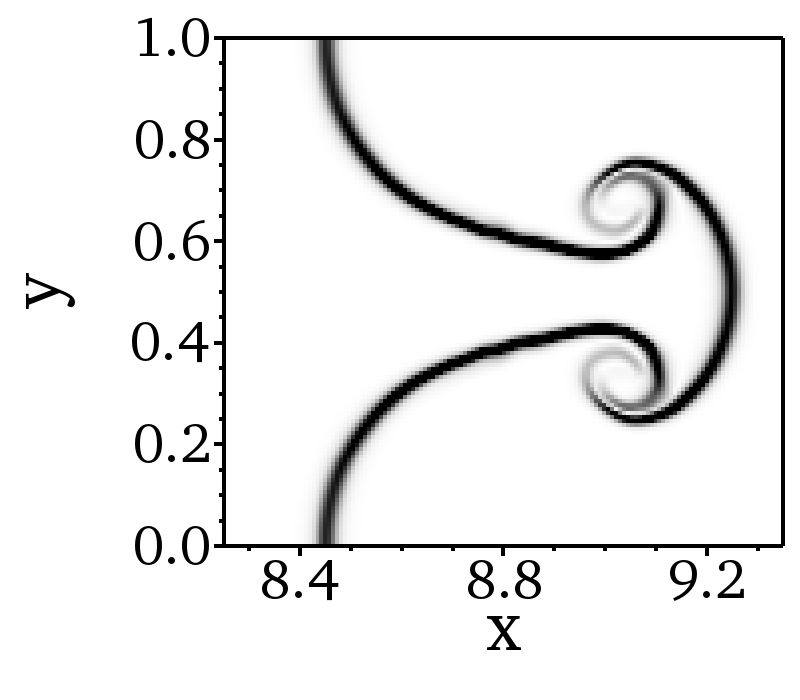}
\label{fig:2D_Richtmyer_Meshkov_8_25_WCNS5_Z}}
\subfigure[$t=11.00$]{%
\includegraphics[width=0.25\textwidth]{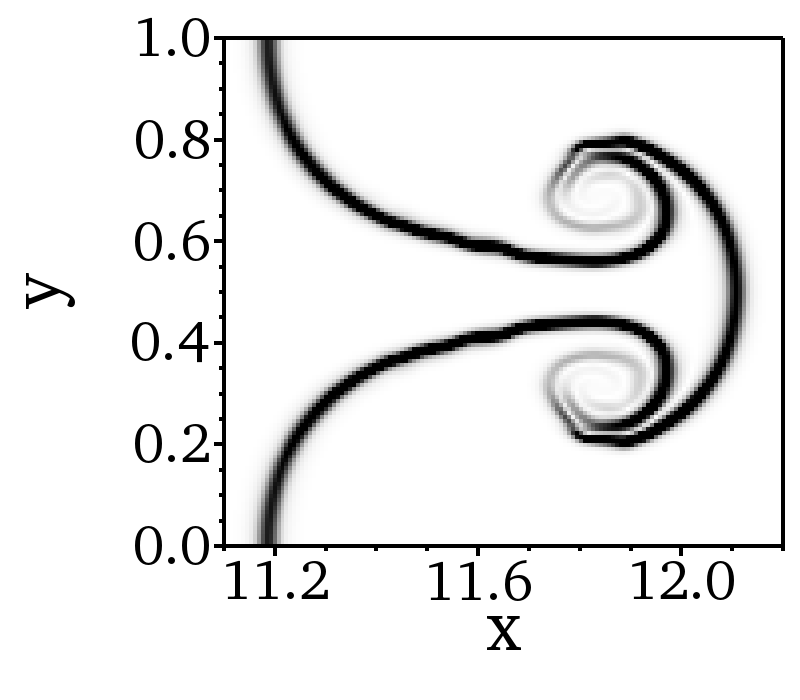}
\label{fig:2D_Richtmyer_Meshkov_11_10_WCNS5_Z}}

\subfigure[$t=5.50$]{%
\includegraphics[width=0.25\textwidth]{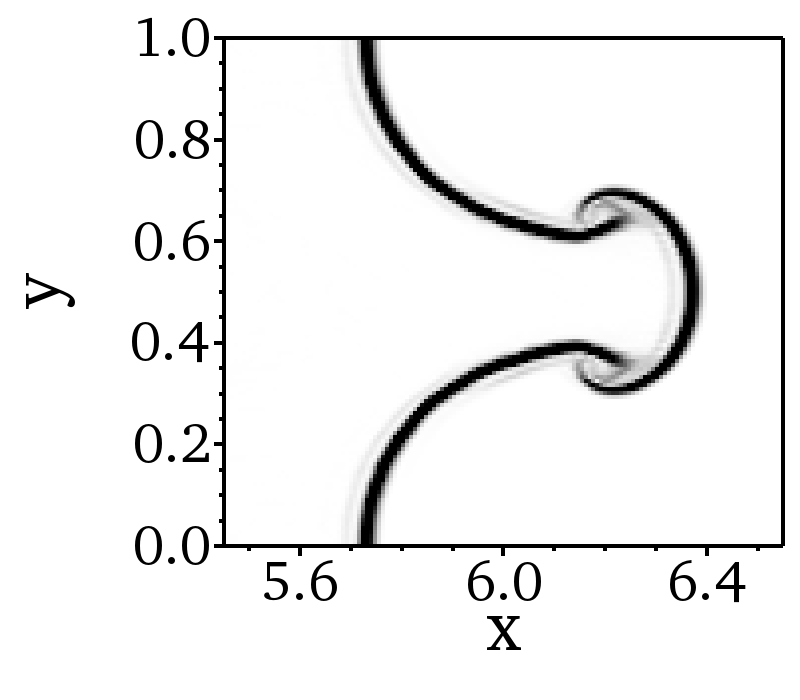}
\label{fig:2D_Richtmyer_Meshkov_5_50_WCNS6_CU_M2}}
\subfigure[$t=8.25$]{%
\includegraphics[width=0.25\textwidth]{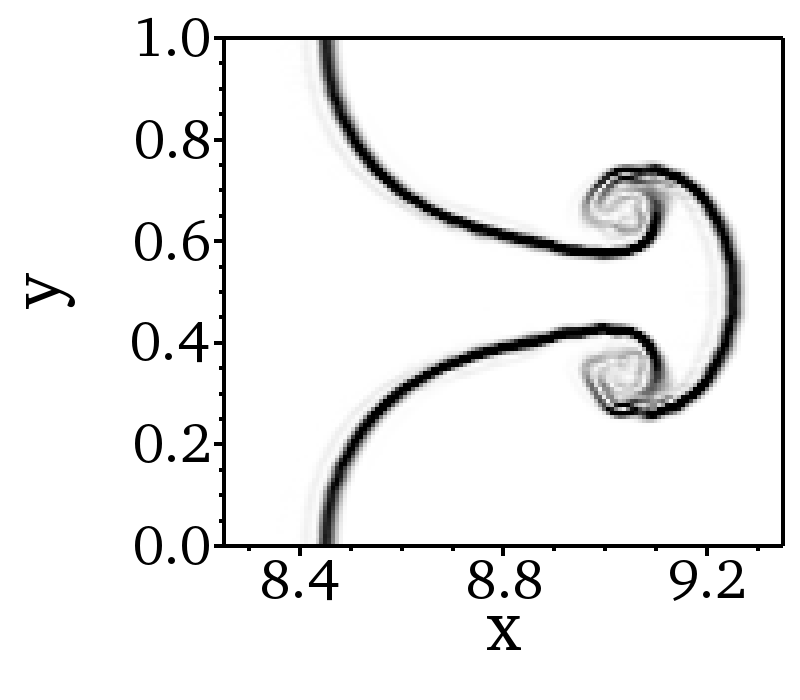}
\label{fig:2D_Richtmyer_Meshkov_8_25_WCNS6_CU_M2}}
\subfigure[$t=11.00$]{%
\includegraphics[width=0.25\textwidth]{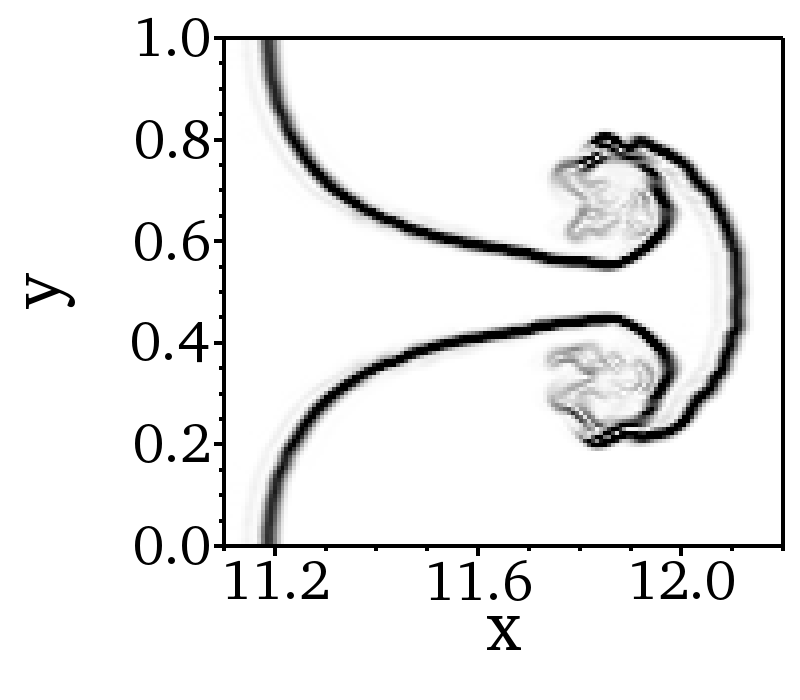}
\label{fig:2D_Richtmyer_Meshkov_11_10_WCNS6_CU_M2}}

\subfigure[$t=5.50$]{%
\includegraphics[width=0.25\textwidth]{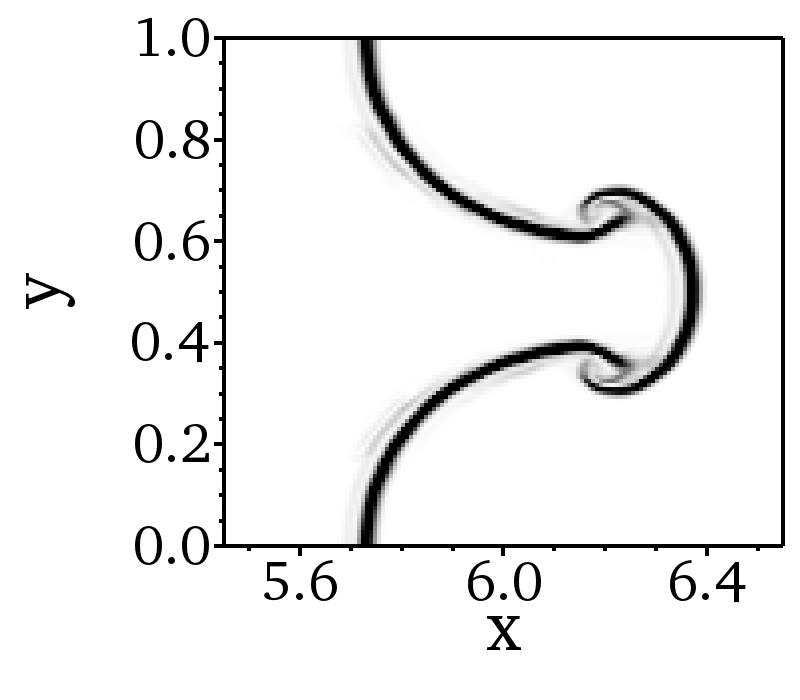}
\label{fig:2D_Richtmyer_Meshkov_5_50_WCNS6_LD}}
\subfigure[$t=8.25$]{%
\includegraphics[width=0.25\textwidth]{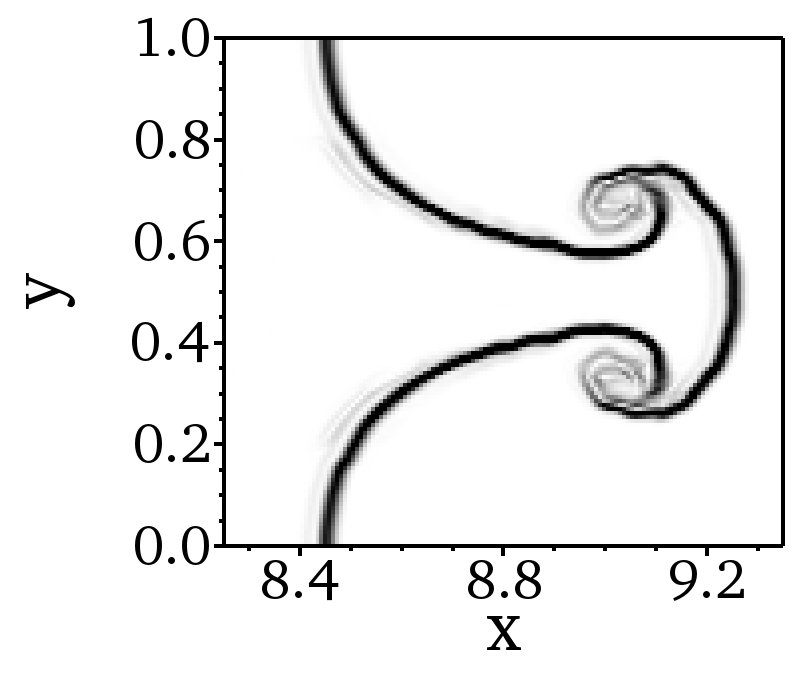}
\label{fig:2D_Richtmyer_Meshkov_8_25_WCNS6_LD}}
\subfigure[$t=11.00$]{%
\includegraphics[width=0.25\textwidth]{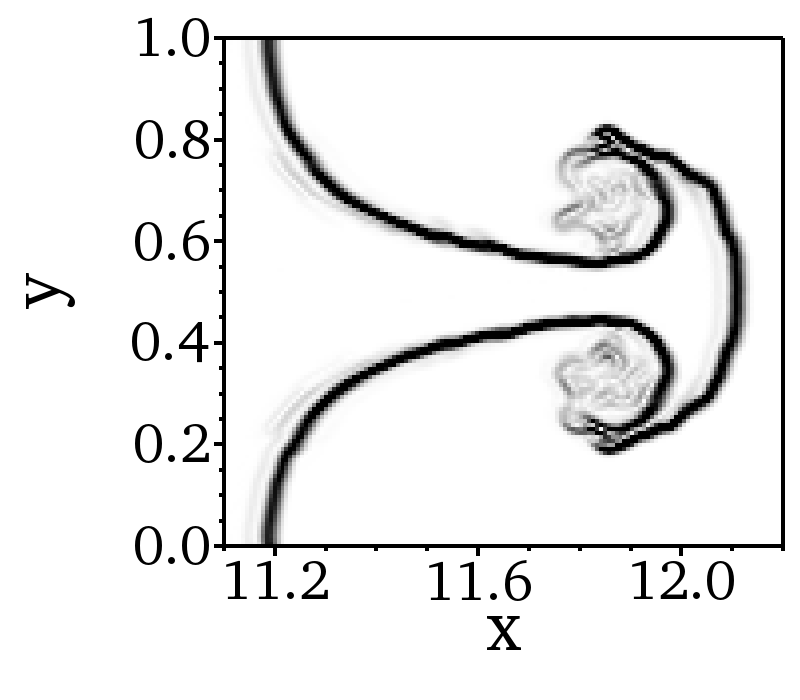}
\label{fig:2D_Richtmyer_Meshkov_11_10_WCNS6_LD}}

\caption{Nonlinear function of normalized density gradient magnitude, $\phi$, for the Richtmyer--Meshkov instability problem. Contours are from 1 to 1.7 at different times using different schemes. Top row: WCNS5-JS; second row: WCNS5-Z; third row: WCNS6-CU-M2; bottom row: WCNS6-LD.}
\label{fig:2D_Richtmyer_Meshkov_instability_density_gradient}
\end{figure}

Figure~\ref{fig:2D_Richtmyer_Meshkov_instability_density_gradient} shows the time evolution of a nonlinear function of density gradient magnitude, $\phi = \exp \left( \left| \nabla \rho \right| / \left| \nabla \rho \right|_{\max} \right)$ from different schemes at times $t = 5.50$, $8.25$, and $11.00$. The perturbed interface starts to deform nonlinearly after the shock wave hits the interface due to the baroclinic generation of vorticity. As the time grows, a spike is formed when the heavier fluid ($SF_6$) penetrates into the lighter fluid (air). WCNS5-Z generates a thinner material interface between the two fluids over WCNS5-JS. However, both WCNS6-CU-M2 and WCNS6-LD produce even thinner material interfaces between the two fluids over WCNS5-Z. They also show equivalent level of improvements with regards to the resolution of the two rolled up vortices over the fifth order WCNS's because of smaller numerical dissipation introduced by both schemes in smooth regions.

\subsubsection{Two-dimensional shock-cylinder interaction}

Another 2D multi-species problem is a problem of shock-cylinder interaction by Shankar et al.~\cite{shankar2010numerical} with the domain size of $\left[ 0, 6.5D \right] \times \left[ 0, 1.78D \right]$. Initially a helium cylinder of size $D$ is placed at location $\left[ 3.5D, 0.89D \right]$ in stationary pre-shock air. A Mach 1.22 normal shock is launched at position $x = 4.5D$ and moves to the left to interact with the cylinder. After the shock has interacted with the cylinder, the interface between the helium and air deforms due to the baroclinic torque. This problem can simultaneously test different methods' capabilities in capturing material interface, shock and shear instability along the material interface.

Initial conditions and boundary conditions prescribed in a similar way as Shankar et al.~\cite{shankar2010numerical} are used. The initial conditions are given by:
\begin{equation*}
	\left( \rho, u, v, p, \gamma \right)
    =
    \begin{cases}
    	\left(1, 0, 0, 1/1.4, 1.4 \right), &\mbox{for pre-shock air}, \\
        \left(1.3764, -0.3336, 0, 0, 1.5698/1.4, 1.4 \right), &\mbox{for post-shock air}, \\
        \left(0.1819, 0, 0, 1/1.4, 1.648 \right), &\mbox{for helium cylinder}.  \\
    \end{cases}
\end{equation*}

\begin{figure}[!ht]
 \centering
	\includegraphics[width=0.9\textwidth]{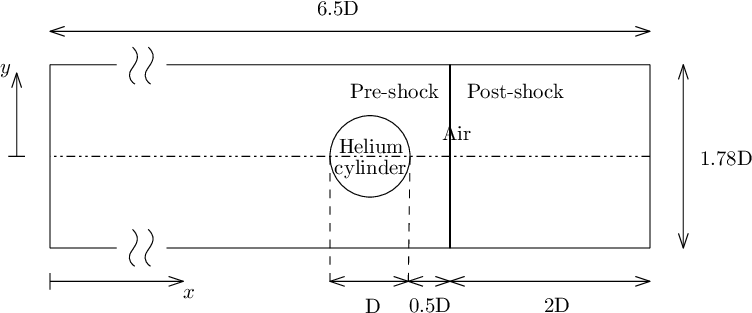}
	\caption{Schematic diagram of initial flow field and computational domain of the shock-cylinder interaction problem.}
    \label{fig:sb_settings}
\end{figure}

\noindent The initial material interface is also smoothed like the 2D Richtmyer--Meshkov instability problem with equation~\eqref{eq:smoothing_interface} and $C_{i} = 3$. Slip-wall boundary conditions are applied on both the upper and lower boundaries. Both left and right boundary conditions are extrapolated from interior solutions. The initial flow field and computational domain are shown in figure~\ref{fig:sb_settings}. Three levels of mesh resolutions $650 \times 178$, $1300 \times 356$, and $2600 \times 712$ are employed with $D = 1$. The corresponding grid spacings of the three mesh resolutions are $\Delta x = \Delta y = 0.01$, $0.005$, and $0.0025$ respectively. The simulations are run with constant $\textnormal{CFL}=0.5$.

\begin{figure}[!ht]
\centering
\subfigure[$t=3.25$]{%
\includegraphics[width=0.315\textwidth]{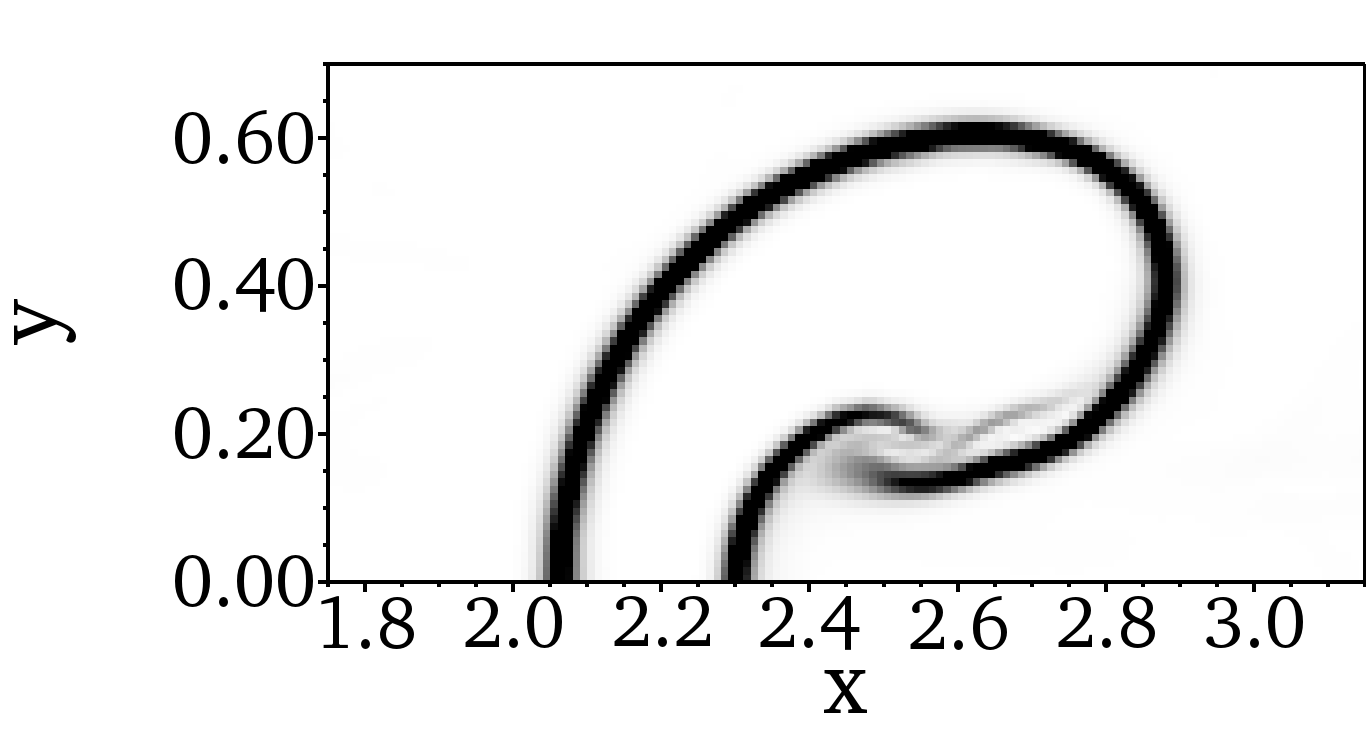}
\label{fig:2D_shock_cylinder_coarse_3_25_WCNS5_JS}}
\subfigure[$t=4.90$]{%
\includegraphics[width=0.315\textwidth]{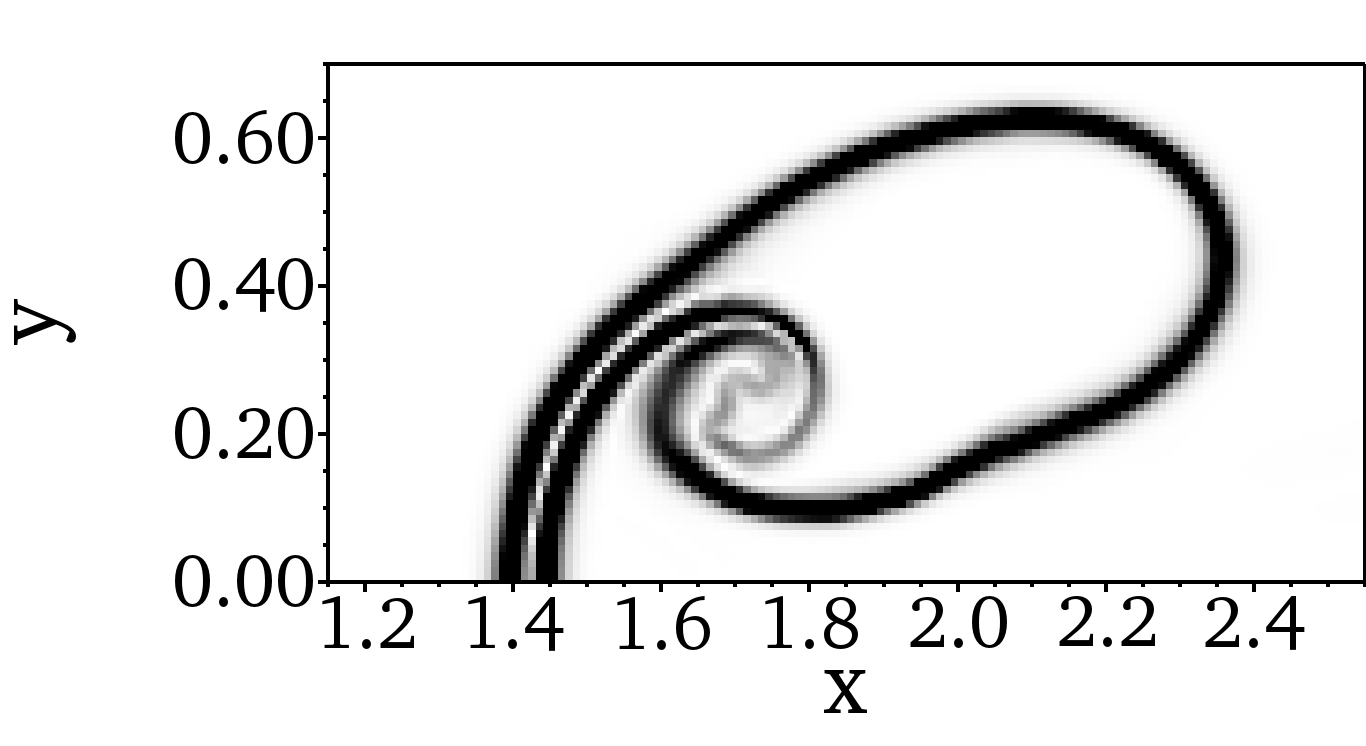}
\label{fig:2D_shock_cylinder_coarse_4_90_WCNS5_JS}}
\subfigure[$t=6.95$]{%
\includegraphics[width=0.315\textwidth]{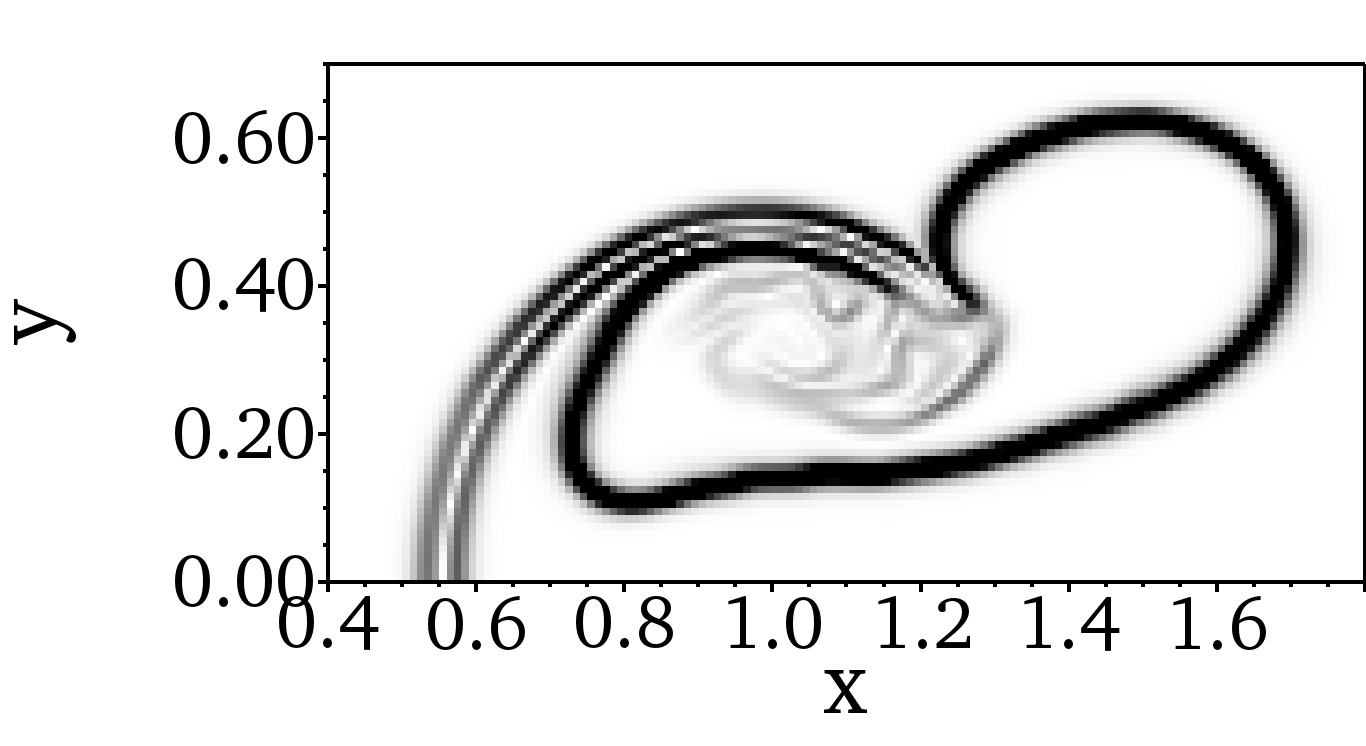}
\label{fig:2D_shock_cylinder_coarse_6_95_WCNS5_JS}}

\subfigure[$t=3.25$]{%
\includegraphics[width=0.315\textwidth]{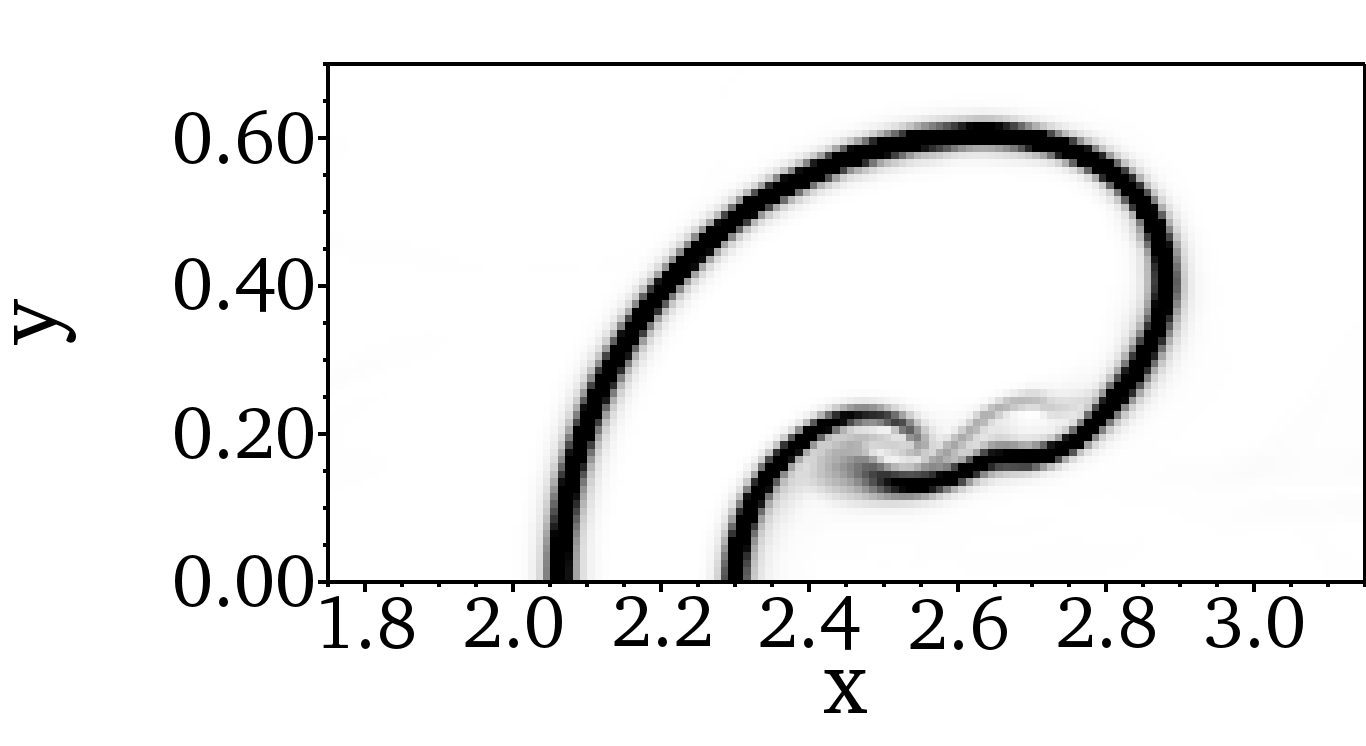}
\label{fig:2D_shock_cylinder_coarse_3_25_WCNS5_Z}}
\subfigure[$t=4.90$]{%
\includegraphics[width=0.315\textwidth]{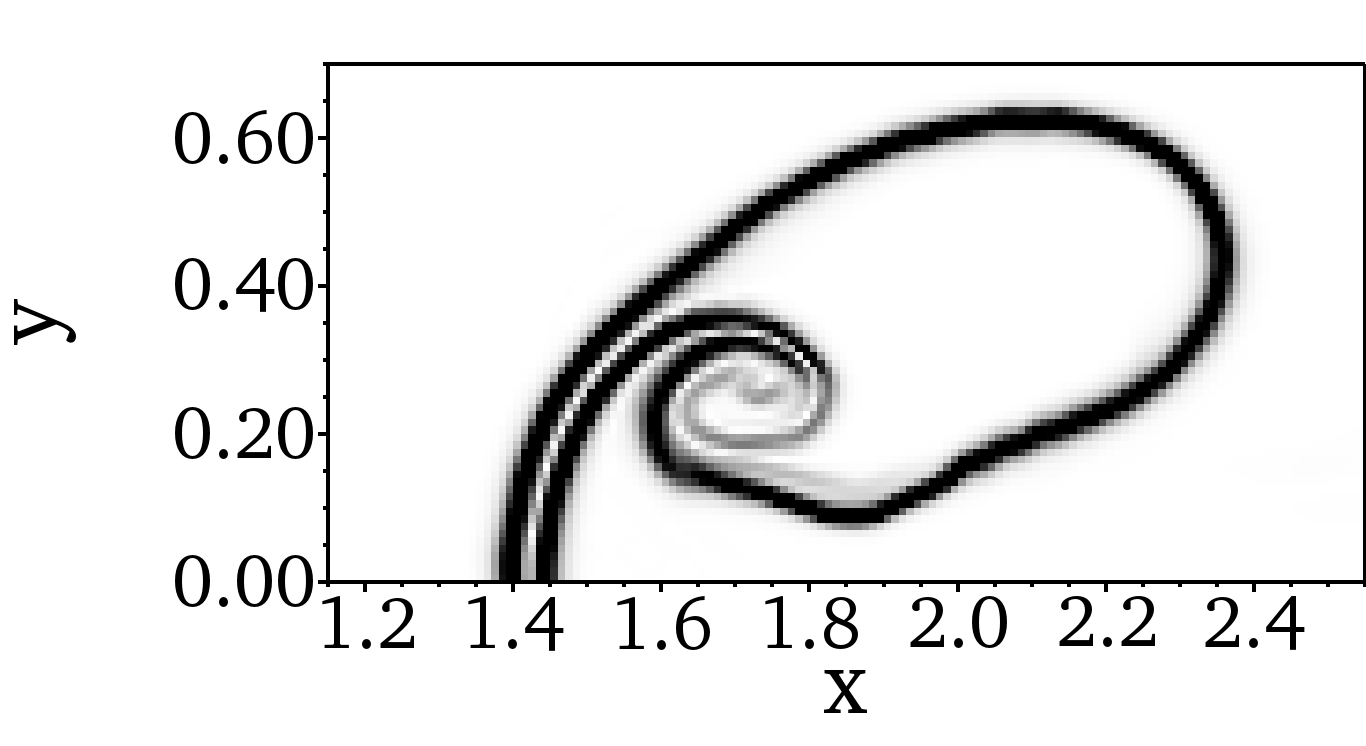}
\label{fig:2D_shock_cylinder_coarse_4_90_WCNS5_Z}}
\subfigure[$t=6.95$]{%
\includegraphics[width=0.315\textwidth]{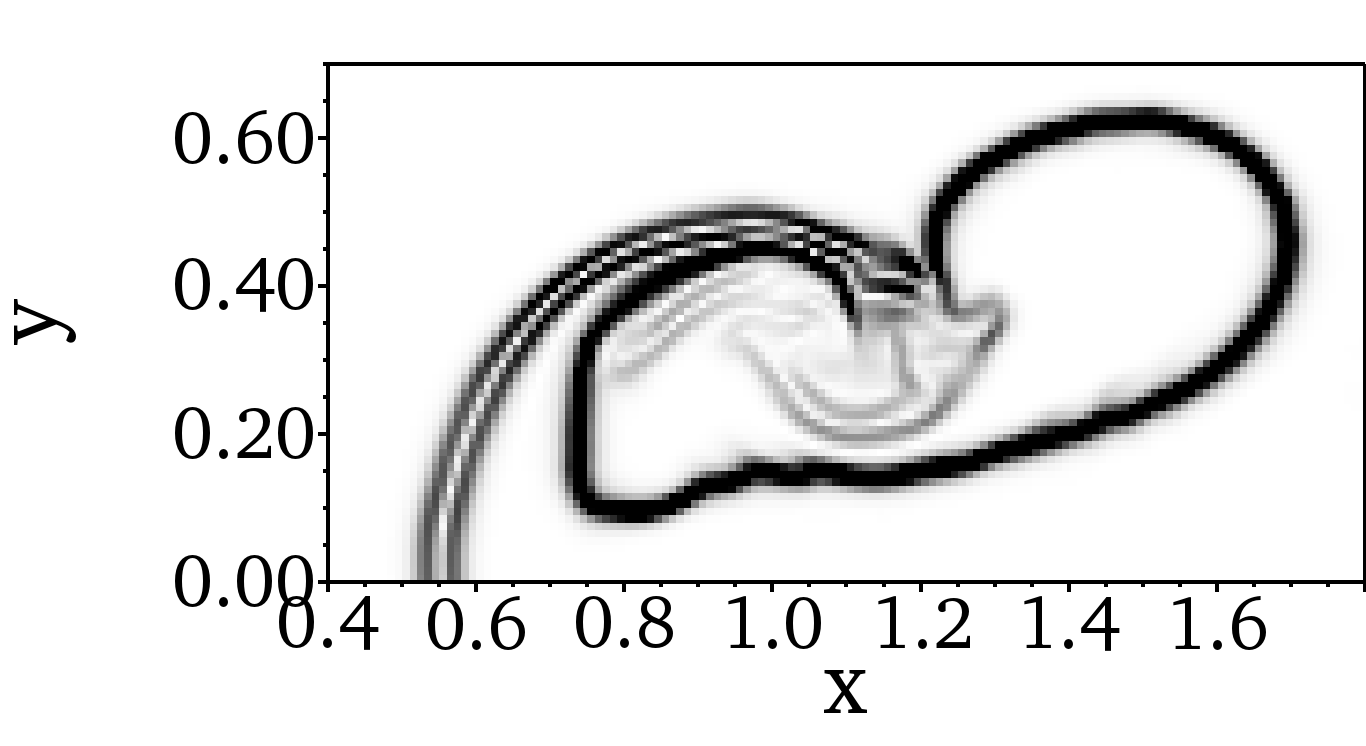}
\label{fig:2D_shock_cylinder_coarse_6_95_WCNS5_Z}}

\subfigure[$t=3.25$]{%
\includegraphics[width=0.315\textwidth]{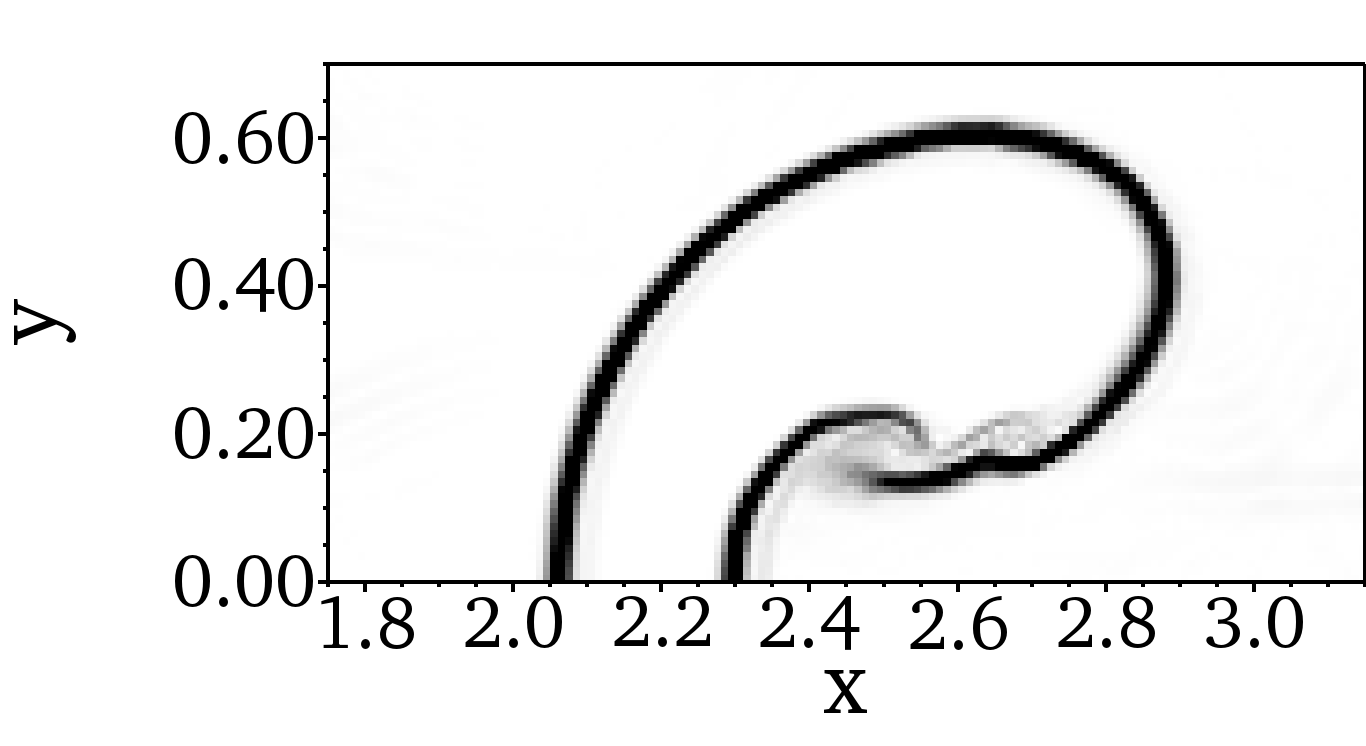}
\label{fig:2D_shock_cylinder_coarse_3_25_WCNS6_CU_M2}}
\subfigure[$t=4.90$]{%
\includegraphics[width=0.315\textwidth]{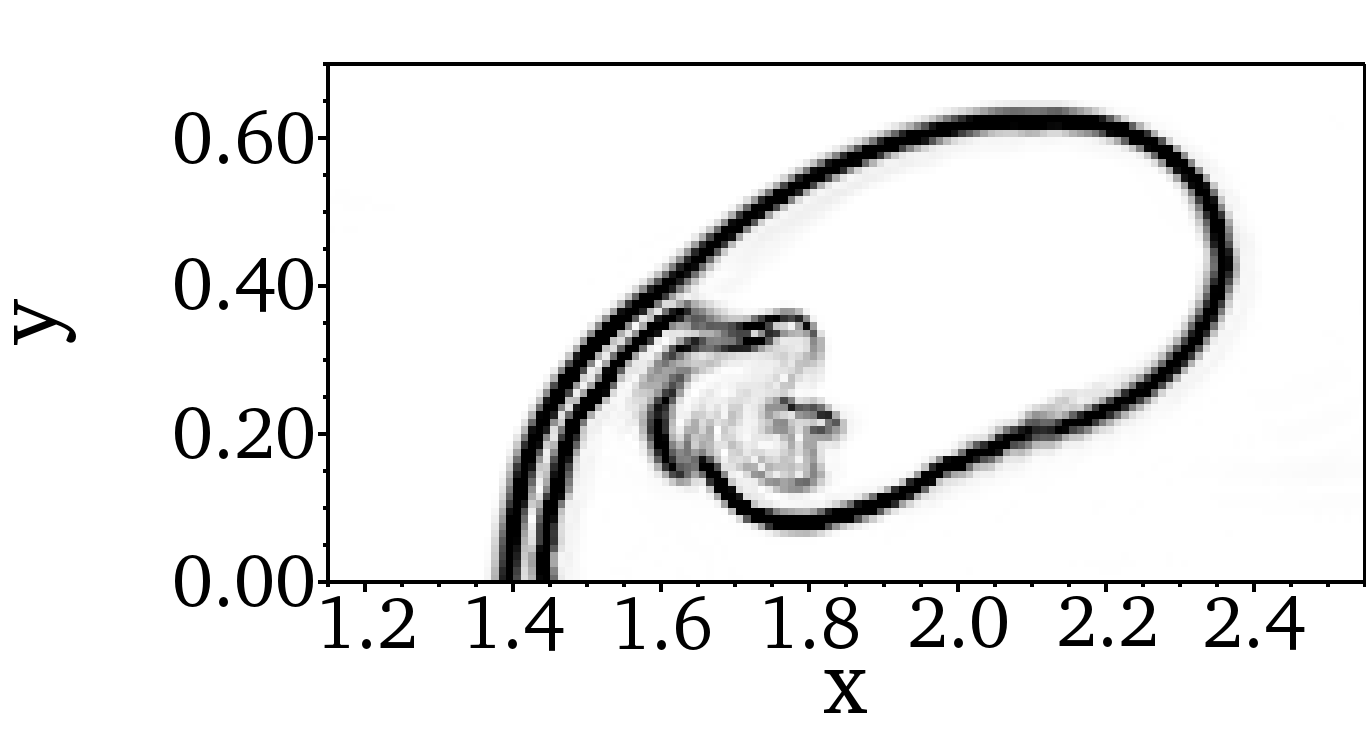}
\label{fig:2D_shock_cylinder_coarse_4_90_WCN6_CU_M2}}
\subfigure[$t=6.95$]{%
\includegraphics[width=0.315\textwidth]{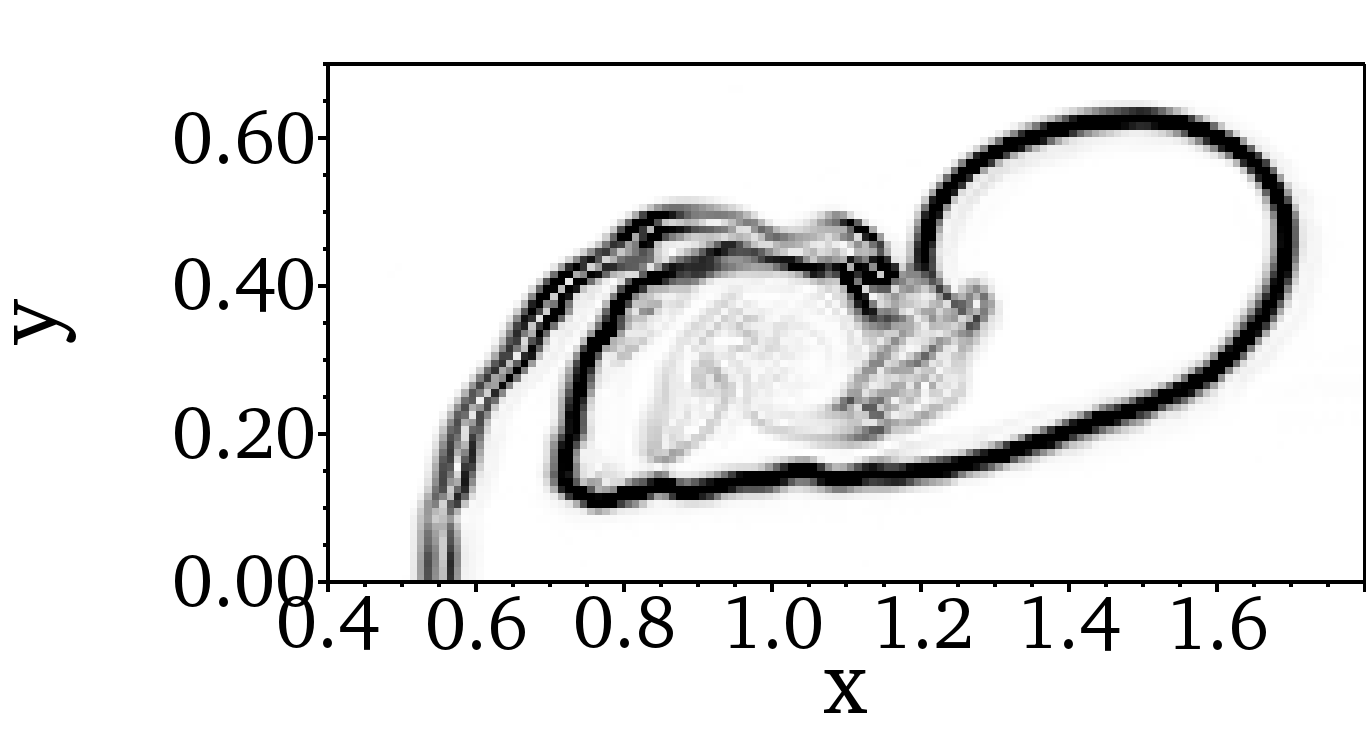}
\label{fig:2D_shock_cylinder_coarse_6_95_WCNS6_CU_M2}}

\subfigure[$t=3.25$]{%
\includegraphics[width=0.315\textwidth]{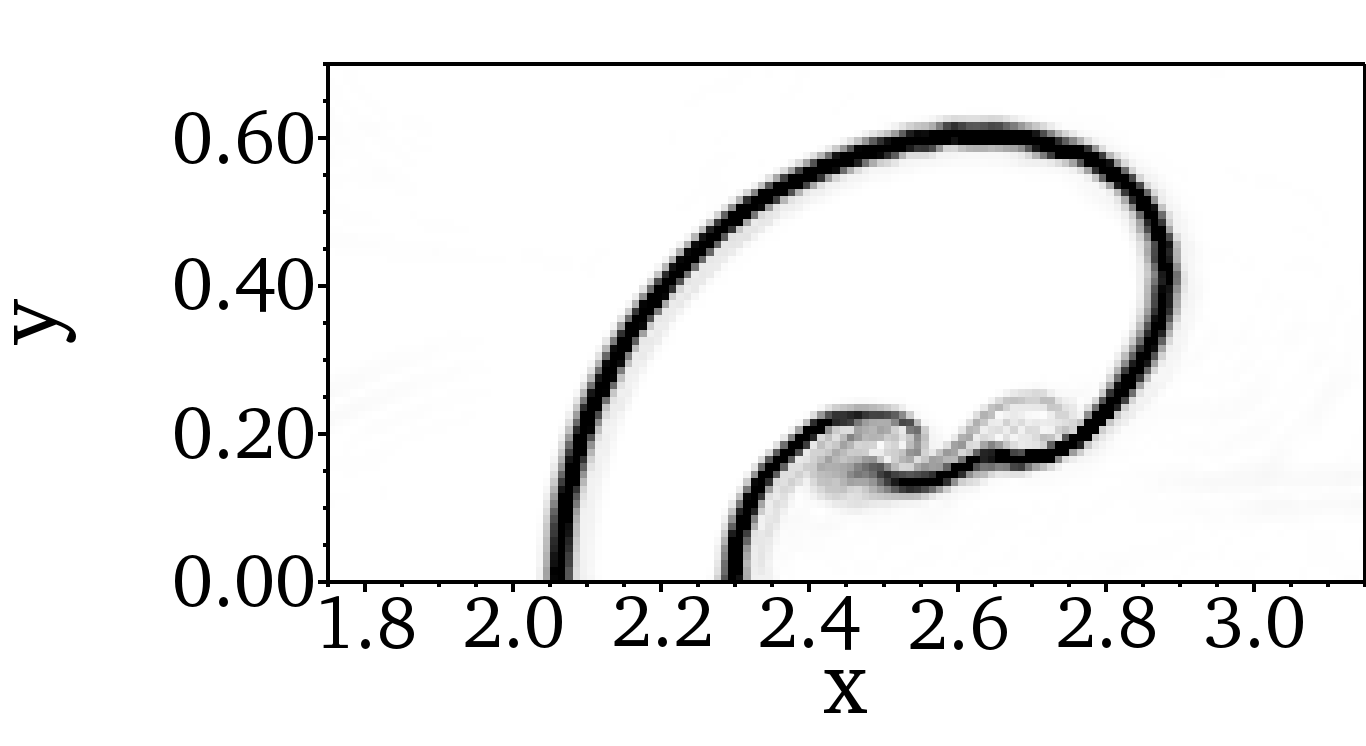}
\label{fig:2D_shock_cylinder_coarse_3_25_WCNS6_LD}}
\subfigure[$t=4.90$]{%
\includegraphics[width=0.315\textwidth]{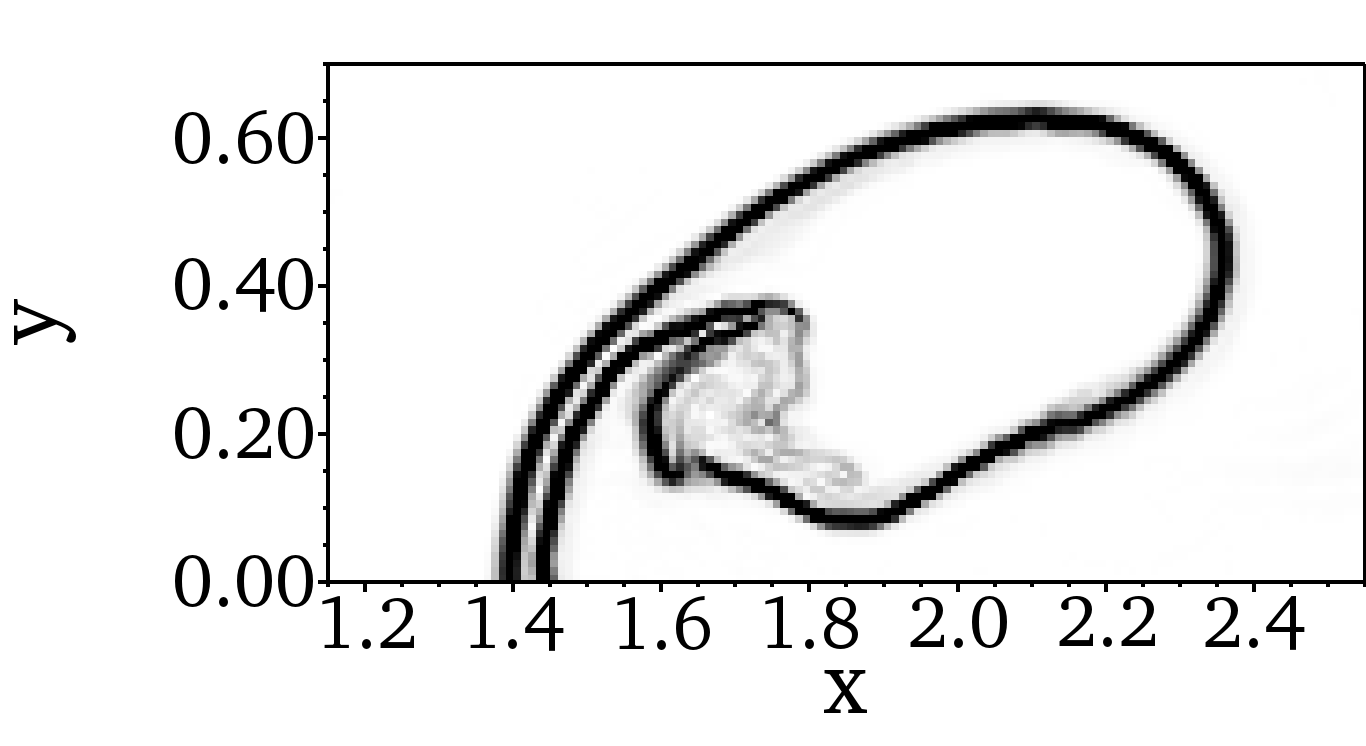}
\label{fig:2D_shock_cylinder_coarse_4_90_WCN6_LD}}
\subfigure[$t=6.95$]{%
\includegraphics[width=0.315\textwidth]{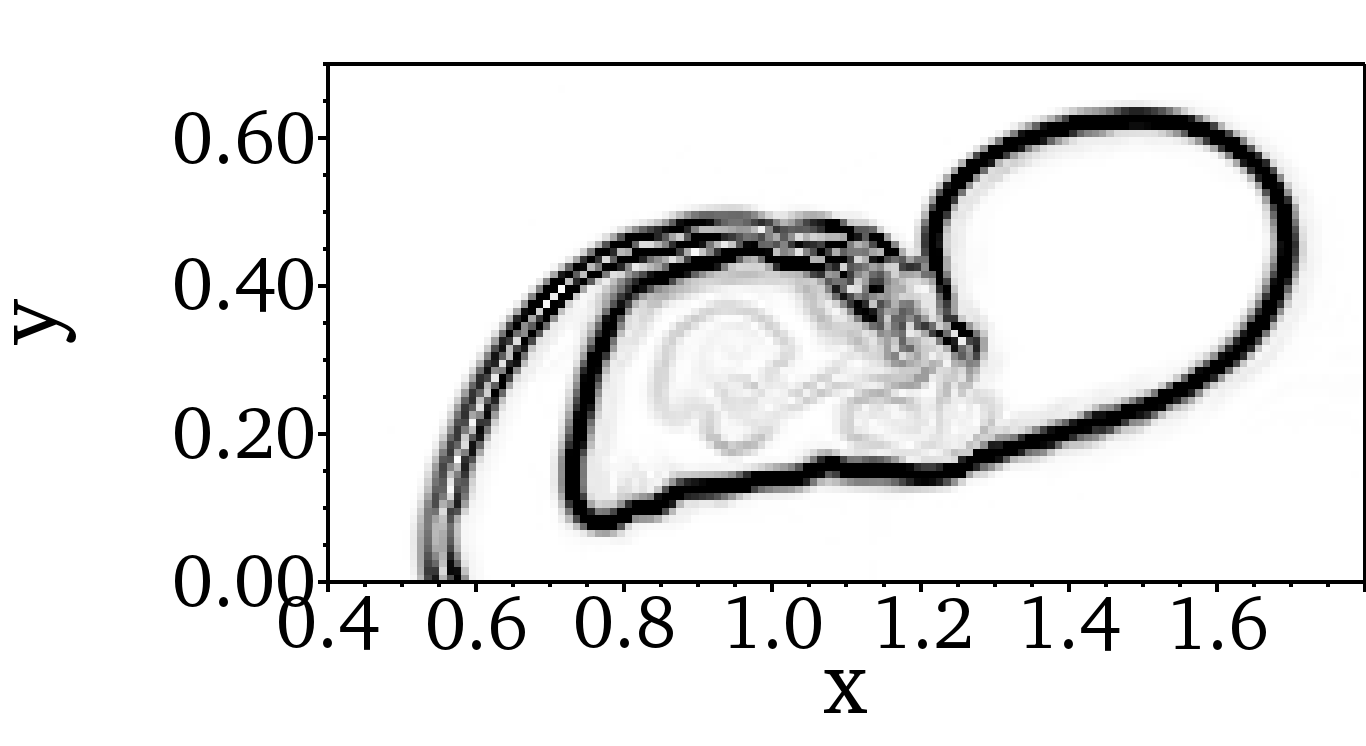}
\label{fig:2D_shock_cylinder_coarse_6_95_WCNS6_LD}}

\caption{Nonlinear function of normalized density gradient magnitude, $\phi$, for the shock-cylinder interaction problem. Contours are from 1 to 1.7 at different times using different schemes. Grid spacings are $\Delta x = \Delta y = 1/100$. Top row: WCNS5-JS; second row: WCNS5-Z; third row: WCNS6-CU-M2; bottom row: WCNS6-LD.}
\label{fig:2D_shock_cylinder_density_gradient_coarse}
\end{figure}

\begin{figure}[!ht]
\centering
\subfigure[$t=3.25$]{%
\includegraphics[width=0.315\textwidth]{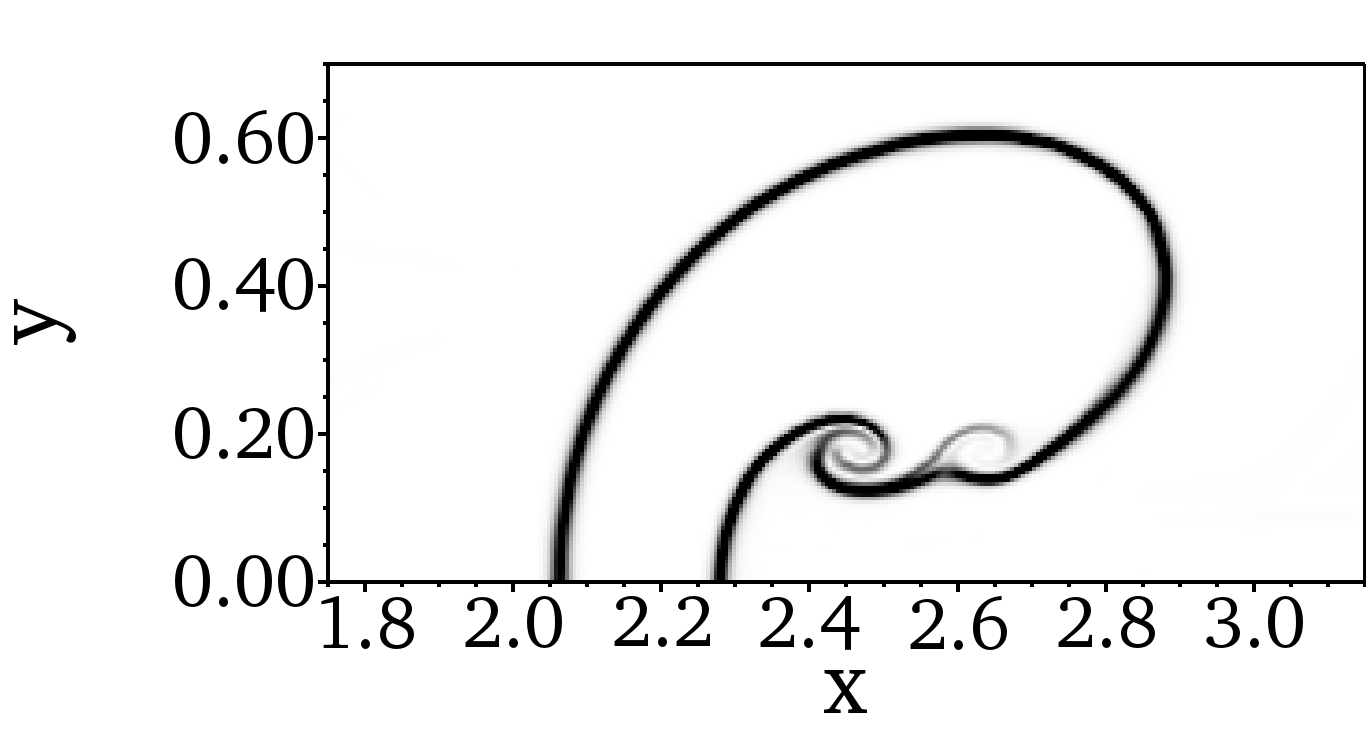}
\label{fig:2D_shock_cylinder_medium_3_25_WCNS5_JS}}
\subfigure[$t=4.90$]{%
\includegraphics[width=0.315\textwidth]{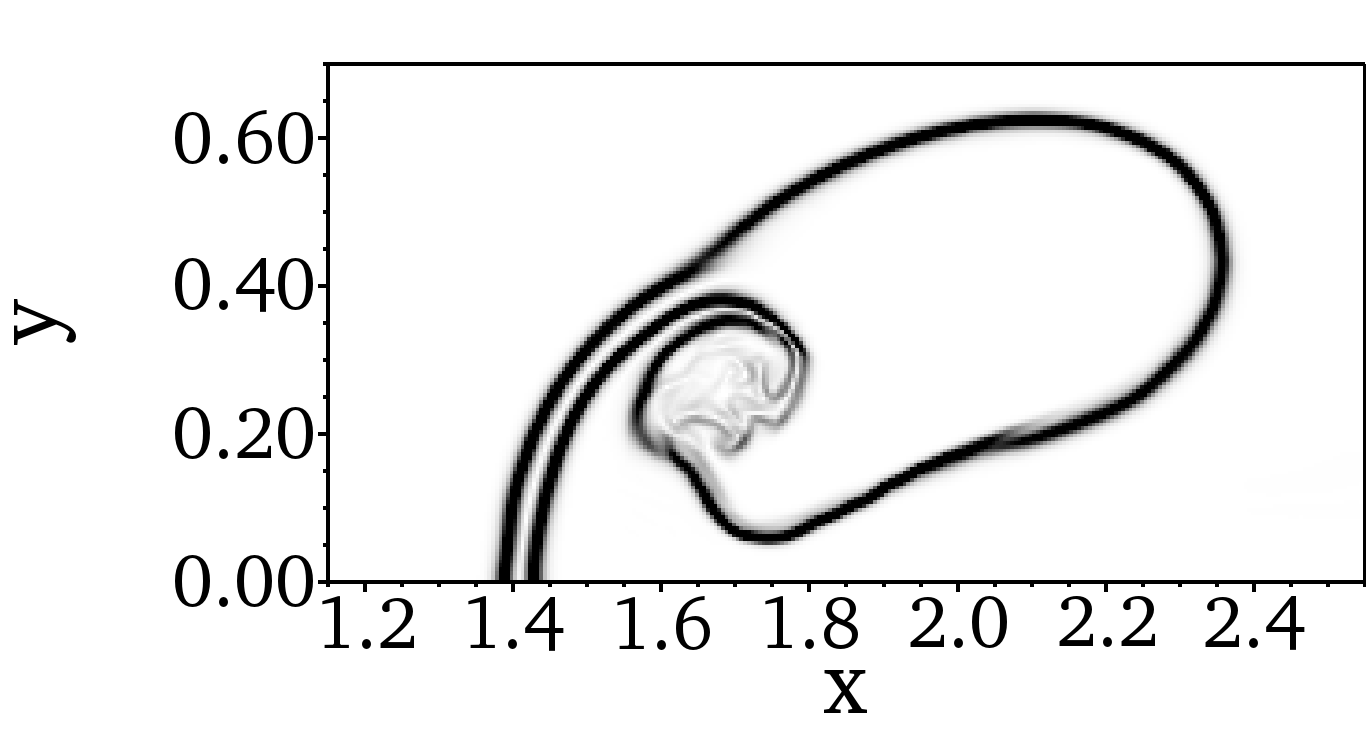}
\label{fig:2D_shock_cylinder_medium_4_90_WCNS5_JS}}
\subfigure[$t=6.95$]{%
\includegraphics[width=0.315\textwidth]{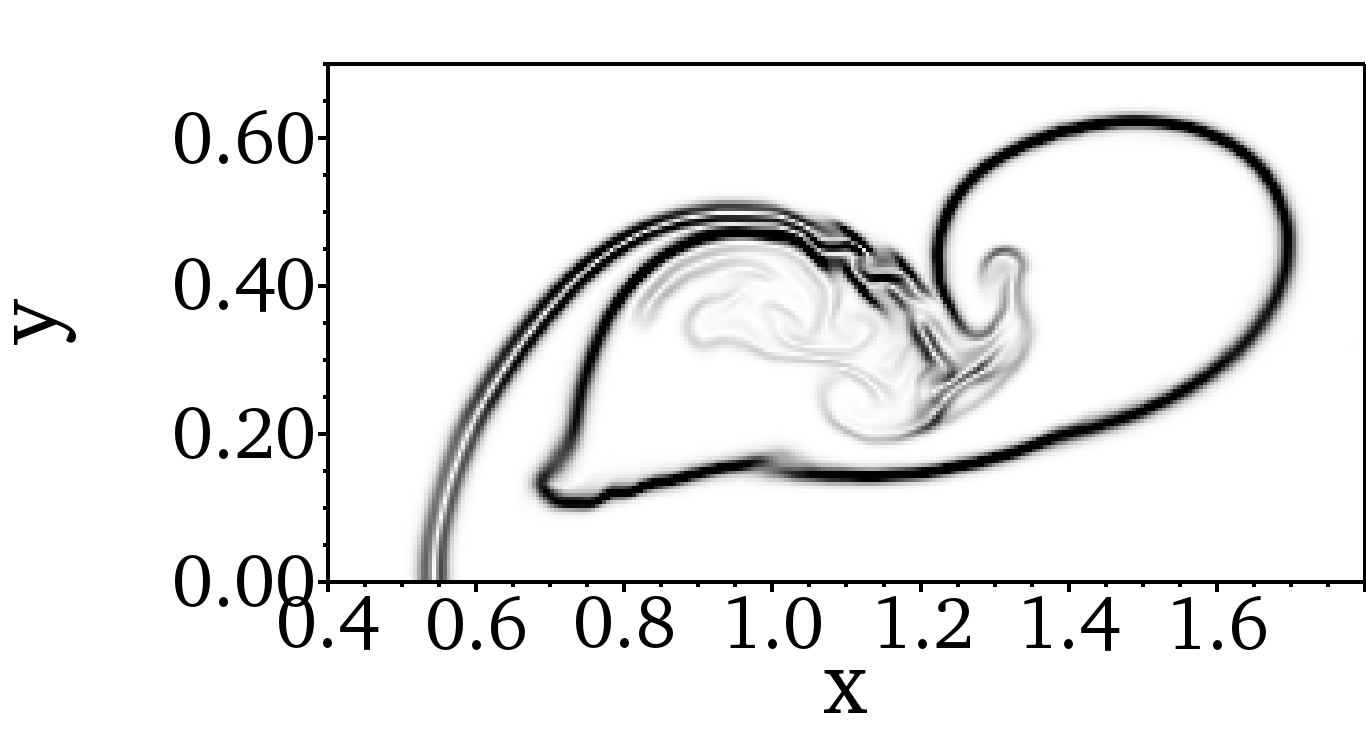}
\label{fig:2D_shock_cylinder_medium_6_95_WCNS5_JS}}

\subfigure[$t=3.25$]{%
\includegraphics[width=0.315\textwidth]{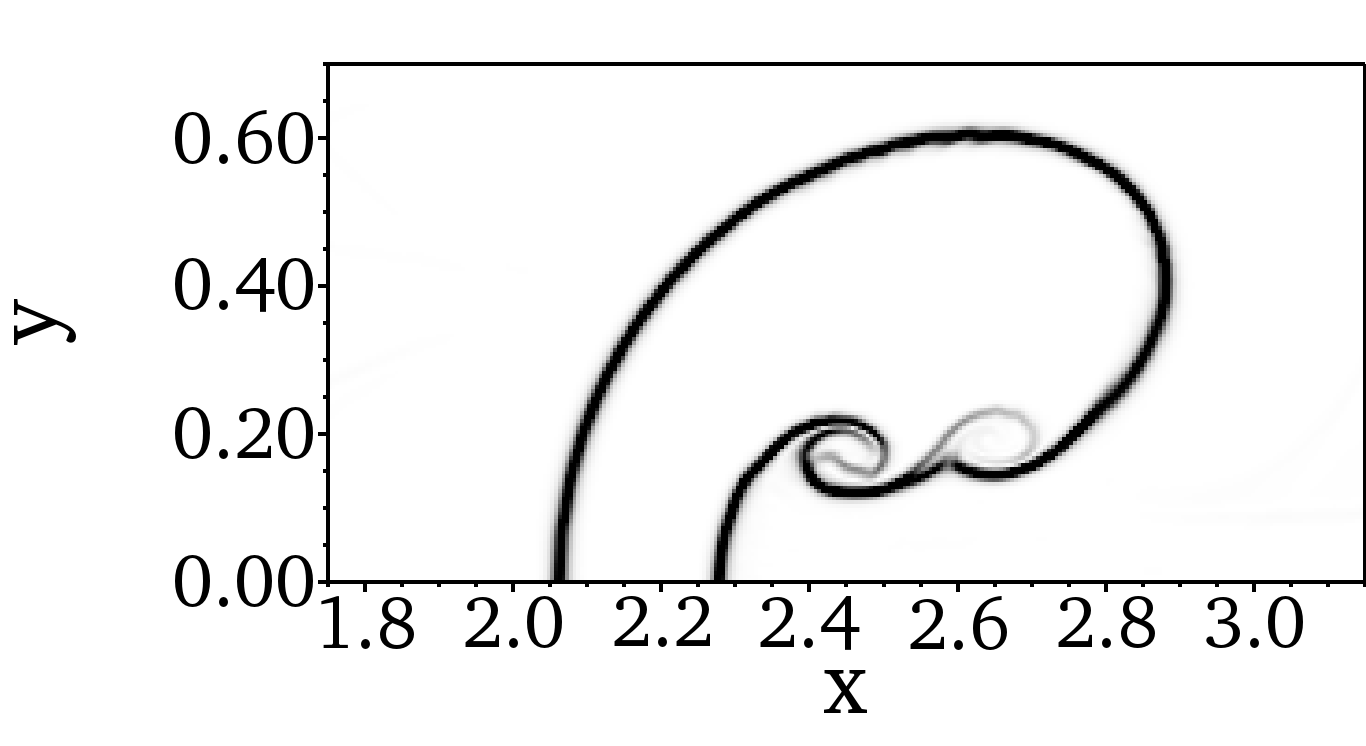}
\label{fig:2D_shock_cylinder_medium_3_25_WCNS5_Z}}
\subfigure[$t=4.90$]{%
\includegraphics[width=0.315\textwidth]{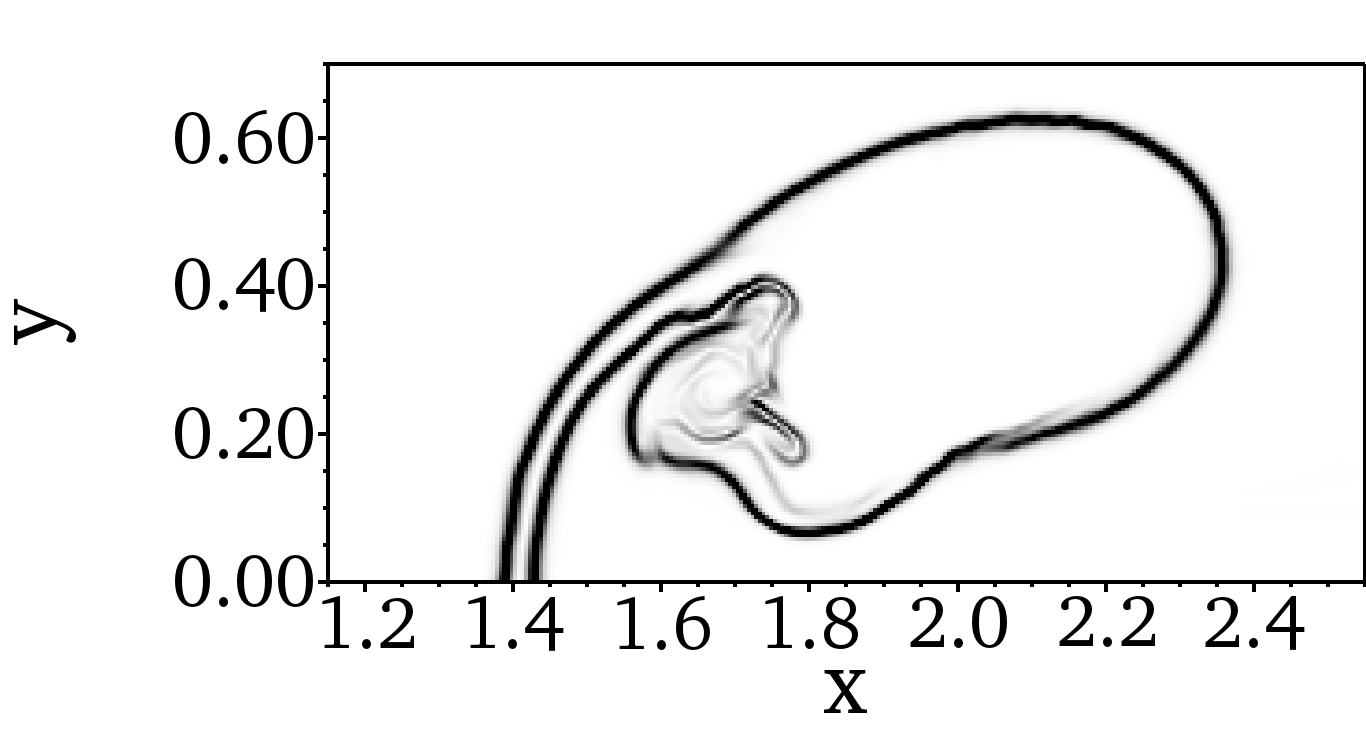}
\label{fig:2D_shock_cylinder_medium_4_90_WCNS5_Z}}
\subfigure[$t=6.95$]{%
\includegraphics[width=0.315\textwidth]{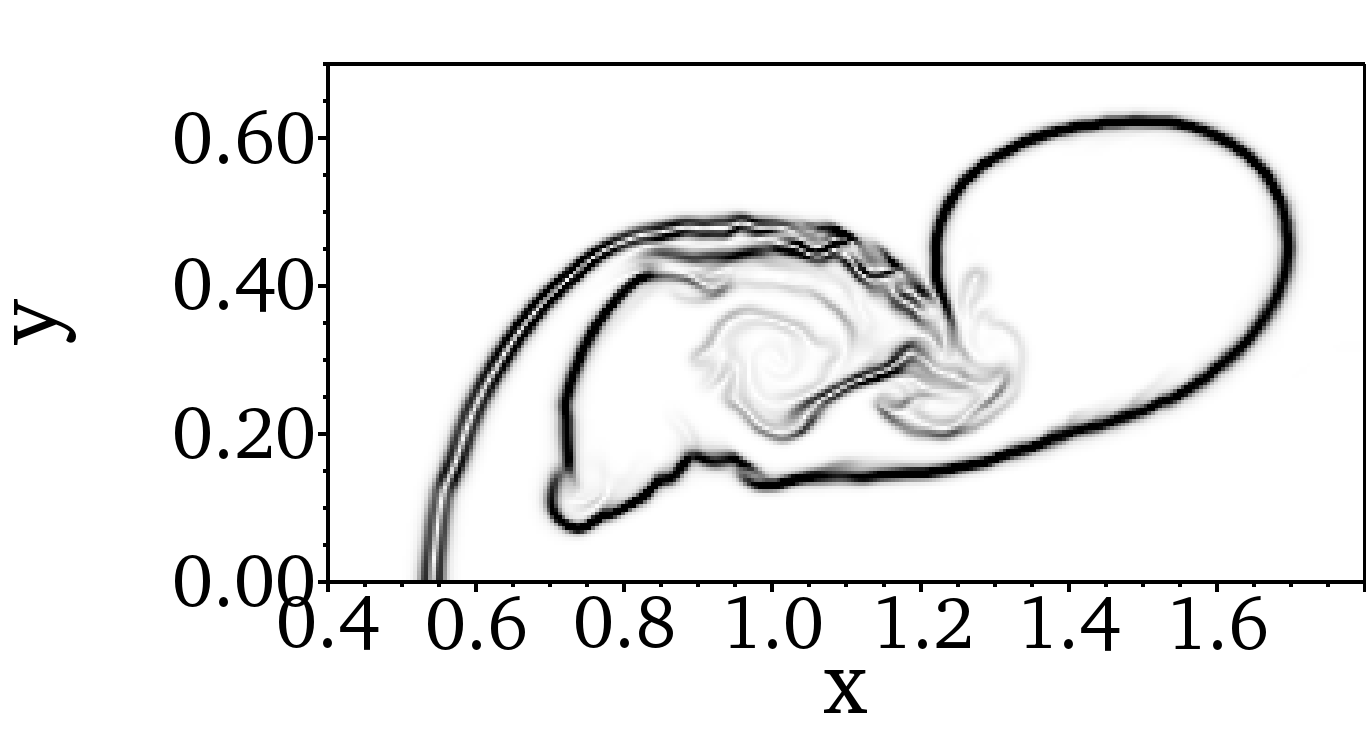}
\label{fig:2D_shock_cylinder_medium_6_95_WCNS5_Z}}

\subfigure[$t=3.25$]{%
\includegraphics[width=0.315\textwidth]{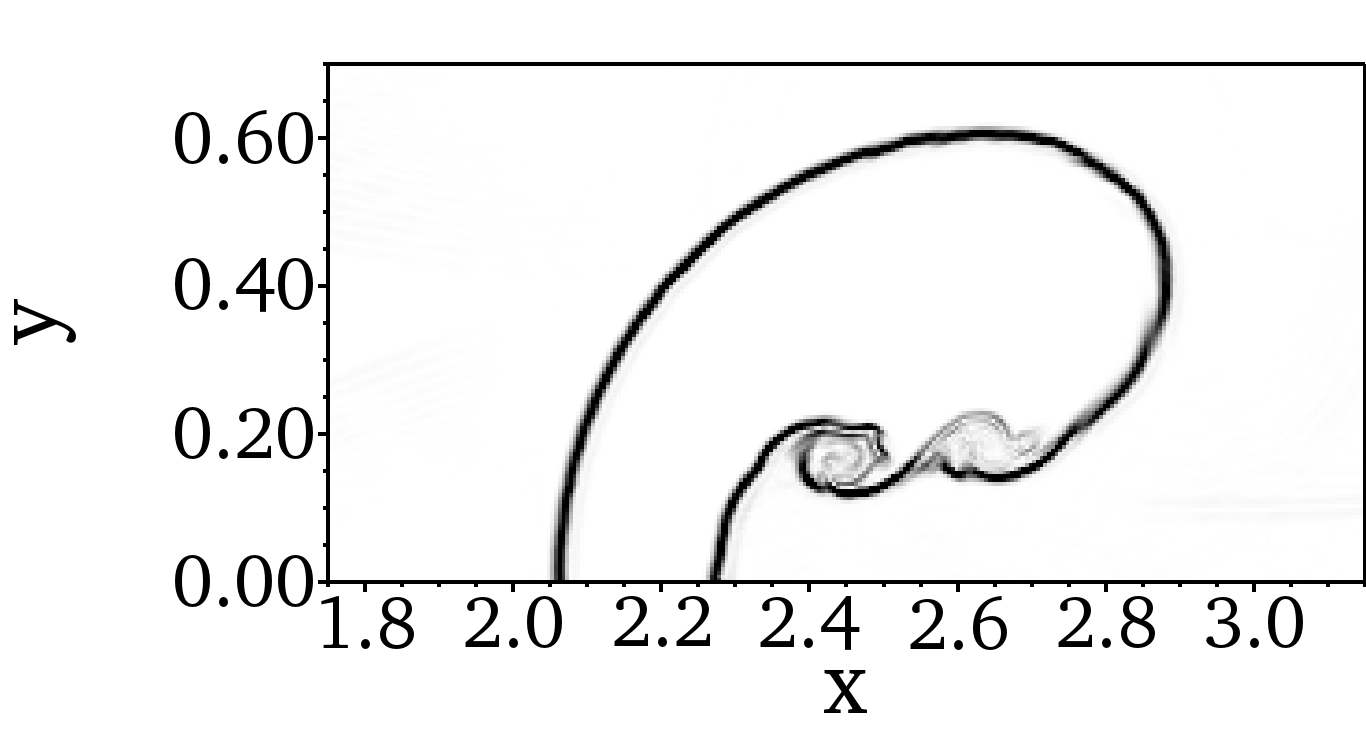}
\label{fig:2D_shock_cylinder_medium_3_25_WCNS6_CU_M2}}
\subfigure[$t=4.90$]{%
\includegraphics[width=0.315\textwidth]{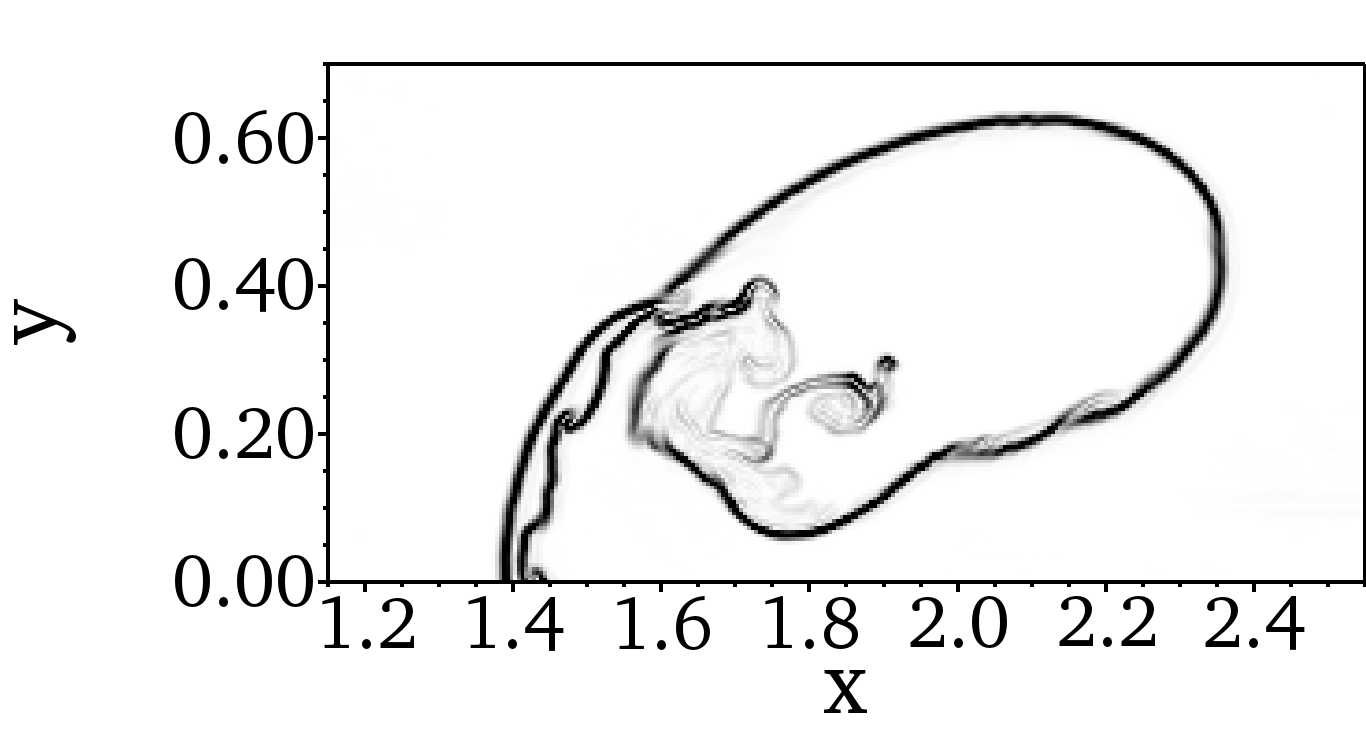}
\label{fig:2D_shock_cylinder_medium_4_90_WCN6_CU_M2}}
\subfigure[$t=6.95$]{%
\includegraphics[width=0.315\textwidth]{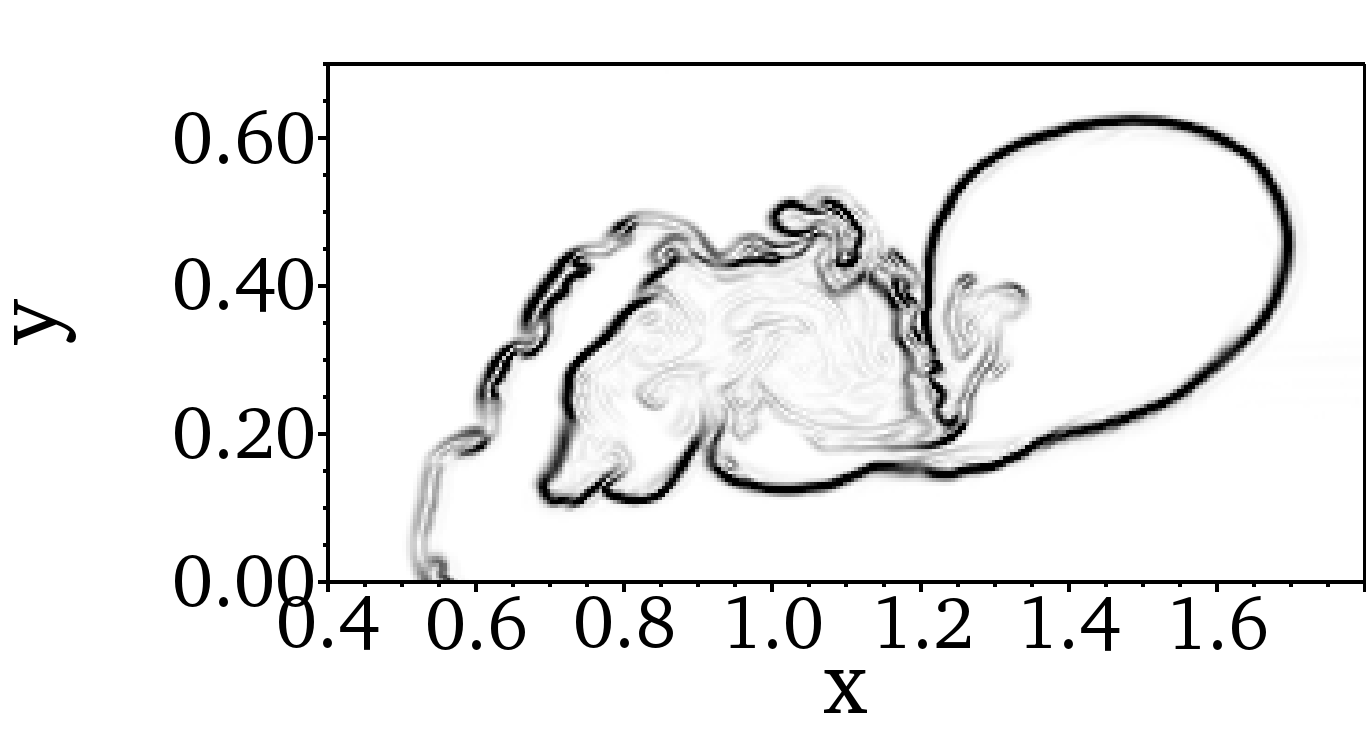}
\label{fig:2D_shock_cylinder_medium_6_95_WCNS6_CU_M2}}

\subfigure[$t=3.25$]{%
\includegraphics[width=0.315\textwidth]{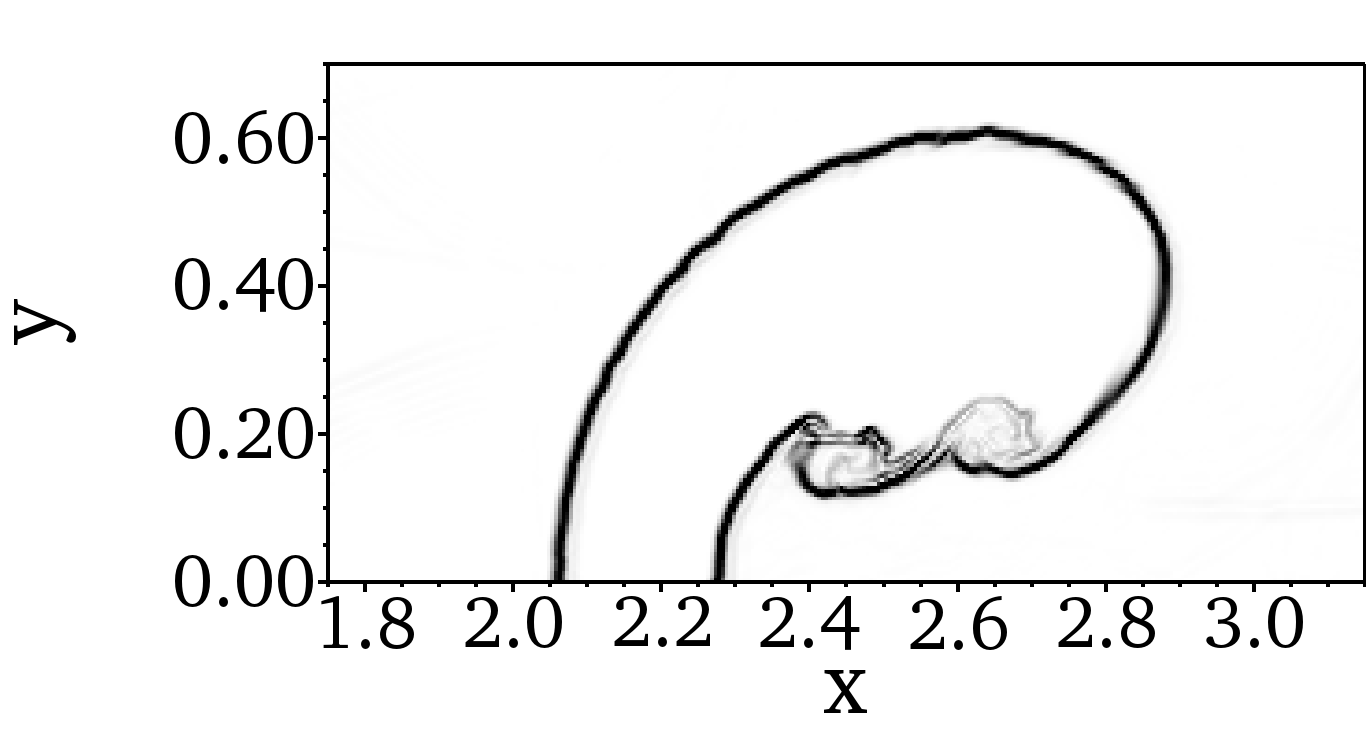}
\label{fig:2D_shock_cylinder_medium_3_25_WCNS6_LD}}
\subfigure[$t=4.90$]{%
\includegraphics[width=0.315\textwidth]{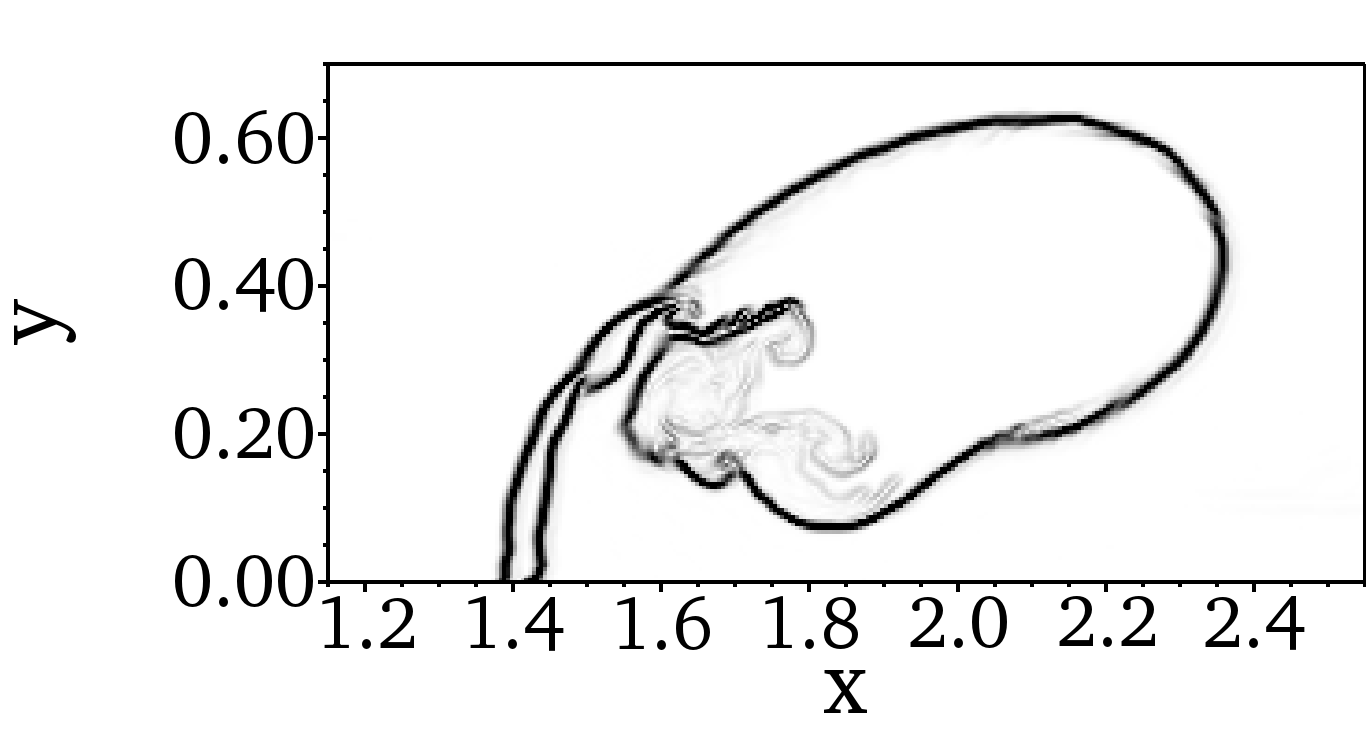}
\label{fig:2D_shock_cylinder_medium_4_90_WCN6_LD}}
\subfigure[$t=6.95$]{%
\includegraphics[width=0.315\textwidth]{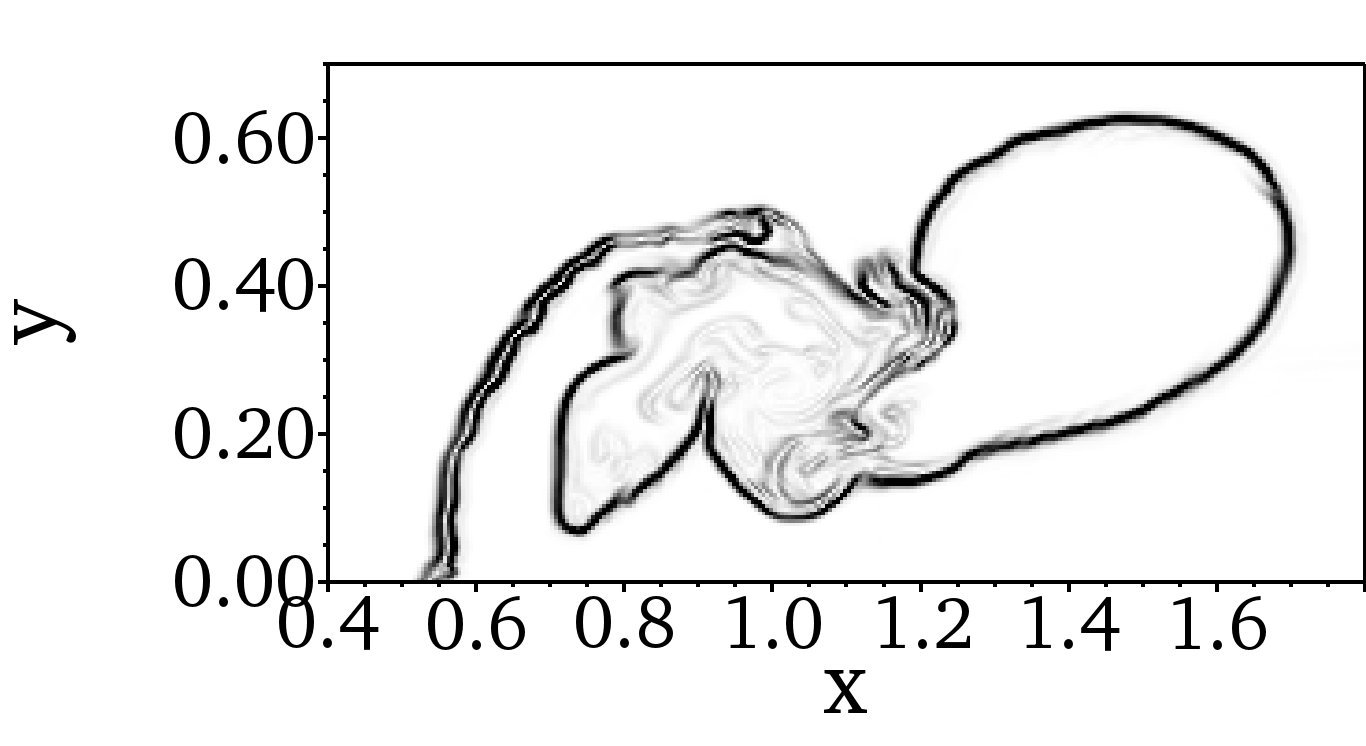}
\label{fig:2D_shock_cylinder_medium_6_95_WCNS6_LD}}

\caption{Nonlinear function of normalized density gradient magnitude, $\phi$, for the shock-cylinder interaction problem. Contours are from 1 to 1.7 at different times using different schemes. Grid spacings are $\Delta x = \Delta y = 1/200$. Top row: WCNS5-JS; second row: WCNS5-Z; third row: WCNS6-CU-M2; bottom row: WCNS6-LD.}
\label{fig:2D_shock_cylinder_density_gradient_medium}
\end{figure}

\begin{figure}[!ht]
\centering
\subfigure[$t=3.25$]{%
\includegraphics[width=0.315\textwidth]{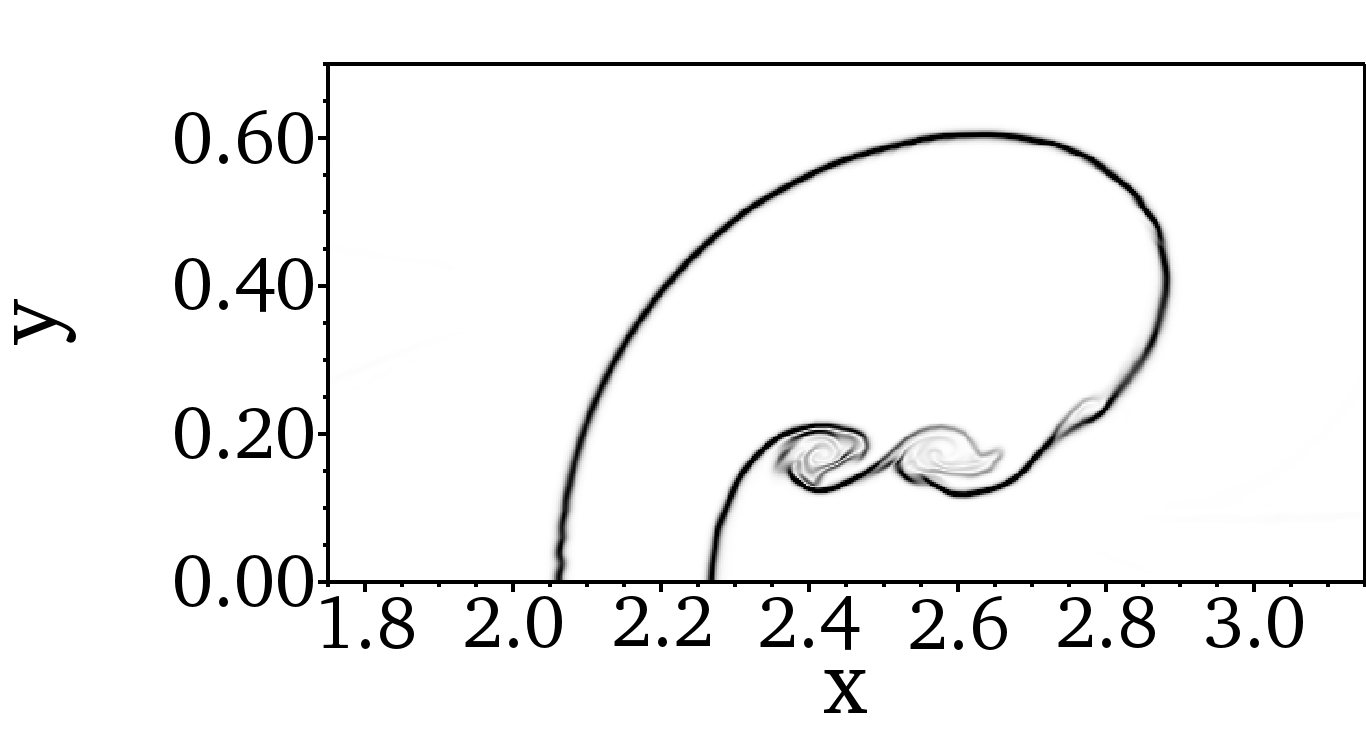}
\label{fig:2D_shock_cylinder_fine_3_25_WCNS5_JS}}
\subfigure[$t=4.90$]{%
\includegraphics[width=0.315\textwidth]{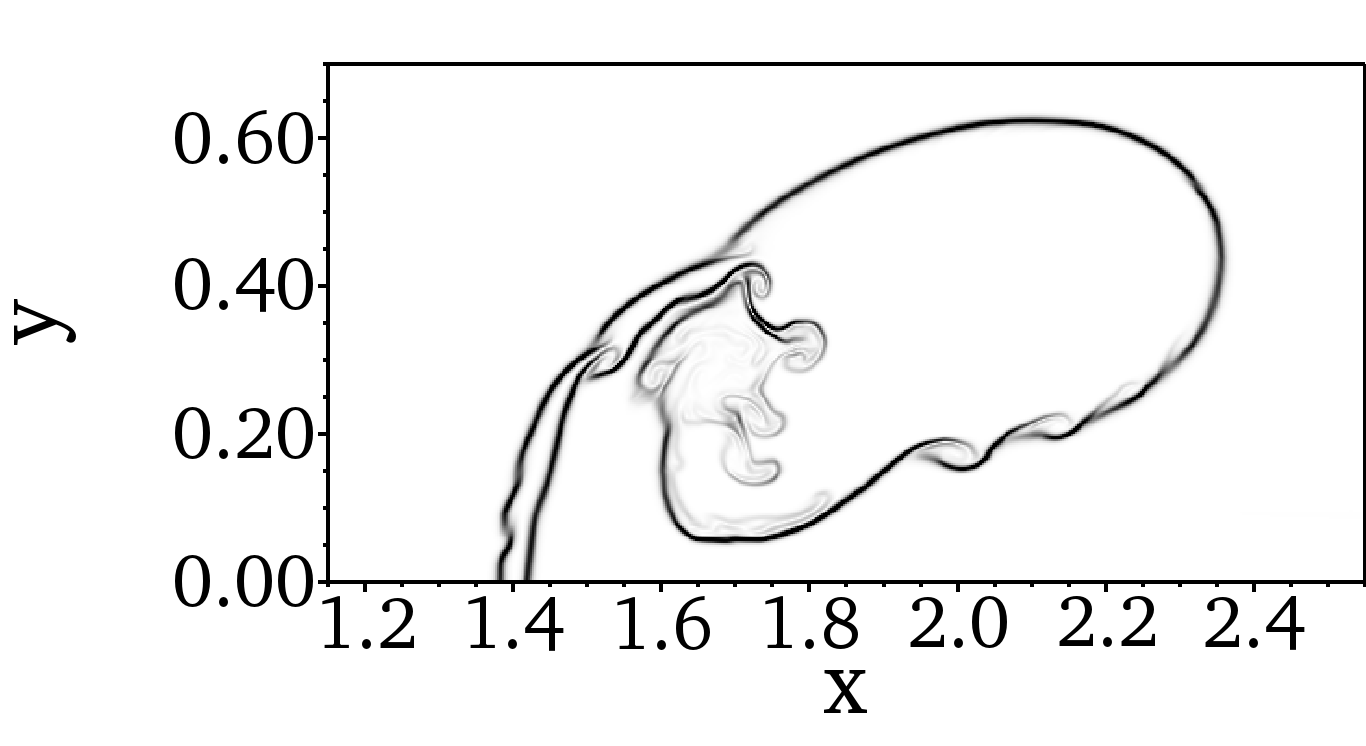}
\label{fig:2D_shock_cylinder_fine_4_90_WCNS5_JS}}
\subfigure[$t=6.95$]{%
\includegraphics[width=0.315\textwidth]{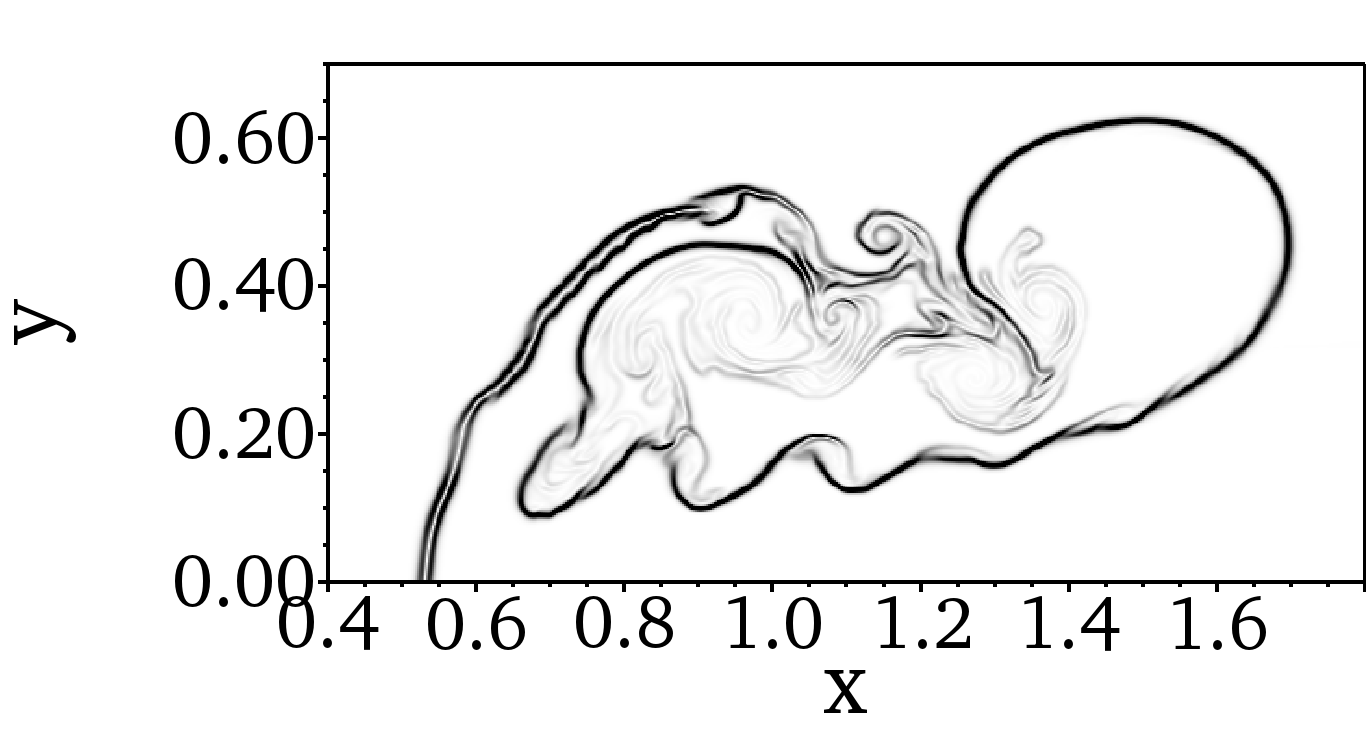}
\label{fig:2D_shock_cylinder_fine_6_95_WCNS5_JS}}

\subfigure[$t=3.25$]{%
\includegraphics[width=0.315\textwidth]{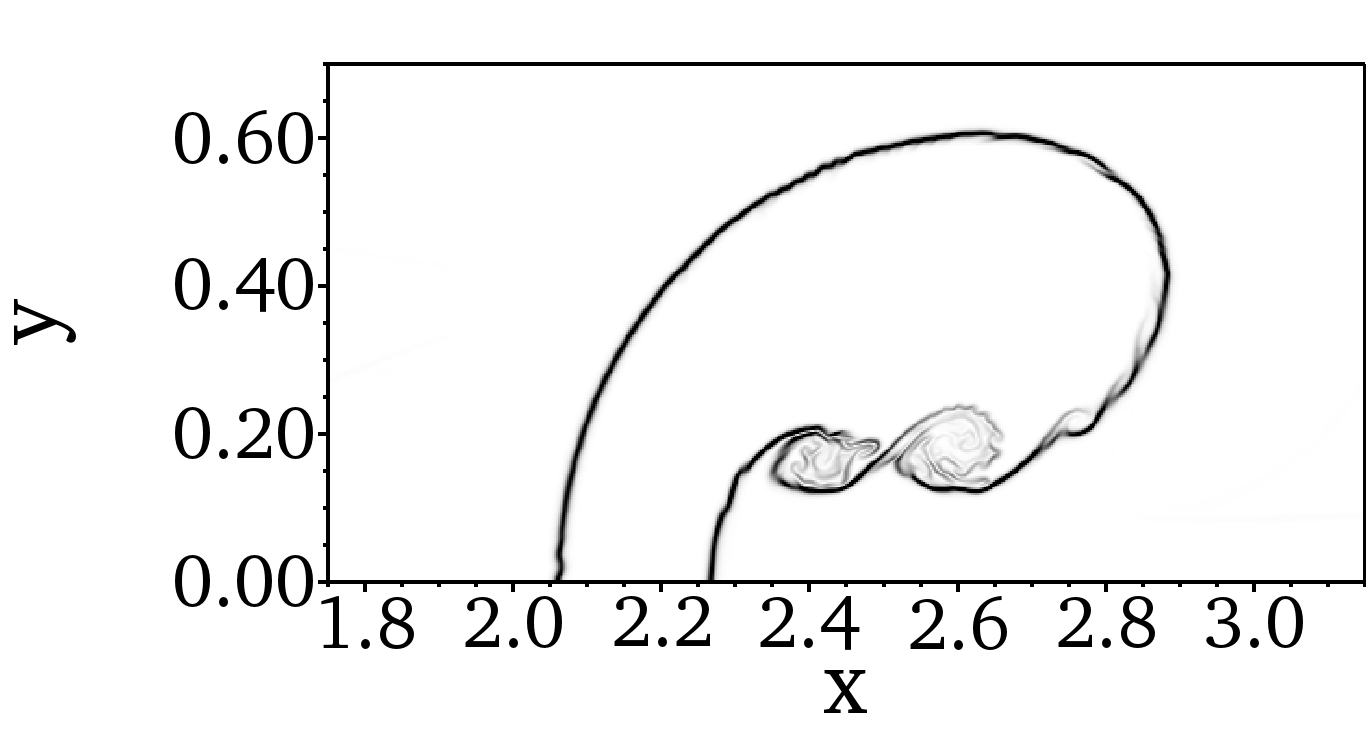}
\label{fig:2D_shock_cylinder_fine_3_25_WCNS5_Z}}
\subfigure[$t=4.90$]{%
\includegraphics[width=0.315\textwidth]{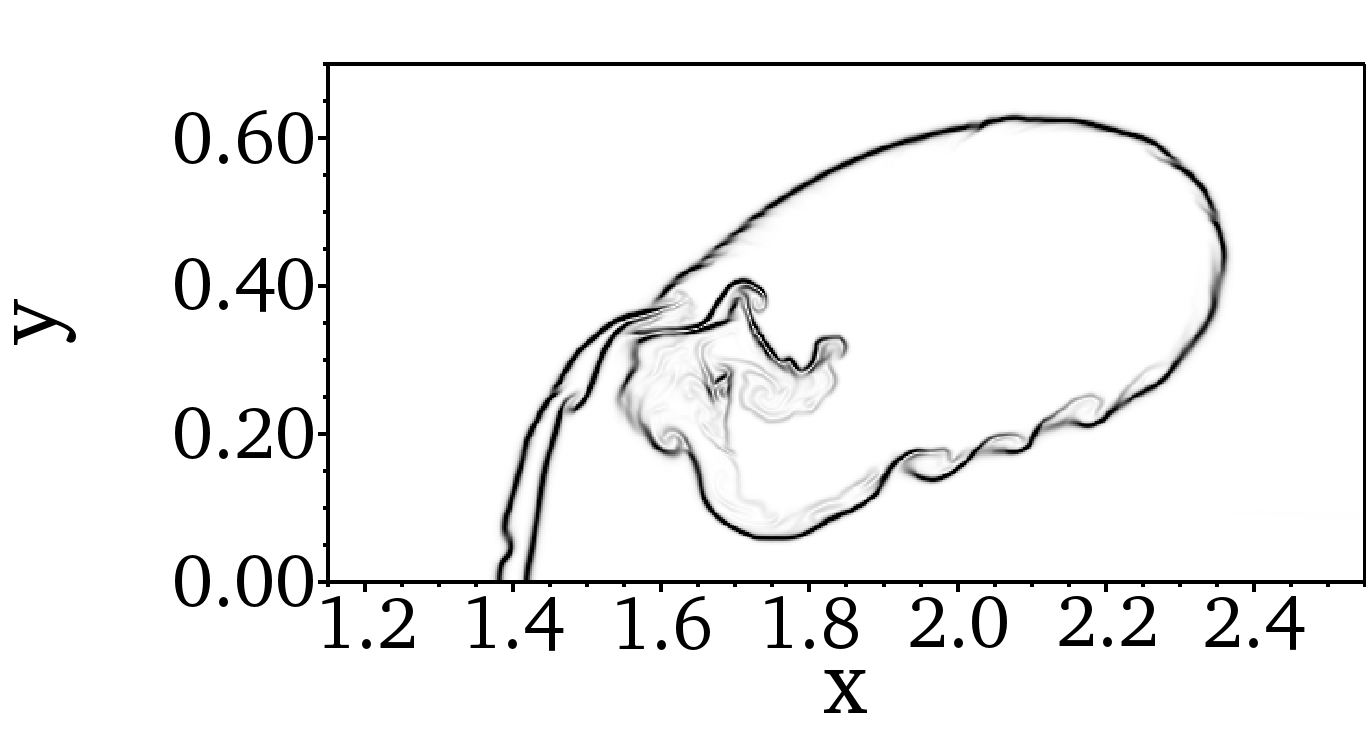}
\label{fig:2D_shock_cylinder_fine_4_90_WCNS5_Z}}
\subfigure[$t=6.95$]{%
\includegraphics[width=0.315\textwidth]{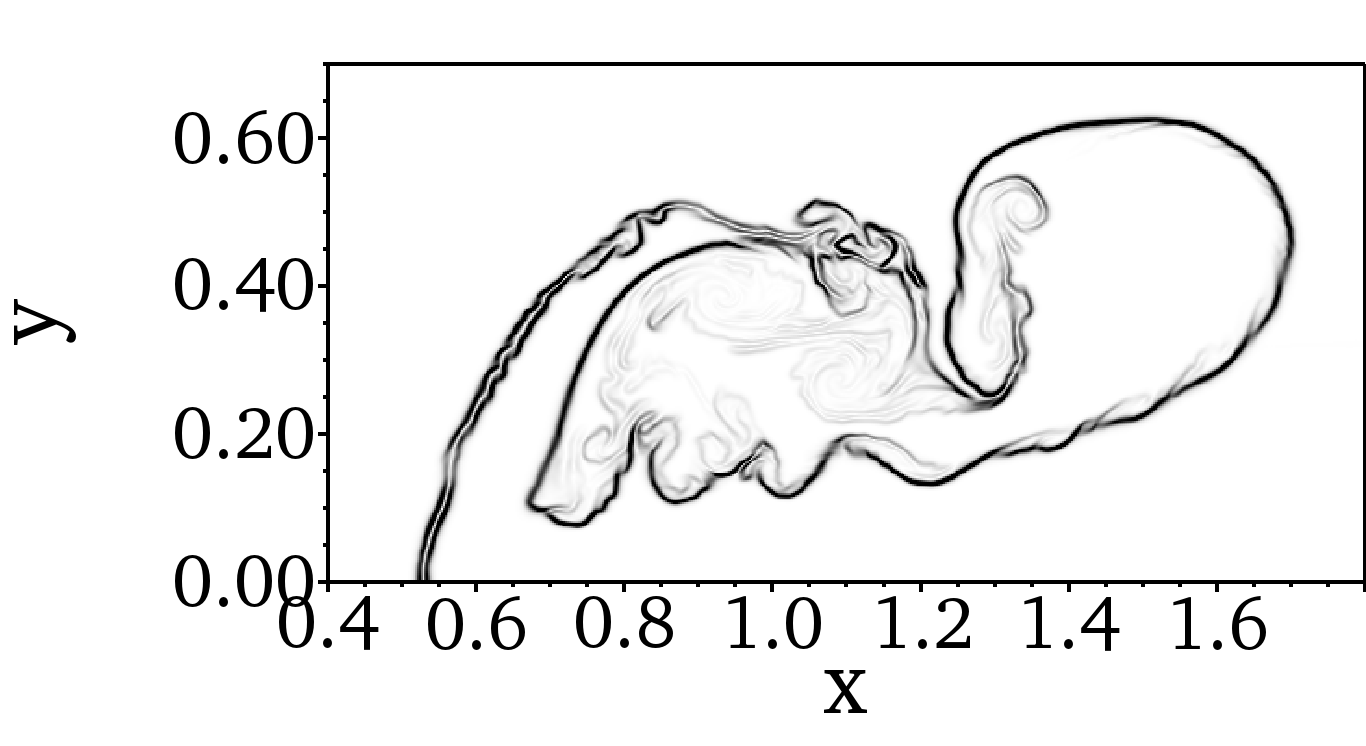}
\label{fig:2D_shock_cylinder_fine_6_95_WCNS5_Z}}

\subfigure[$t=3.25$]{%
\includegraphics[width=0.315\textwidth]{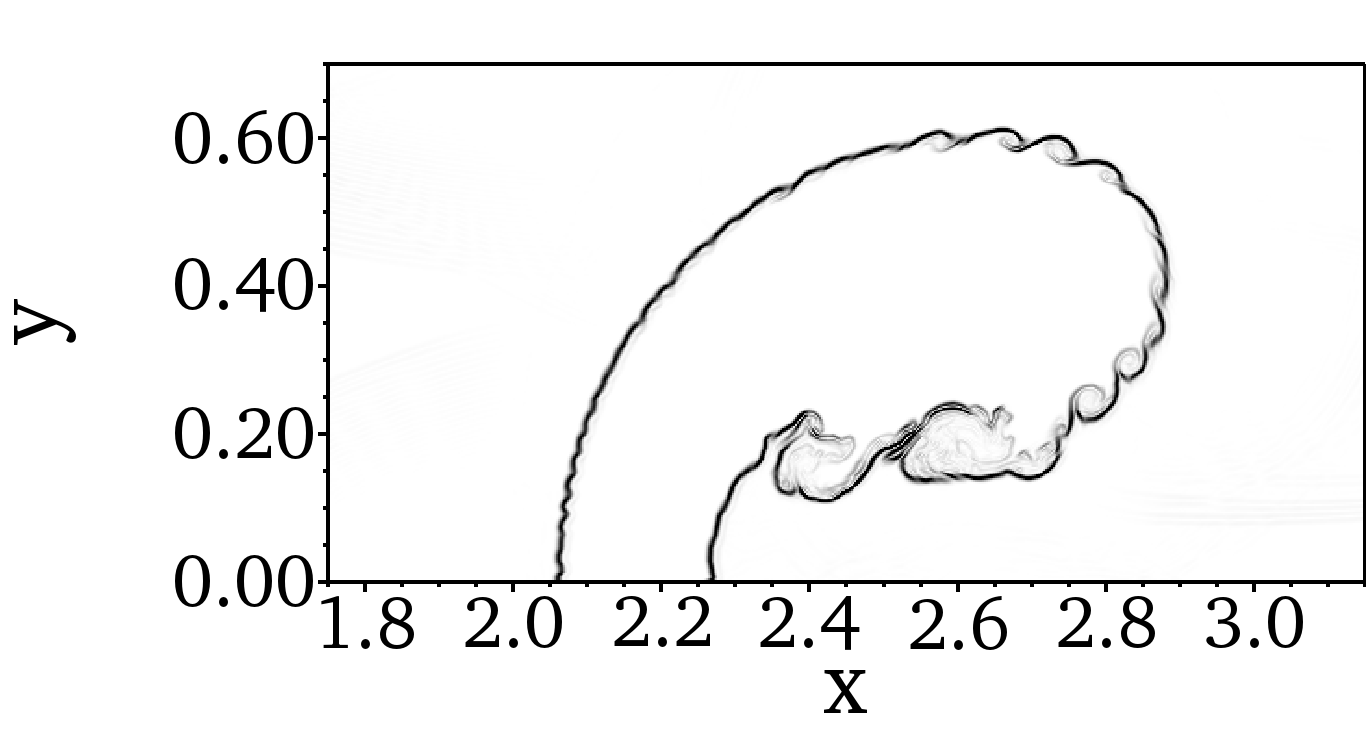}
\label{fig:2D_shock_cylinder_fine_3_25_WCNS6_CU_M2}}
\subfigure[$t=4.90$]{%
\includegraphics[width=0.315\textwidth]{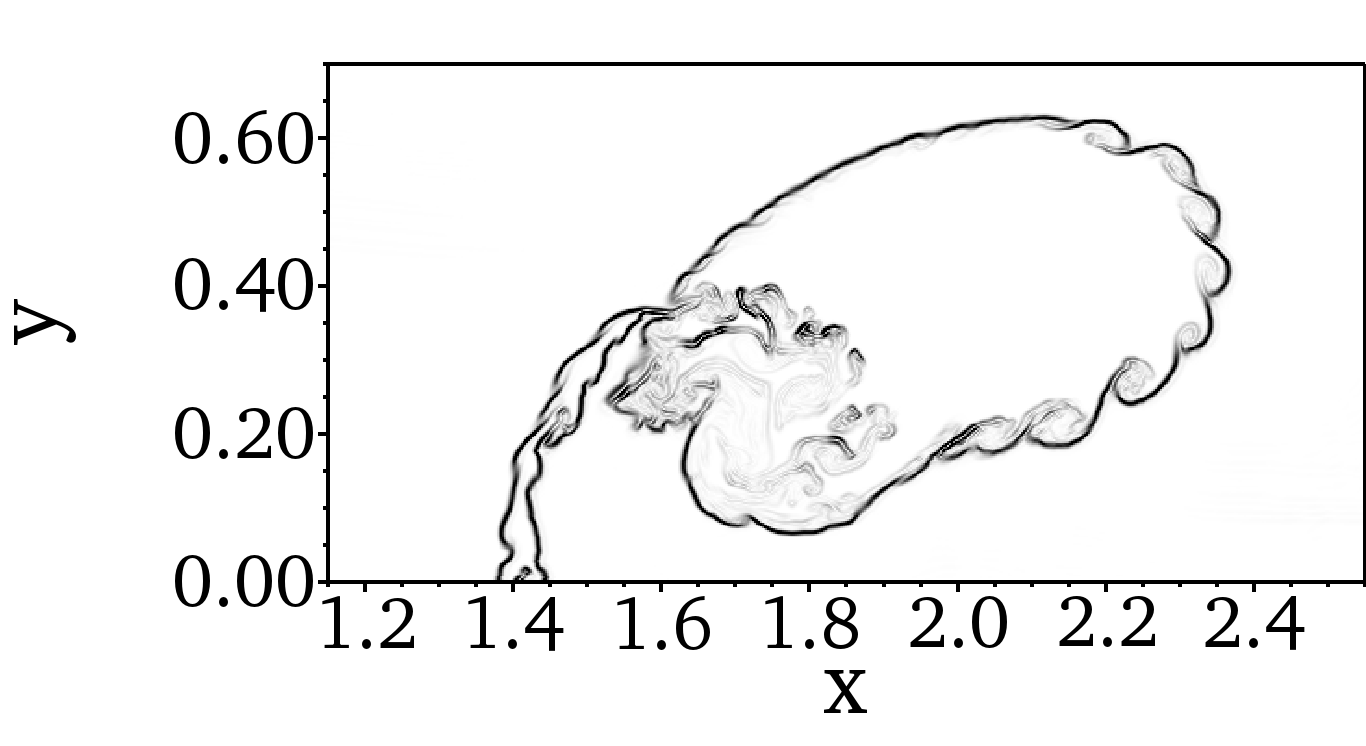}
\label{fig:2D_shock_cylinder_fine_4_90_WCN6_CU_M2}}
\subfigure[$t=6.95$]{%
\includegraphics[width=0.315\textwidth]{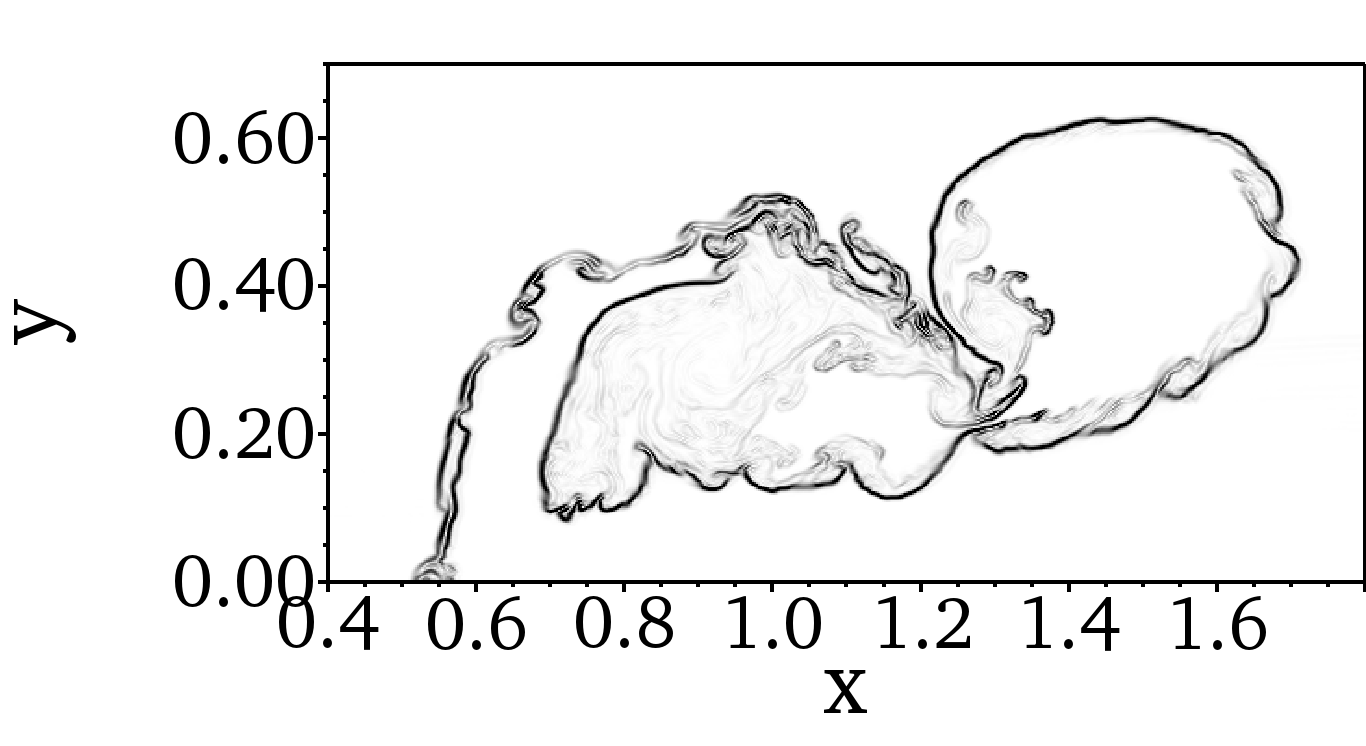}
\label{fig:2D_shock_cylinder_fine_6_95_WCNS6_CU_M2}}

\subfigure[$t=3.25$]{%
\includegraphics[width=0.315\textwidth]{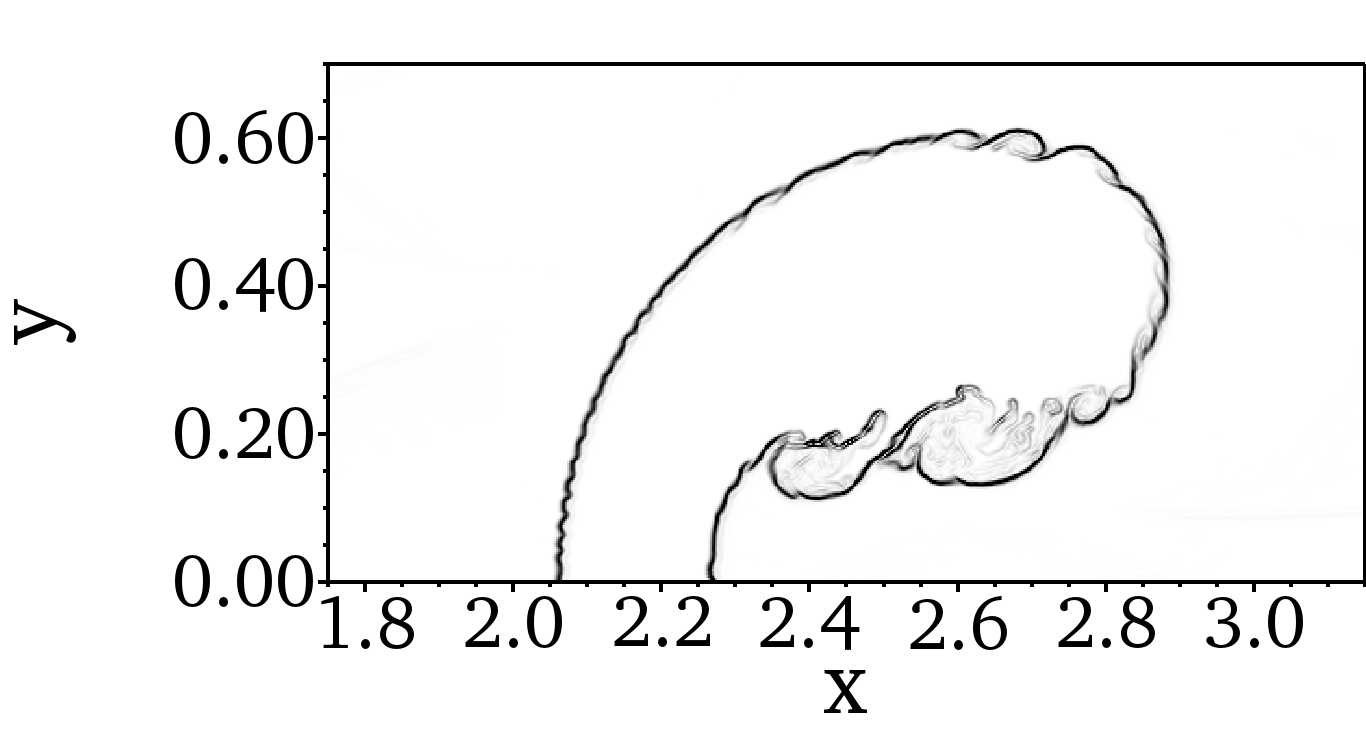}
\label{fig:2D_shock_cylinder_fine_3_25_WCNS6_LD}}
\subfigure[$t=4.90$]{%
\includegraphics[width=0.315\textwidth]{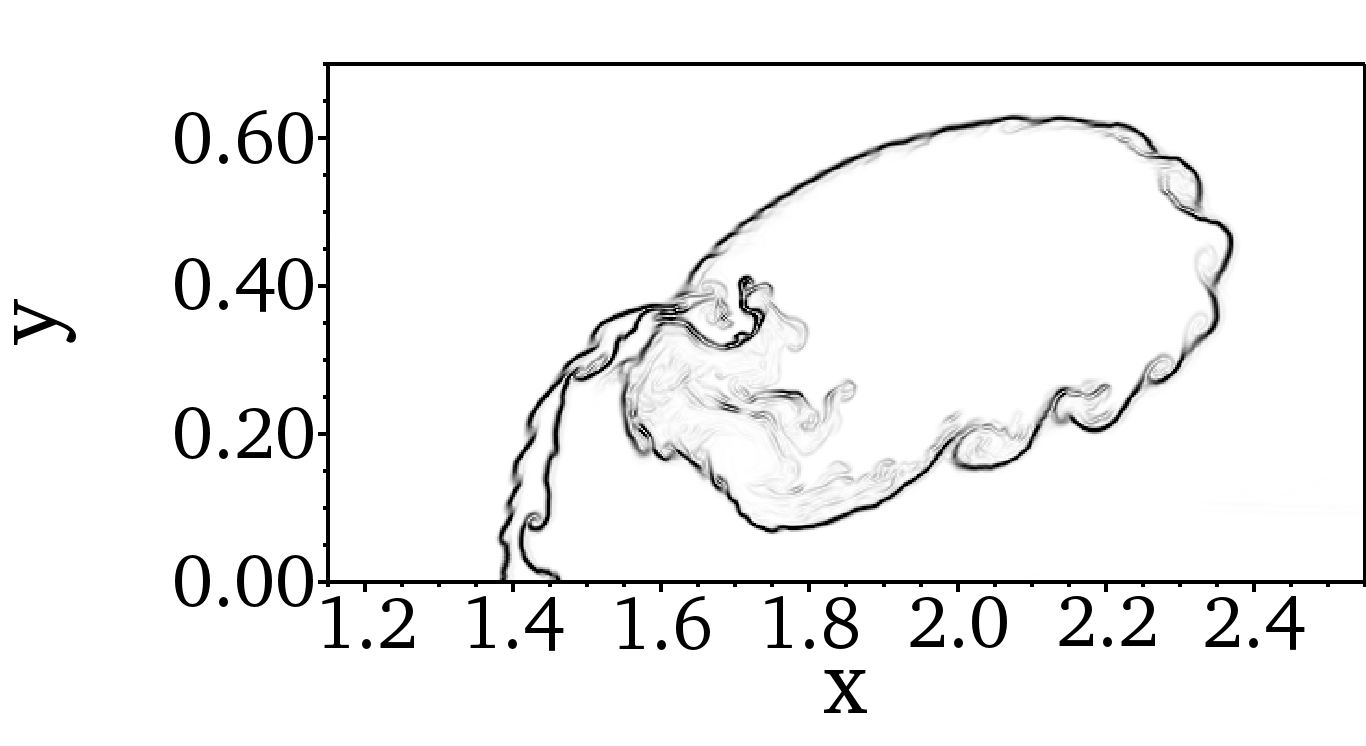}
\label{fig:2D_shock_cylinder_fine_4_90_WCN6_LD}}
\subfigure[$t=6.95$]{%
\includegraphics[width=0.315\textwidth]{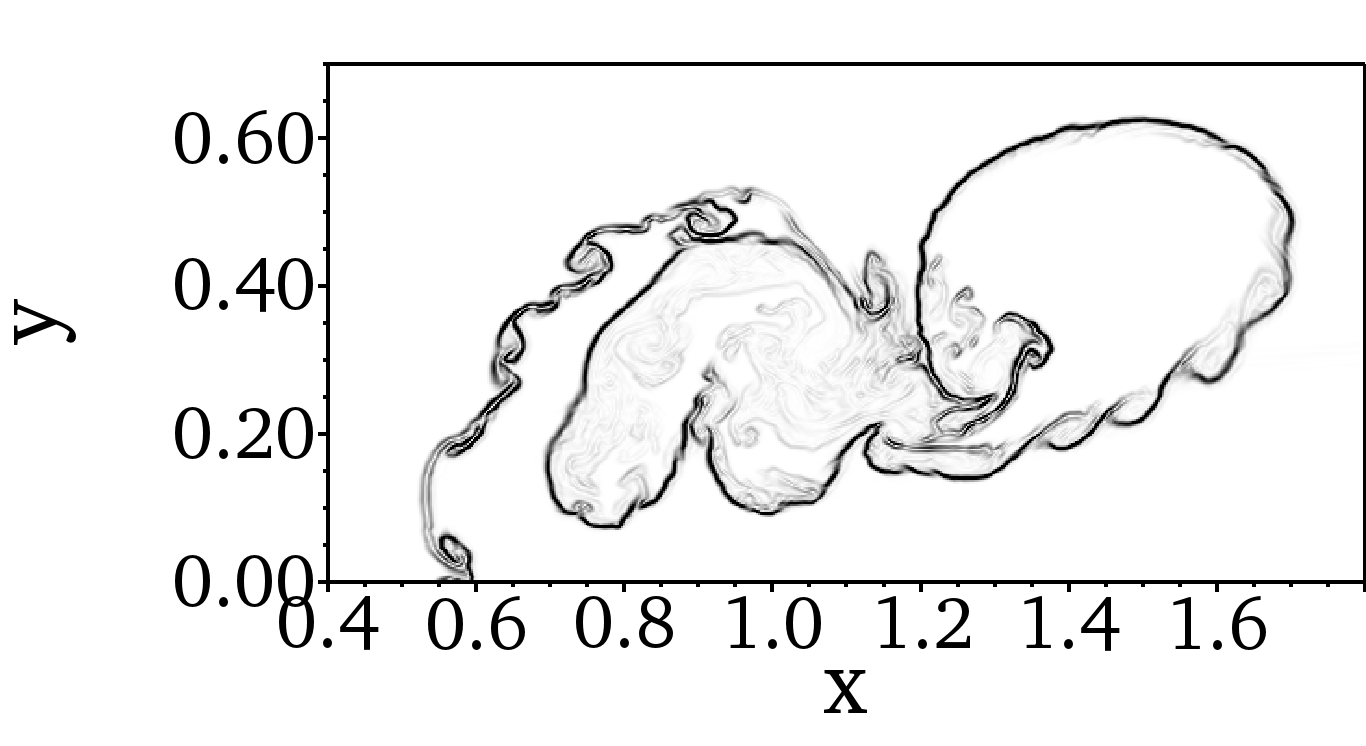}
\label{fig:2D_shock_cylinder_fine_6_95_WCNS6_LD}}

\caption{Nonlinear function of normalized density gradient magnitude, $\phi$, for the shock-cylinder interaction problem. Contours are from 1 to 1.7 at different times using different schemes. Grid spacings are $\Delta x = \Delta y = 1/400$. Top row: WCNS5-JS; second row: WCNS5-Z; third row: WCNS6-CU-M2; bottom row: WCNS6-LD.}
\label{fig:2D_shock_cylinder_density_gradient_fine}
\end{figure}

\begin{figure}[!ht]
\centering
\includegraphics[width=0.82\textwidth]{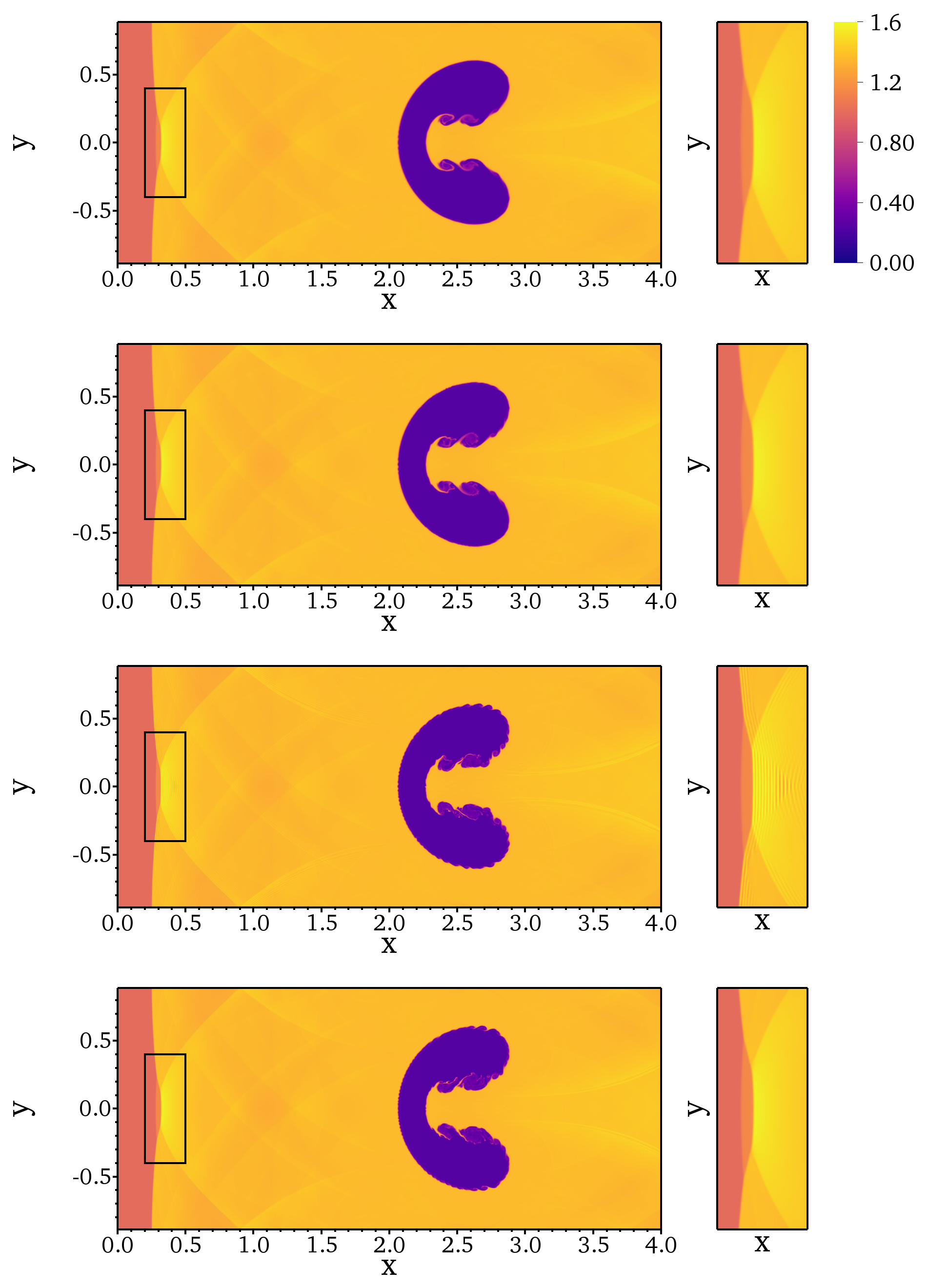}
\caption{Density plots of various schemes for the shock-cylinder interaction problem at $t=3.25$. Top row: WCNS5-JS; second row: WCNS5-Z; third row: WCNS6-CU-M2; bottom row: WCNS6-LD. Grid spacings are $\Delta x = \Delta y = 1/400$. Boxed regions are enlarged and shown on the right.}
\label{fig:compare_2D_shock_cylinder_density}
\end{figure}

Figures~\ref{fig:2D_shock_cylinder_density_gradient_coarse}, \ref{fig:2D_shock_cylinder_density_gradient_medium}, and \ref{fig:2D_shock_cylinder_density_gradient_fine} respectively show the time evolution of the helium cylinder structures from different mesh resolutions at times $t = 3.25$, $4.90$, and $6.95$. Contours of the nonlinear function of density gradient magnitude, 
\newline $\phi = \left( \left| \nabla \rho \right| / \left| \nabla \rho \right|_{\max} \right)$ are shown. Since inviscid simulations are conducted without physical dissipation, we shouldn't expect the solutions to converge with increasing mesh resolutions. Therefore, we can see that as we refine the grid, features in smaller scale are produced. The resolution of the small-scale features depends on the numerical dissipation introduced by the schemes. From the figures, we can see that both WCNS6-CU-M2 and WCNS6-LD are numerically less dissipative to capture the small-scale features up to the same level of resolution over WCNS5-JS and WCNS5-Z. The WCNS5-JS is too numerically dissipative to produce the secondary instabilities at the material interface even with the highest mesh resolution. The improved WCNS5-Z is slightly better than WCNS5-JS in resolving secondary instabilities. Both sixth order WCNS's can add numerical dissipation more locally at the material interfaces to capture the discontinuities more sharply compared with WCNS5-JS and WCNS5-Z. Figure~\ref{fig:compare_2D_shock_cylinder_density} shows the density plots from various schemes at $t = 3.25$. From the figures, we can notice that WCNS5-JS, WCNS5-Z, and WCNS6-LD can capture the left-propagating incident shock wave stably. However, spurious oscillations found in the solutions of WCNS6-CU-M2 near the two triple points of the normal shock indicate that CU-M2 interpolation does not add sufficient numerical dissipation to stabilize solutions around the incident shock.

\section{Conclusions}

\noindent In this paper, the localized dissipation scheme WCNS6-LD was tested in a series of single- and multi-species test problems. Compared with the upwind-biased schemes WCNS5-JS and WCNS5-Z, most tests indicated that WCNS-LD and WCNS6-CU-M2 have similar levels of improvement to preserve smooth but fluctuating features over WCNS5-JS and WCNS-Z. However, unlike WCNS6-CU-M2 that may introduce insufficient numerical dissipation to stabilize discontinuities and high-wavenumber features, WCNS6-LD is as robust as the improved WCNS5-Z in capturing shock waves and contact discontinuities without any noticeable spurious oscillations. This suggests that the proposed localized dissipative interpolation is the most suitable interpolation method for use with WCNS among all the other tested methods for simulating chaotic or turbulent compressible flows involving shock waves and material interfaces. In future work we plan to report results on turbulent flow simulations conducted with the newly proposed WCNS6-LD scheme.

\section{Appendices}

\subsection{Characteristic decomposition}
The choice of variables for WENO interpolation is very important to avoid spurious oscillations across discontinuities. Johnsen et al.~\cite{johnsen2006implementation} showed that if conservative variables are chosen for WENO reconstruction, spurious oscillations will appear at material interfaces. They suggested to interpolate primitive variables in order to maintain pressure and velocity equilibria across interfaces. Furthermore, they recommended that characteristic variables projected from primitive variables should be used for WENO reconstruction and interpolation to avoid the interaction of discontinuities in different characteristic fields. To illustrate how the primitive variables are converted into characteristic variables, we follow Coralic et al.~\cite{coralic2014finite} by first rewriting the conservative equations given by equation~\eqref{eq:conservative_form} in the quasi-linear primitive form:
\begin{equation} \label{eq:quasi-conservative_eqn}
	\frac{\partial{\bm{V}}}{\partial{t}} + \bm{A}(\bm{V})\frac{\partial{\bm{V}}}{\partial{x}} + \bm{B}(\bm{V}) \frac{\partial{\bm{V}}}{\partial{y}} + \bm{C}(\bm{V}) \frac{\partial{\bm{V}}}{\partial{z}} = 0,
\end{equation}
where $\bm{V}$ is the vector of primitive variables, $\bm{A} = \partial{\bm{F}} / \partial{\bm{V}}$, $\bm{B} = \partial{\bm{G}} / \partial{\bm{V}}$, and $\bm{C} = \partial{\bm{H}} / \partial{\bm{V}}$ are the Jacobian matrices. The source term is neglected as it does not affect the characteristic decomposition process. For single-species flow:
\begin{flalign}
	\quad \bm{V} &= \begin{bmatrix} 
       				\rho \\
                    u \\
                    v \\
                    w \\
                    p \\
       			 \end{bmatrix}, \enskip
	\bm{A} = \begin{bmatrix} 
       				u & \rho & 0 & 0 & 0  \\
       				0 & u & 0 & 0 & \frac{1}{\rho}  \\
                    0 & 0 & u & 0 & 0  \\
                    0 & 0 & 0 & u & 0  \\
                    0 & \rho c^2 & 0 & 0 & u  \\
       			 \end{bmatrix}, \enskip
	\bm{B} = \begin{bmatrix}
       				v & 0 & \rho & 0 & 0  \\
                    0 & v & 0 & 0 & 0  \\
                    0 & 0 & v & 0 & \frac{1}{\rho} \\
                    0 & 0 & 0 & v & 0 \\
                    0 & 0 & \rho c^2 & 0 & v  \\
       			 \end{bmatrix}, &&\\\nonumber
    \bm{C} &= \begin{bmatrix}
    				w & 0 & 0 & \rho & 0 \\
                    0 & w & 0 & 0 & 0 \\
                    0 & 0 & w & 0 & 0 \\
                    0 & 0 & 0 & w & \frac{1}{\rho} \\
                    0 & 0 & 0 & \rho c^2 & w \\
    			 \end{bmatrix}. &&
\end{flalign}

\noindent For two-species flow with five-equation model:
\begin{flalign}
	\quad \bm{V} &= \begin{bmatrix} 
       				Z_1 \rho_1 \\
       				Z_2 \rho_2 \\
                    u \\
                    v \\
                    w \\
                    p \\
                    Z_1 \\
       			 \end{bmatrix}, \quad
	\bm{A} = \begin{bmatrix} 
       				u & 0 & Z_1 \rho_1 & 0 & 0 & 0 & 0  \\
       				0 & u & Z_2 \rho_2 & 0 & 0 & 0 & 0  \\
                    0 & 0 & u & 0 & 0 & \frac{1}{\rho} & 0  \\
                    0 & 0 & 0 & u & 0 & 0 & 0  \\
                    0 & 0 & 0 & 0 & u & 0 & 0  \\
                    0 & 0 & \rho c^2 & 0 & 0 & u & 0  \\
                    0 & 0 & 0 & 0 & 0 & 0 & u  \\
       			 \end{bmatrix}, &&\\\nonumber
	\bm{B} &= \begin{bmatrix} 
       				v & 0 & 0 & Z_1 \rho_1 & 0 & 0 & 0  \\
       				0 & v & 0 & Z_2 \rho_2 & 0 & 0 & 0  \\
                    0 & 0 & v & 0 & 0 & 0 & 0  \\
                    0 & 0 & 0 & v & 0 & \frac{1}{\rho} & 0  \\
                    0 & 0 & 0 & 0 & v & 0 & 0 \\
                    0 & 0 & 0 & \rho c^2 & 0 & v & 0  \\
                    0 & 0 & 0 & 0 & 0 & 0 & v  \\
       			 \end{bmatrix}, \quad
    \bm{C} = \begin{bmatrix}
    				w & 0 & 0 & 0 & Z_1 \rho_1 & 0 & 0 \\
                    0 & w & 0 & 0 & Z_2 \rho_2 & 0 & 0 \\
                    0 & 0 & w & 0 & 0 & 0 & 0 \\
                    0 & 0 & 0 & w & 0 & 0 & 0 \\
                    0 & 0 & 0 & 0 & w & \frac{1}{\rho} & 0 \\
                    0 & 0 & 0 & 0 & \rho c^2 & w & 0 \\
                    0 & 0 & 0 & 0 & 0 & 0 & w \\
    			 \end{bmatrix}. &&
\end{flalign}

\noindent The eigenvectors of the Jacobian matrices in the quasi-conservative equations (\ref{eq:quasi-conservative_eqn}) have to be determined first in order to transform primitive variables to characteristic variables. The eigenvalue decompositions of the Jacobian matrices are given by:
\begin{equation}
	\bm{A} = \bm{R_{A}} \bm{\Lambda_{A}} \bm{R}_{\bm{A}}^{-1}, \quad
    \bm{B} = \bm{R_{B}} \bm{\Lambda_{B}} \bm{R}_{\bm{B}}^{-1}, \quad
    \bm{C} = \bm{R_{C}} \bm{\Lambda_{C}} \bm{R}_{\bm{C}}^{-1},
\end{equation}

\noindent where $\bm{R_{A}}$, $\bm{R_{B}}$, and $\bm{R_{C}}$ are matrices whose columns are the right eigenvectors of matrices $\bm{A}$, $\bm{B}$, and $\bm{C}$ respectively. $\bm{\Lambda_{A}}$, $\bm{\Lambda_{B}}$, and $\bm{\Lambda_{C}}$ are matrices whose diagonal elements are the corresponding eigenvalues. Since characteristic decomposition applies in one coordinate direction at a time, only the decomposition in the $x$ direction is illustrated here. For single-species flow:
\begin{flalign}
	\bm{R_{A}} &= \begin{bmatrix}
    				\frac{1}{c^2} & 1 & 0 & 0 & \frac{1}{c^2}  \\
                    -\frac{1}{\rho c} & 0 & 0 & 0 & \frac{1}{\rho c}  \\
                    0 & 0 & 1 & 0 & 0 \\
                    0 & 0 & 0 & 1 & 0 \\
                    1 & 0 & 0 & 0 & 1  \\
       			   \end{bmatrix}, \quad
    \bm{R}_{\bm{A}}^{-1} = \begin{bmatrix}
    				0 & -\frac{\rho c}{2} & 0 & 0 & \frac{1}{2}  \\
       				1 & 0 & 0 & 0 & -\frac{1}{c^2}  \\
                    0 & 0 & 1 & 0 & 0  \\
                    0 & 0 & 0 & 1 & 0 \\
                    0 & \frac{\rho c}{2} & 0 & 0 & \frac{1}{2}  \\
       			   \end{bmatrix}, &&\\\nonumber
    \bm{\Lambda_{A}} &= \begin{bmatrix}
    							u - c & 0 & 0 & 0 & 0 \\
                                0 & u & 0 & 0 & 0 \\
                                0 & 0 & u & 0 & 0 \\
                                0 & 0 & 0 & u & 0 \\
                                0 & 0 & 0 & 0 & u + c \\
                             \end{bmatrix}. &&
\end{flalign}

\noindent For two-species flow with five-equation model:
\begin{flalign}
	\bm{R_{A}} &= \begin{bmatrix}
    				-\frac{Z_1 \rho_1}{2c} & 1 & 0 & 0 & 0 & 0 & \frac{Z_1 \rho_1}{2c}  \\
       				-\frac{Z_2 \rho_2}{2c} & 0 & 1 & 0 & 0 & 0 & \frac{Z_2 \rho_2}{2c}  \\
                    \frac{1}{2} & 0 & 0 & 0 & 0 & 0 & \frac{1}{2}  \\
                    0 & 0 & 0 & 1 & 0 & 0 & 0 \\
                    0 & 0 & 0 & 0 & 1 & 0 & 0 \\
                    -\frac{\rho c}{2} & 0 & 0 & 0 & 0 & 0 & \frac{\rho c}{2}  \\
                    0 & 0 & 0 & 0 & 0 & 1 & 0  \\
       			   \end{bmatrix}, \enskip
    \bm{R}_{\bm{A}}^{-1} = \begin{bmatrix}
    				0 & 0 & 1 & 0 & 0 & -\frac{1}{\rho c} & 0  \\
       				1 & 0 & 0 & 0 & 0 & -\frac{Z_1 \rho_1}{\rho c^2} & 0  \\
                    0 & 1 & 0 & 0 & 0 & -\frac{Z_2 \rho_2}{\rho c^2} & 0  \\
                    0 & 0 & 0 & 1 & 0 & 0 & 0 \\
                    0 & 0 & 0 & 0 & 1 & 0 & 0 \\
                    0 & 0 & 0 & 0 & 0 & 0 & 1  \\
                    0 & 0 & 1 & 0 & 0 & \frac{1}{\rho c} & 0  \\
       			   \end{bmatrix}, &&\\\nonumber
    \bm{\Lambda_{A}} &= \begin{bmatrix}
    							u - c & 0 & 0 & 0 & 0 & 0 & 0 \\
                                0 & u & 0 & 0 & 0 & 0 & 0 \\
                                0 & 0 & u & 0 & 0 & 0 & 0 \\
                                0 & 0 & 0 & u & 0 & 0 & 0 \\
                                0 & 0 & 0 & 0 & u & 0 & 0 \\
                                0 & 0 & 0 & 0 & 0 & u & 0 \\
                                0 & 0 & 0 & 0 & 0 & 0 & u + c \\
                             \end{bmatrix}. &&
\end{flalign}

Consider that if we need to approximate the convective flux in the $x$ direction at midpoint between cell nodes $(x_i, y_j, z_k)$ and $(x_{i+1}, y_j, z_k)$, we first compute the projection matrix $\bm{R_{A}}$ frozen at position $(x_{i+\frac{1}{2}}, y_j, z_k)$ with the Roe average or arithmetic average of $\bm{V}_{i,j,k}$ and $\bm{V}_{i+1,j,k}$. In this work, the arithmetic average is used to save computational cost. The primitive variables are then transformed to characteristic variables by the following equation:
\begin{equation}
	\bm{W}_{i,j,k} = \bm{R}^{-1}_{\bm{A}_{i+\frac{1}{2},j,k}} \bm{V}_{i,j,k}.
\end{equation}

\noindent After $\tilde{\bm{W}}_{i+\frac{1}{2},j,k}^L$ and $\tilde{\bm{W}}_{i+\frac{1}{2},j,k}^R$ are obtained from the WENO interpolation, the primitive variables can be recovered by projecting the characteristic variables back to physical fields:
\begin{equation}
\begin{aligned}
	\tilde{\bm{V}}_{i+\frac{1}{2},j,k}^L &= \bm{R}_{\bm{A}_{i+\frac{1}{2},j,k}} \tilde{\bm{W}}_{i+\frac{1}{2},j,k}^L, \\
    \tilde{\bm{V}}_{i+\frac{1}{2},j,k}^R &= \bm{R}_{\bm{A}_{i+\frac{1}{2},j,k}} \tilde{\bm{W}}_{i+\frac{1}{2},j,k}^R.
\end{aligned}
\end{equation}

\noindent The HLLC-HLL Riemann solver is used to compute the flux $\tilde{\bm{F}}_{i+\frac{1}{2},j,k}$ with $\tilde{\bm{V}}_{i+\frac{1}{2},j,k}^L$ and $\tilde{\bm{V}}_{i+\frac{1}{2},j,k}^R$.

\subsection{HLLC and HLL fluxes}
The HLLC-HLL Riemann solver~\cite{huang2011cures} approximates the convective flux by hybridizing the HLLC~\cite{toro1994restoration} and HLL~\cite{harten1983upstream} fluxes. For simplicity, only the HLLC and HLL flux approximations in the $x$ direction are illustrated in this section. The HLLC flux in $x$ direction is given by:
\begin{equation}
	\mathbf{F}_{\textnormal{HLLC}} = \frac{1+\sign(s_{*})}{2} \left[ \bm{F}_L + s_{-} \left( \bm{Q}_{*L} - \bm{Q}_{L} \right) \right] + \frac{1-\sign(s_{*})}{2} \left[ \bm{F}_{R} + s_{+} \left( \bm{Q}_{*R} - \bm{Q}_{R} \right) \right],
\end{equation}
where $L$ and $R$ are the left and right states respectively. With $K = L$ or $R$, the star state for single-species flow is defined as:
\begin{equation}
		\bm{Q}_{*K} = \chi_{*K}
    \begin{bmatrix}
    	\rho_K \\
        \rho_K s_* \\
        \rho_K v_K \\
        \rho_K w_K \\
        E_k + \left( s_* - u_K \right) \left( \rho_K s_* + \frac{p_K}{s_K - u_K} \right)
    \end{bmatrix}.
\end{equation}
For two-species flow with five-equation model:
\begin{equation}
	\bm{Q}_{*K} = \chi_{*K}
    \begin{bmatrix}
    	\left( Z_1 \rho_1 \right)_K \\
        \left( Z_2 \rho_2 \right)_K \\
        \rho_K s_* \\
        \rho_K v_K \\
        \rho_K w_K \\
        E_K + \left( s_* - u_K \right) \left( \rho_K s_* + \frac{p_K}{s_K - u_K} \right) \\
        {Z_{1}}_K
    \end{bmatrix}.
\end{equation}

\noindent $\chi_{*K}$ is defined as:
\begin{equation}
    \chi_{*K} = \frac{s_K - u_K}{s_K - s_*}.
\end{equation}

\noindent We use the waves speeds suggested by Einfeldt et al.~\cite{einfeldt1991godunov}:
\begin{equation}
	s_{-} = \min{\left( 0, s_L \right)}, \quad s_{+} = \max{\left( 0, s_R \right)},
\end{equation}
and
\begin{equation}
	s_{L} = \min{\left( \bar{u} - \bar{c}, u_L - c_L \right)}, \quad s_{R} = \max{\left( \bar{u} + \bar{c}, u_R + c_R \right)},
\end{equation}
where $\bar{u}$ and $\bar{c}$ are the averages from the left and right states. Arithmetic averages are used in this paper. Following Batten et al.~\cite{batten1997choice}, the wave speed in the star region is given by:
\begin{equation}
	s_{*} = \frac{p_R - p_L + \rho_L u_L \left( s_L - u_L \right) - \rho_R u_R \left( s_R - u_R \right)}{\rho_L \left( s_L - u_L \right) - \rho_R \left( s_R - u_R \right) }.
\end{equation}

\noindent The HLL Riemann solver was proposed by Harten et al.~ \cite{harten1983upstream} and the HLL flux is given by:
\begin{equation}
	\mathbf{F}_{\textnormal{HLL}} =
	\begin{cases}
    	\bm{F}_L , &\mbox{if } s_L \geq 0, \\
        \frac{s_R \bm{F}_L - s_L \bm{F}_R + s_R s_L \left( \bm{Q}_R - \bm{Q}_L \right) }{s_R - s_L} &\mbox{if } s_L \leq 0 \leq s_R, \\
        \bm{F}_R , &\mbox{if } s_R \leq 0.
    \end{cases}
\end{equation}

\subsection{Approximation of velocity at mid-point between cell nodes} \label{velocity_appendix}
The approximated velocity components at any midpoints between cell nodes are consistent with the HLLC flux. For instance, the $x$ component of the velocity is given by:
\begin{equation}
\begin{aligned}
	\tilde{u}_{i+\frac{1}{2},j,k} =& \frac{1+\sign(s_*)}{2} \left[ u_L + s_{-} \left( \chi_{*L} - 1 \right) \right] + \frac{1-\sign(s_*)}{2} \left[ u_R + s_{+} \left( \chi_{*R} - 1 \right) \right].
\end{aligned}
\end{equation}
The $y$ and $z$ components of the velocity can be computed similarly.

\section{Acknowledgements}
\noindent The authors acknowledge the Institute for Computational and Mathematical Engineering of Stanford University for providing computing time of the high performance MPI clusters. We also gratefully acknowledge Mr. Akshay Subramaniam for valuable discussions.

\section*{References}

% produces the bibliography section when processed by BibTeX
\bibliography{references.bib}

\end{document}